\documentclass[twocolumn]{aastex6}
\usepackage{amsmath}
\usepackage{natbib}
\usepackage{graphicx}

\shorttitle{Warm Debris Disks}
\shortauthors{Ballering et al.}

\begin{document}

\title{WHAT SETS THE RADIAL LOCATIONS OF WARM DEBRIS DISKS?}

\author{Nicholas P. Ballering, George H. Rieke, Kate Y. L. Su, Andr\'as G\'asp\'ar}
\affil{Steward Observatory, University of Arizona, 933 North Cherry Avenue, Tucson, AZ 85721, USA}
\email{ballerin@email.arizona.edu}

\begin{abstract}
The architectures of debris disks encode the history of planet formation in these systems. Studies of debris disks via their spectral energy distributions (SEDs) have found infrared excesses arising from cold dust, warm dust, or a combination of the two. The cold outer belts of many systems have been imaged, facilitating their study in great detail. Far less is known about the warm components, including the origin of the dust. The regularity of the disk temperatures indicates an underlying structure that may be linked to the water snow line. If the dust is generated from collisions in an exo-asteroid belt, the dust will likely trace the location of the water snow line in the primordial protoplanetary disk where planetesimal growth was enhanced. If instead the warm dust arises from the inward transport from a reservoir of icy material farther out in the system, the dust location is expected to be set by the current snow line. We analyze the SEDs of a large sample of debris disks with warm components. We find that warm components in single-component systems (those without detectable cold components) follow the primordial snow line rather than the current snow line, so they likely arise from exo-asteroid belts. While the locations of many warm components in two-component systems are also consistent with the primordial snow line, there is more diversity among these systems, suggesting additional effects play a role.
\end{abstract}


\section{Introduction}
\label{sec:introduction}

A debris disk comprises a remnant population of planetesimals on circumstellar orbits and the dust generated by their collisional destruction. While observations of protoplanetary disks show planetary systems in the early stages of formation, debris disks reveal the properties of more mature systems. The spatial structure of a debris disk traces the architecture of the planetary system because planets remove planetesimals from their vicinity. The interpretation of debris disk observations involves connecting the properties of the dust to those of the unseen planetesimals and planets. For recent reviews of debris disk science, see \citet{wyatt2008} and \citet{matthews2014}.

Hundreds of spatially unresolved debris disks have been characterized by the infrared excess observed in the SEDs of their systems---the thermal emission of the debris disk dust. This excess can often be modeled simply with one or two blackbody functions, and it typically takes the form of a cold component ($<$130 K), a warm component ($\sim$190 K), or both \citep{morales2011,ballering2013,chen2014}.\footnote{\citet{su2014} identified five dust components that a debris disk can possess, which, in addition to the warm and cold components described here, also include: a blowout halo of small grains outside of the cold belt \citep{augereau2001,su2005}; exozodiacal dust that is hotter and nearer to the star than the warm dust and emits at $\sim$10 $\micron$ \citep{kennedy2013,ballering2014}; and very hot dust emitting in the near-IR \citep{absil2013,ertel2014} likely composed of nanograins trapped in the stellar magnetic field \citep{rieke2016}.} \citet{kennedy2014} concluded that for most systems the warm and cold components arise from radially distinct distributions of dust (as opposed to being co-located and having different temperatures due to different grain properties).

The cold components are the best-studied parts of debris disk systems. They reside far enough (i.e. $>$ tens of au) from their host stars that some have been resolved, revealing a belt analogous to the Kuiper belt in the solar system. The nature of the warm components is less certain, as they reside closer to the star and cannot easily be spatially resolved. For example, the nearby (7.7 pc) star Fomalhaut hosts a well-studied cold belt that has been resolved at several wavelengths \citep{kalas2005,acke2012,boley2012,macgregor2017}. From analyses of its SED and infrared images, Fomalhaut also hosts a warm component \citep{stapelfeldt2004,su2013}, but obtaining resolved images of this warm component to confirm its properties remains difficult \citep{su2016}. In this study, we draw conclusions about warm components by analyzing the SEDs of a large sample of sources and examining how their properties vary with the properties of their host stars. 

The origin of the warm dust is heavily debated in the literature. Given that these components originate from zones likely well-populated by planets, one might expect the belts to be so strongly disturbed that all traces of their origins are erased. However, \citet{morales2011} found a similarity in the warm belt temperatures among stars of different masses\footnote{\citet{kennedy2014}, on the other hand, found that for stars with effective temperatures greater than $\sim$8500 K there are also a number of notably warmer disks (see their Figure 6). Nevertheless, in this paper we analyze the locations of the dust belts, rather than their temperatures (which depend both on the dust location and grain sizes.)}, suggesting a common underlying structure, possibly related to the water snow line. There are two general hypotheses regarding the source of such structures: (1) the dust is produced in-situ by the collisional processing of a belt of parent bodies analogous to the asteroid belt in the solar system, or (2) the dust is transported inward from an outer reservoir of cold planetesimals. As we will describe later, both of these possibilities predict that the locations of warm components will be set by the snow line (i.e. where water ice condensation/sublimation occurs). However, these hypotheses differ as to whether it is the \textit{primordial} snow line or the \textit{current} snow line that sets the warm dust location. These two snow lines predict different relations between the location of the warm dust and the mass of the host star. By examining the observed trend between warm dust location and stellar mass ($M_\star$), we can determine which snow line (primordial or current) was responsible for setting the dust location, and thus which hypothesis for the origin of the dust is favored.       

\subsection{Hypothesis 1: In-situ Production and the Primordial Snow Line}

If the warm dust is produced in-situ from an exo-asteroid belt, it is expected to occur near the primordial snow line. Several mechanisms predict an enhancement of solid material and planetesimal formation at (or near) the snow line in a protoplanetary disk. Water vapor diffusing outward through the disk will condense at the snow line and increase the density of solid material there \citep{stevenson1988}. Icy, roughly meter-size bodies from the outer disk will migrate inward due to gas drag and will sublimate at the snow line, creating an enhancement of vapor and solids \citep{cuzzi2004}. The increase in the dust-to-gas ratio at the snow line can create a region of lower turbulence in the disk. This leads to lower collisional velocities, promoting planetesimal growth, and creates a gas pressure maximum that traps inward migrating solids \citep{kretke2007,brauer2008}. The resulting population of planetesimals may become the parent bodies in the debris disk, or this may trigger the formation of a giant planet that will stir the planetesimals around it and prevent them from coalescing into a planet. The continued stirring will cause planetesimal collisions that yield warm dust for an extended period of time. This is similar to how the gravitational influence of Jupiter stirs the asteroid belt in the solar system \citep[e.g.][]{petit2001}. In any case, the location of the resulting warm dust is linked to the primordial snow line.

The temperature in the midplane of a protoplanetary disk is set primarily by viscous heating. \citet{min2011} give the following relation for the location of the primordial snow line:
\begin{equation}
\label{eq:primordialline}
    r_\text{SL} \propto {M_\star}^{1/3} \dot{M}^{4/9} {\kappa_R}^{2/9} f^{-2/9} \alpha^{-2/9} {T_\text{ice}}^{-10/9},
\end{equation}
where $\dot{M}$ is the mass accretion rate, $\kappa_R$ is the Rosseland mean opacity, $f$ is the gas-to-dust ratio, $\alpha$ is the turbulent mixing strength, and $T_\text{ice}$ is the ice sublimation temperature. Of these parameters, only $\dot{M}$ is believed to vary significantly with stellar mass and thus is relevant for estimating the form of the $r_\text{SL}$--$M_\star$ relation. The mass accretion rate has been found to vary with stellar mass as $\dot{M} \propto M_\star^2$ over a large range of stellar masses, including the masses of the stars in our sample \citep{calvet2004,muzerolle2005,natta2006}. This implies that $r_\text{SL} \propto {M_\star}^{1.2}$. As we will emphasize later, this relation is shallower than that for the current snow line. Other investigations into the location of the primordial snow line also predict the $r_\text{SL}$--${M_\star}$ relation to be significantly shallower than the current snow line relation \citep{kennedy2008,martin2013}.  

\subsection{Hypothesis 2: Inward Transport and the Current Snow Line}

If, instead, the warm dust is transported inward from an outer reservoir during the present debris disk phase, it is expected to reside at the current snow line. There are two plausible mechanisms for the inward transport. The first mechanism is analogous to that described by \citet{nesvorny2010}, who found that most of the warm dust in the inner region of the solar system originates from the disruption of Jupiter family comets (JFCs). JFCs are dynamically controlled by Jupiter and have orbits with significantly lower eccentricities and smaller aphelia than Halley-type comets or long-period comets.  Simulations show that JFCs likely originate in the Kuiper belt and are dynamically passed inward by the giant planets \citep{levison1997}, with some JFCs arriving on asteroid-like orbits inside that of Jupiter \citep{rickman2017}. When JFC orbits cross the current snow line, they begin to sublimate and eventually disintegrate, releasing dust. Simulations of generic planetary systems also show that a chain of several planets is required to efficiently transport planetesimals inward from an outer reservoir \citep{bonsor2012,bonsor2014}. This is consistent with the idea that the region between the snow line and the cold belt is maintained by one or more planets \citep{su2013}.

For the second mechanism, dust generated by collisions in the outer parent body belt flows inward due to Poynting-Robertson (P-R) drag and stellar wind drag. While \citet{wyatt2005} argued that most disks we can detect are collision-dominated rather than drag-dominated (that is, grains are destroyed by mutual collisions faster than they can move inward), \citet{kennedy2015} noted that some inward transport is inevitable unless planets are present interior to the cold belt to remove the inflowing dust. Since these grains originate in the outer part of the system, they may contain a mixture of icy and refractory material. When the grains reach the snow line, the ices sublimate, reducing the grain size and consequently increasing the ratio of the radiation force to the gravitational force on the grain ($\beta$). This halts the grain's inward motion and eventually causes it to be expelled outward. The net result is a pile-up of grains at the location of the current snow line \citep{kobayashi2008}. However, numerical simulations \citep{vanlieshout2014} indicate that the inward flow via this mechanism may be inadequate to maintain the amount of warm dust required for detectable infrared excesses with \textit{Spitzer}, even with the snow line pile-up.

The location of the current snow line is determined by the incident stellar flux, so it scales as $r_\text{SL} \propto L_\star^{1/2}$, where $L_\star$ is the stellar luminosity. Combining this with $L_\star \propto M_\star^4$, the typical relation between stellar luminosity and mass, yields $r_\text{SL} \propto M_\star^2$. Importantly, the current snow line relation is steeper than the primordial snow line relation (index of 2.0 versus 1.2). 

\subsection{Overview}

In this paper, we analyze the SEDs of a sample of debris disks with warm components and infer the stellocentric locations of the warm dust. The dust location derived solely from an SED is subject to many uncertainties and cannot be determined absolutely for any given system. Therefore we focus our attention on the relative behavior of dust location with stellar mass, while holding all other parameters (e.g., grain materials) constant---except for the minimum grain size, which is known to vary systematically with stellar properties. We examine the $r_\text{dust}$--$M_\star$ trend and compare the slope with those predicted by the primordial and current snow lines, providing insight into which snow line sets the dust location. From this we can deduce the the origin of the warm dust components.

\section{Methods}

\subsection{Target Selection}
\label{sec:targets}

For our sample, we used the systems with a warm component found by \citet{ballering2013}, where ``warm" was defined as warmer than 130 K. All of these systems have data available from the Multiband Imaging Photometer for \textit{Spitzer} (MIPS; \citealp{rieke2004}) at 24 and 70 $\micron$ and from the {\it Spitzer} Infrared Spectrograph (IRS; \citealp{houck2004}).

Throughout the analysis, we separated the systems with only a warm component from those that also possess a detected cold component. The systems without a detected cold component should provide less ambiguous results, since these warm components could not arise from particles moving inward from cold belts via P-R drag, although they could still arise from comets originating in cold components that are below the current detection limit \citep{wyatt2007b}. In addition, modeling systems with a single dust component is less complicated and the results have less uncertainty.

\citet{ballering2014} discovered silicate emission features in the IRS spectra of 22 of these systems. These features revealed the presence of exozodiacal dust, which is believed to reside at a different location than the typical warm component. \citet{ballering2014} found that, besides the exozodiacal dust, an additional colder component was also required to fit the full IRS spectra of these sources. Whether this remaining excess consists of one or multiple components is difficult to determine. Thus, to ensure a pure sample of warm components, we excluded these 22 targets. If, however, the dust components giving rise to these features are a natural extension of standard warm components, then excluding these targets may introduce a bias to our sample.

We also excluded HIP 32435 (HD 53842) because the IRS data may have been contaminated by background sources \citep{donaldson2012}. We removed additional targets in the course of our fitting procedure, as described in Section \ref{sec:fitting}. The remaining 83 targets used for our analysis are listed in Table \ref{table:warmtargetlist}.

\subsection{Stellar Properties}

The stellar temperature ($T_\star$), luminosity, and distance from Earth ($d$) of most of our targets were taken from \citet{mcdonald2012}, who derived $T_\star$ and $L_\star$ by fitting the visible and near-IR photometry of these systems with stellar SED models. We then obtained the stellar mass ($M_\star$) from $L_\star$ using the (broken) power-law relation by \citet{eker2015}. \citet{mcdonald2012} assumed a 10\% uncertainty on the photometry they used to derive $L_\star$, so we also assumed a 10\% uncertainty on $L_\star$. Combining this with the intrinsic 25--38\% scatter in the $L_\star$ values \citet{eker2015} used to derive their relations yields a $\sim$6-10\% uncertainty on our $M_\star$ values. For the targets not listed in \citet[][denoted with an ``a" after the target name in Table \ref{table:warmtargetlist}]{mcdonald2012}, we inferred their stellar properties from their $V-K$ color using the tabulated values maintained online\footnote{\url{http://www.pas.rochester.edu/~emamajek/EEM_dwarf_UBVIJHK_colors_Teff.txt}} by E. Mamajek as an expanded and updated version of Table 5 in \citet{pecaut2013}.

We required a model spectrum of the stellar photosphere for each of our targets, both for modeling the photospheric contribution to the observed SED and for calculating the temperature of dust grains when generating model spectra of the dust emission. We used an ATLAS9 \citep{castelli2004} photosphere model with log $g$ = 4.0, solar metallicity, and $T_\star$ closest to that for each target (at most a difference of 125 K). These photosphere model spectra were modeled only out to 160 $\micron$, so for completeness we extended them to 10,000 $\micron$ by extrapolating with a Rayleigh-Jeans power-law. We normalized the integrated spectra to $L_\star$ for each target. (Although during the fitting process we allowed the amplitude of the model photosphere to vary by a small amount to improve agreement with the photometry; see Section \ref{sec:fitting}.)  

\subsection{IRS Data}

While many infrared excesses have been identified from photometric measurements alone \citep{rieke2005,su2006,wyatt2008,matthews2014,sierchio2014}, accurately measuring the temperature/location of the emitting dust requires the wide spectral coverage offered by IRS.

We obtained low-resolution IRS spectra for our targets from the LR7 release of The Combined Atlas of Sources with \textit{Spitzer} IRS Spectra\footnote{The Combined Atlas of Sources with \textit{Spitzer} IRS Spectra (CASSIS) is a product of the IRS instrument team, supported by NASA and JPL. http://cassis.sirtf.com/} (CASSIS; \citealp{lebouteiller2011}). Both long-low orders (LL1: 19.5--38.0 $\micron$; LL2: 14.0--21.3 $\micron$) were available for all targets, and one or both of the short-low orders (SL1: 7.4--14.5 $\micron$; SL2: 5.2--7.7 $\micron$) were also available for most of the targets. The IRS Astronomical Observation Requests numbers (AORs) for our targets are given in Table \ref{table:warmtargetlist}.

We removed outlying points more than 3$\sigma$ away from a third-degree polynomial fit to the measurements in each spectral order. To remove offsets between orders, we multiplied the LL1, SL1, and SL2 flux density values by correction factors (determined by eye), to align them with the LL2 order and to each other. The choice to pin the other orders to LL2 was arbitrary but had no effect on the results because, as described in Section \ref{sec:fitting}, the amplitude of the whole IRS spectrum was varied as part of the fitting process. These correction factors (designated $x_\text{LL1}$, $x_\text{SL1}$, and $x_\text{SL2}$) are listed in Table \ref{table:warmtargetlist}. During this process we opted to remove HIP 79631 from our sample because the offsets between the orders were much greater than for any other target, suggesting the data may be unreliable.

\subsection{IR and Sub-mm Photometry}

In addition to the IRS data, we included in our SEDs photometry from MIPS at 24 and 70 $\micron$ plus additional photometry at wavelengths $\geq$70 $\micron$ from the literature. These data are listed in Table \ref{table:photometry}. Upper limits are at the 3$\sigma$ level.

\subsection{Modeling Dust Emission}
\label{sec:models}

Careful restriction of the model characteristics was necessary to avoid degeneracies that would undermine our ability to determine the trend of warm dust location with stellar mass. We assumed that the dust lies in a circular ring. To test the sensitivity of our conclusions to the particular ring geometry we used two different models: one with a constant ring width (independent of radius) and a second with a ring width that was a constant fraction of the ring radius. We discuss the first set of models here and the second set in Section \ref{sec:testassumptions} where we explore the robustness of our results. For these models, we assumed that the ring has $r_\text{out} = r_\text{in}$ + 2 au ($r$ is the stellocentric distance). We modeled the surface number density of grains as $\Sigma(r) \propto r^{-p}$ with $p$ = 1, but varying this parameter within reasonable bounds had virtually no effect on our results. We assumed a power-law grain size distribution ($n(a) \propto a^{-q}$ where $a$ is the grain radius) with size index $q=3.65$ \citep{gaspar2012}, minimum grain size $a_\text{min}=a_\text{BOS}$ ($a_\text{BOS}$ is the blowout size), and maximum grain size $a_\text{max}=1000$ $\micron$. Grains larger than $a_\text{max}$ contribute negligibly to the overall emission. We assumed a grain composition of 60\% astronomical silicates and 40\% organic refractory material by volume \citep{ballering2016}. To test how robust our conclusions are to grain composition, we report on similar models with grains composed entirely of astronomical silicates in Section \ref{sec:testassumptions}. The only free parameters in our dust belt models were $r_\text{in}$ and the total mass of dust.

Fixing the minimum grain size to the predicted blowout size was an important step in the modeling because $a_\text{BOS}$ varies systematically with stellar type and $a_\text{min}$ can have a substantial effect on the derived dust location (a smaller value of $a_\text{min}$ leads to a larger inferred $r_\text{in}$). That is, we do not simply link the dust temperature and location using the blackbody temperature. The choice to use $a_\text{min}$ = $a_\text{BOS}$ is justified theoretically and also empirically. Detailed fits to debris disk spectra exhibiting silicate emission features were able to constrain both the dust location and minimum grain size; these fits showed the minimum spherical grain size to be consistent with the blowout size \citep{ballering2014}. Furthermore, \citet{booth2013} compared the sizes of cold debris disks, as measured from their resolved \textit{Herschel} images and as derived from fitting blackbody functions to their SEDs, and found that the differences could be largely explained by models assuming $a_\text{min}$ = $a_\text{BOS}$, although in a similar analysis \citet{pawellek2014} found $a_\text{min}$ values somewhat larger than $a_\text{BOS}$.

The blowout size is the largest grain size for which $\beta>0.5$, where $\beta$ is the ratio of the radiation force to the gravitational force on a grain. For spherical grains, $\beta$ is given by
\begin{equation}
\label{eq:beta}
\beta = \frac{3L_\star}{16\pi GM_\star ac\rho}\frac{\int_0^\infty Q_\text{pr}(\lambda,a) F_{\lambda \star}(\lambda) \,\mathrm{d}\lambda}{\int_0^\infty F_{\lambda \star}(\lambda) \,\mathrm{d}\lambda},
\end{equation}
where $F_{\lambda \star}(\lambda)$ is the stellar flux density, $Q_\text{pr}(\lambda,a)$ is the radiation pressure efficiency on the grain, $G$ is the gravitational constant, $c$ is the speed of light, and $\rho$ = 2.34 g cm$^{-3}$ is the grain density (2.7 g cm$^{-3}$ for the astronomical silicates and 1.8 g cm$^{-3}$ for the organic refractory material). We computed $Q_\text{pr}(\lambda,a)$ and the absorption efficiency, $Q_\text{abs}(\lambda,a)$ (needed to model the dust emission as discussed later) from the optical constants using the Mie theory code \texttt{miex} \citep{wolf2004}. The optical constants of this grain composition mixture are given in Table 3 of \citet{ballering2016}. $a_\text{BOS}$ for each target is given in Table \ref{table:warmtargetlist}.

To compute the model dust emission, we needed the temperature of the grains ($T_\text{dust}$) as a function of their radial location and size. We calculated this by computing 
\begin{equation}
\label{eq:Tdust}
    r(T_\text{dust},a) = \frac{1}{4 \pi}\sqrt{\frac{\int Q_\text{abs}(\lambda,a) L_{\lambda \star}(\lambda) \,\mathrm{d}\lambda}{\int Q_\text{abs}(\lambda,a) B_\lambda(\lambda,T_\text{dust}) \,\mathrm{d}\lambda}}
\end{equation}
then inverting it to solve for $T_\text{dust}(r,a)$. $L_{\lambda \star}(\lambda)$ is the stellar spectral luminosity, and $B_\lambda(\lambda,T_\text{dust})$ is the Planck function. Equation \ref{eq:Tdust} is derived from balancing the heating and cooling power on the grain. Finally, we calculated the emission spectrum from each grain,
\begin{equation}
\label{eq:emission}
    F_\nu(\lambda,r,a) = \left(\frac{a}{d}\right)^2 Q_\text{abs}(\lambda,a) \pi B_\nu(\lambda,T_\text{dust}),
\end{equation}
and combined these spectra into a single spectrum weighted by the spatial distribution and grain size distribution of the model.

\subsection{Fitting Models to the Observed SEDs}
\label{sec:fitting}

We fit the observed SED of each target, including photometric points at $V$, $J$, $H$, and $K$, the IRS spectrum, and the additional photometry listed in Table \ref{table:photometry}. The stellar photosphere accounts for virtually all of the observed flux density in the visible and near-IR, but it also contributes significantly at longer wavelengths. We modeled the photospheric contribution to the SED as described in Section \ref{sec:targets}, and we allowed small adjustments in the photospheric level to optimize the fits. To fit the excess emission from the debris disk, we used three different models: (1) a single modified blackbody (described later), which we used to double-check for systems that could be fit best by a single cold component; (2) a single warm dust belt, as described in Section \ref{sec:models}, with the location of the dust ($r_\text{warm} = r_\text{in}$) and the mass of dust in the belt ($M_\text{warm}$) as free parameters; and (3) a warm dust belt plus a modified blackbody, with the blackbody accounting for a possible cold component. We did not fit the cold components with our dust belt model because their locations were not of interest for our analysis.

For the modified blackbody, we followed the formulation used by \citet{kennedy2014}:
\begin{equation}
    F_\nu(\lambda) = c_\text{BB} B_\nu(\lambda,T_\text{cold}) X(\lambda)^{-1},
\end{equation}
where $c_\text{BB}$ is a constant (amplitude), $T_\text{cold}$ is the temperature of the cold component, and
\begin{equation}
    X(\lambda) =
  \begin{cases} 
   1 & \lambda < \lambda_0 \\
   (\lambda/\lambda_0)^{\tilde{\beta}} & \lambda > \lambda_0
  \end{cases}
  .
\end{equation}
The modification to the blackbody, $X(\lambda)$, models the steeper than Rayleigh-Jeans fall off at long wavelengths due to grains not emitting efficiently at wavelengths longer than their size. The free parameters for the modified blackbody were $c_\text{BB}$, $T_\text{cold}$, $\lambda_0$, and $\tilde{\beta}$. In the fitting we required 50 $\micron < \lambda_0 <$ 500 $\micron$, $0 < \tilde{\beta} < 2$, and $T_\text{cold} < $ 130 K when part of a two-component fit.

In the fitting process, we also allowed for a small amplitude adjustment to the IRS data ($c_\text{IRS}$), which we allowed to take values between 0.8 and 1.2. This effectively corrected any systematic calibration error. This procedure yielded good results with c$_\text{IRS}$ constrained at two points: first, the short end of the IRS data needed to match the photosphere model, which in turn had to match the visible/near-IR photometry; second, the IRS data needed to match the MIPS photometry point at 24 $\micron$ in order for the best fit model to pass through both the MIPS and IRS data at this wavelength.

In practice, we performed a grid search over $r_\text{in}$ (in steps of 0.1 au) and $c_\text{IRS}$ (in steps of 0.01). At each point in the grid we then found best values for the rest of the free parameters with a Levenberg-Marquardt algorithm (the Matlab function \texttt{lsqcurvefit}). The best fit was the model that minimized the standard $\chi^2$ metric. When calculating $\chi^2$, we enhanced the weights of the photometry points at $\geq$70 $\micron$ by a factor of 25 to balance their influence on the fit against the large number of points in the IRS spectra. Without this extra weighting, the behavior of the model in the far-IR/sub-mm often reflected an extrapolation from the longest wavelength IRS points, rather than fitting to the data in this wavelength regime. Upper limit photometry measurements were not included in the $\chi^2$ calculation, but we inspected the best-fitting models to ensure they were consistent with these measurements. Parameters $\lambda_0$ and $\tilde{\beta}$ were often not well-constrained by the fitting, except for the targets with accurately measured far-IR/sub-mm photometry.   

We inspected the results of our single blackbody fits to ensure that we only included genuine warm components in our sample. Targets that were fit well by a single blackbody with temperature $<$130 K were discarded from our sample. These included HIPs 544, 2072, 9141, 16852, 17764, 24947, 46843, 51194, 59072, 59960, 61960, 65728, 66065, 90936, and 107649. Some of these targets had single warm components with temperatures just above the 130 K cutoff according to \citet{ballering2013}, but with the additional far-IR photometry and the new fitting procedure used here, they now fell below this cutoff. For others, \citet{ballering2013} had found two components with the warm component being relatively weak, but here we found that the warm component was no longer necessary to fit the IRS data.

We also excluded HIP 6276 because it has an extremely weak warm excess (and no evidence for a cold component) from which we could not place any meaningful constraints on the dust location. Finally, we excluded HIP 53954 and HIP 65109, because any model fit to the IRS data significantly over-predicted the far-IR photometry. A similar steep decline of the disk flux in the far-IR has been noted in a few other disks \citep{ertel2012b}, requiring a very unusual distribution of grain sizes to model.  

We examined the fits of the remaining targets to determine which were best fit with a single warm belt and which required a cold component as well. In many cases, the requirement for a cold component came from the far-IR photometry, with the IRS data fitting well with a single warm belt. We included a cold component when the warm-only model under-predicted the far-IR data by more than 2$\sigma$ (with the offset from multiple far-IR points combined in quadrature) and the addition of a cold component improved the fit. For most systems, the designations agreed with those of \citet{ballering2013}. Five systems that previously had been fit with a single warm component now were fit with two components (HIPs 1473, 1481, 77432, 78045, and 85922), and two systems that previously had been fit with two components were now fit best with a single warm belt (HIP 63836 and HIP 112542). The best fit cold component of HIP 117452 had $T_\text{cold}$ = 130 K (the upper bound allowed by the fit), suggesting this system has an unusually warm cold component. However, the model fit the data well, so a significantly higher value of $T_\text{cold}$ is likely not required. An unusually warm cold component does not, however, impact the need for a separate warm belt, as we found that a single component could not fit all the data.

From our $\chi^2$ metric, we found that the median statistical uncertainties on $r_\text{in}$ were $5\%$ and $13\%$ for the single-component and two-component systems, respectively. We calculated the luminosities of the best fit model components by integrating under the model spectra. We then derived the fractional luminosity of each component: $f_\text{warm} = L_\text{warm}/L_\star$ and $f_\text{cold} = L_\text{cold}/L_\star$. The results of the fitting are given in Tables \ref{table:onecompfits} and \ref{table:twocompfits}. The best fit model SED for each target is shown in Figures \ref{fig:onecompfits} and \ref{fig:twocompfits}. Our final sample had 29 systems with single warm components and 54 systems with two components.

\section{Analysis and Results}
\label{sec:analysis}

With the locations of the warm dust components for our targets found, we next turned our attention to the relationship between the dust location and the stellar mass. We expected the relation to follow a power-law ($r_\text{warm} \propto M_\star^b$), considering the predicted relations for both the primordial and current snow lines take this form. Our goal was to measure the value of the exponent $b$ and see if it aligned with the predicted value for the current or primordial snow line. We did not attempt to compare the absolute values of the measured dust locations to those predicted for the snow lines. The actual location of the primordial snow line is less certain than its predicted relation with stellar mass, considering the uncertainty on the values of all the factors in Equation \ref{eq:primordialline} and the fact that the primordial snow line location likely evolves with time. Nevertheless, \citet{martin2013b} did find that the absolute stellocentric distances of warm debris disks reported in the literature were roughly consistent with the location of the primordial snow line.   

\subsection{Single-component Systems}

We first considered the systems with a single warm component. Figure \ref{fig:scatteronecomp} plots $r_\text{warm}$ versus $M_\star$ and shows clear evidence for a positive trend. To quantify this trend, we fit the function $\log(r_\text{warm}/\text{au}) = a + b \log(M_\star/M_\odot)$ to these points. The best fit values of $b$ and $a$ were computed as
\begin{equation}
\label{eq:bestb}
    b = \frac{\overline{\log(M_\star)\cdot\log(r_\text{warm})} - \overline{\log(M_\star)} \cdot \overline{\log(r_\text{warm})}}{\overline{\log(M_\star)^2} - \overline{\log(M_\star)}^2}
\end{equation} 
and $a = \overline{\log(r_\text{warm})} - b \cdot \overline{\log(M_\star)}$. We found $b$ = 1.08, $a$ = 0.566 (i.e. $r_\text{warm}/\text{au} = 3.68 (M_\star/M_\odot)^{1.08}$). This best fit trend line is plotted with the data points in Figure \ref{fig:scatteronecomp}.

Considering the substantial amount of scatter evident in the data around the trend line, we used a bootstrap procedure to quantify the significance of our derived value of $b$. We used 10,000 trials for the bootstrap procedure. For each trial we randomly selected 29 points (with replacement) from our sample and recomputed the best fit $b$ using Equation \ref{eq:bestb}. Figure \ref{fig:bootstrap} shows the distribution of $b$ values found by the bootstrap procedure. The distribution was fit by a normal distribution with mean = 1.08 and $\sigma$ = 0.21 using the Matlab function \texttt{fitdist}.

We found that $b$ for the targets with a single warm component was consistent (within 0.6$\sigma$) with the value predicted by the primordial snow line (1.2) but was not consistent ($>4.3\sigma$) with the relation predicted by the current snow line (2.0). Thus we conclude that these warm components are more likely to arise from dust produced by exo-asteroid belts than from disintegrating comets or dust dragged inward from an outer belt that is below our current detection limit.

\subsection{Two-component Systems}

Next we examined the $r_\text{warm}$--$M_\star$ trend for the two-component systems, shown in Figure \ref{fig:scattertwocomp}. There was more scatter than for the single-component systems, and no simple relation was evident that represented all the points. For comparison, we added to this plot the best fit trend found for the single-component systems (green dashed line from Figure \ref{fig:scatteronecomp}). We found that many of the warm components in the two-component systems were consistent with this same trend, but there were also several systems both above and below the trend.

\subsection{Scatter around the Single-component Trend}

\begin{figure}
\epsscale{1.2}
\plotone{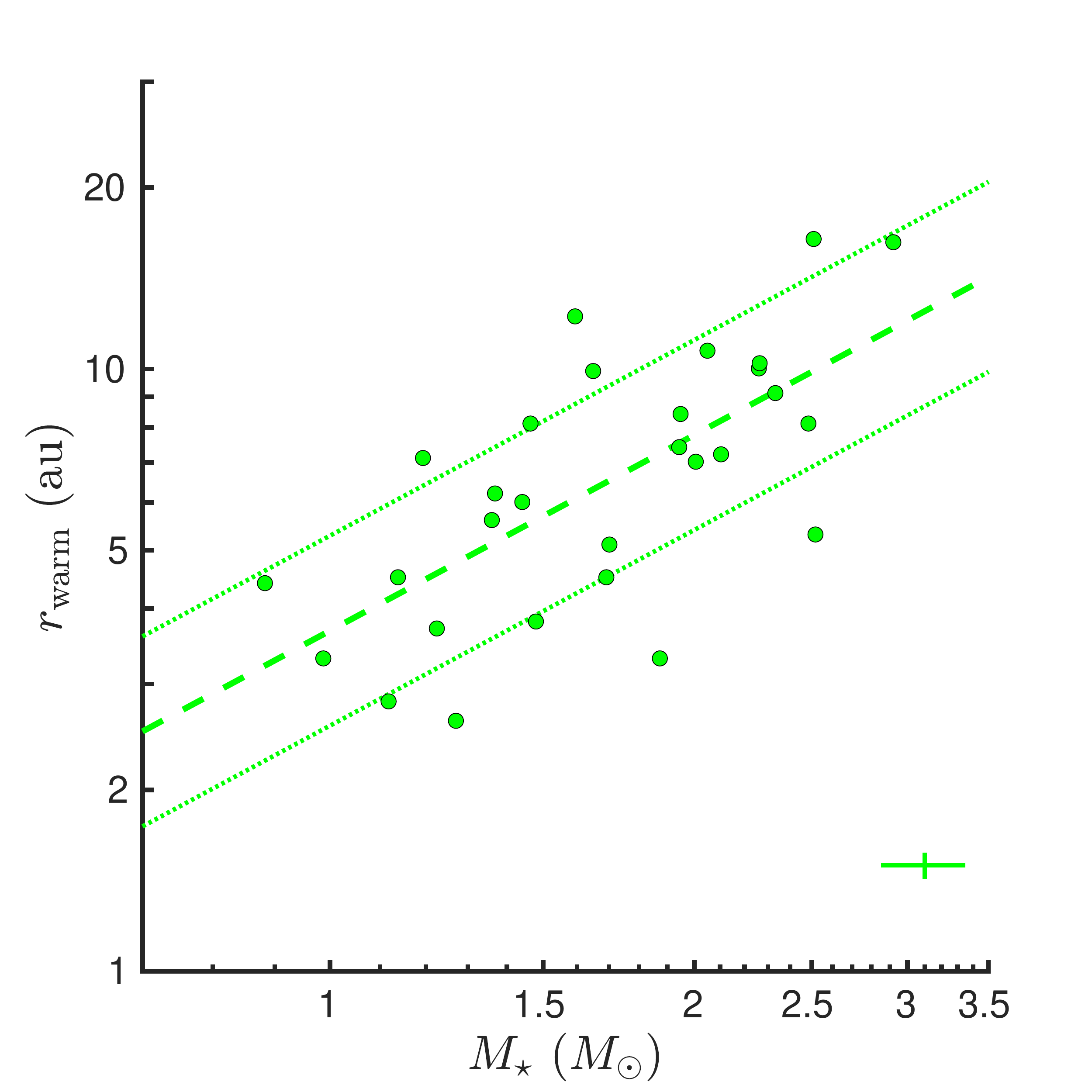}
\caption{Location of the warm dust vs, stellar mass for the 29 targets with a single dust component. The dashed line shows the best fit trend $\log(r_\text{warm}/\text{au}) = 0.566 + 1.08 \log(M_\star/M_\odot)$, equivalent to $r_\text{warm}/\text{au} = 3.68 (M_\star/M_\odot)^{1.08}$. The thin dotted green lines show the measured 1$\sigma$ scatter around the trend ($\pm$ 0.16 dex). A representative error bar is in the lower-right, showing the typical 8$\%$ uncertainty on $M_\star$ and $5\%$ uncertainty on $r_\text{warm}$.}
\label{fig:scatteronecomp}
\end{figure}

\begin{figure}
\epsscale{1.2}
\plotone{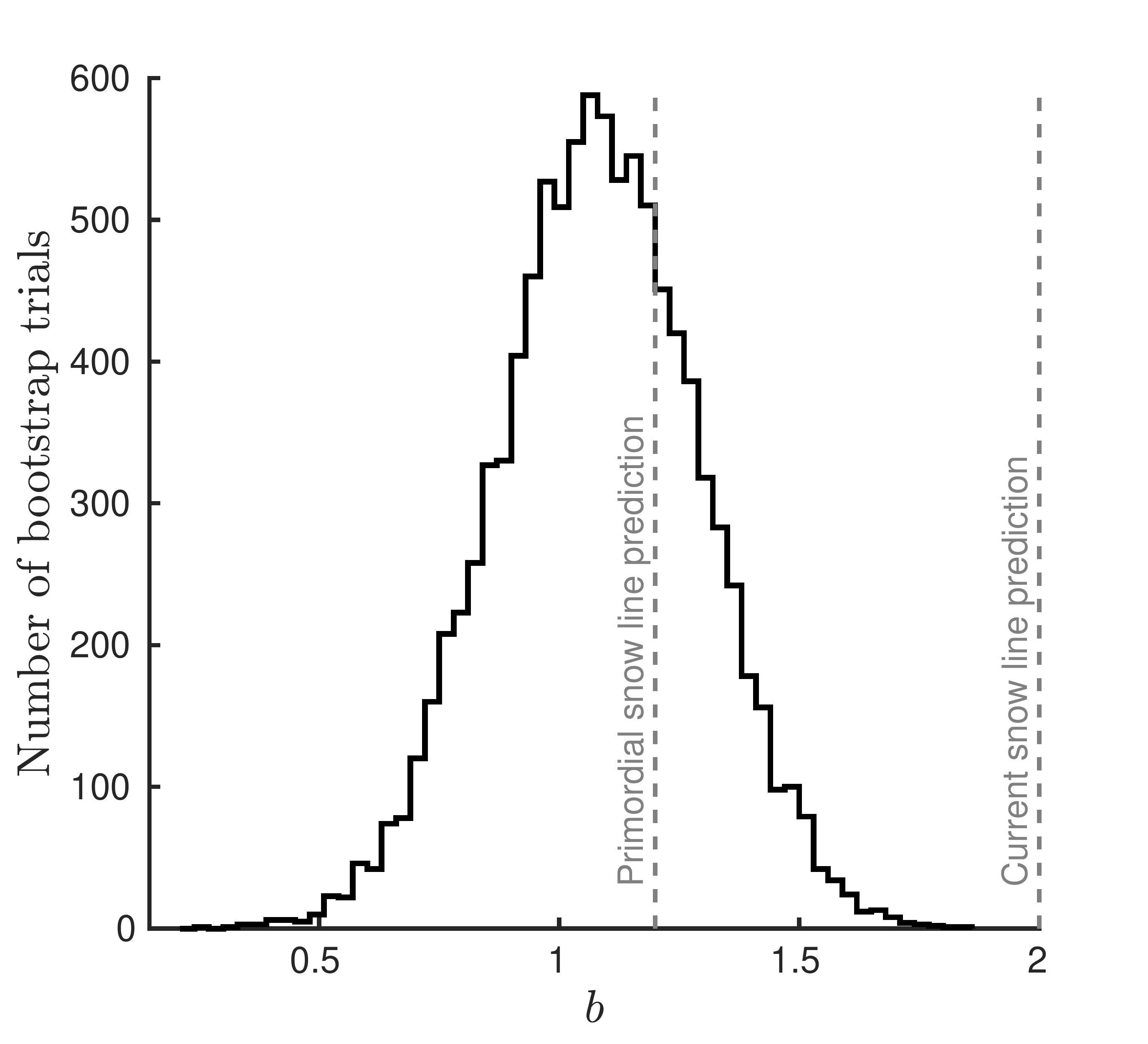}
\caption{Distribution of the results from the bootstrap procedure to estimate the uncertainty on the power-law index ($b$) of the observed $r_\text{warm}$--$M_\star$ relation for systems with a single component. The distribution has mean = 1.08 and $\sigma$ = 0.21. The warm components are thus consistent with being set by the primordial snow line and inconsistent with being set by the current snow line.}
\label{fig:bootstrap}
\end{figure}

\begin{figure}
\epsscale{1.2}
\plotone{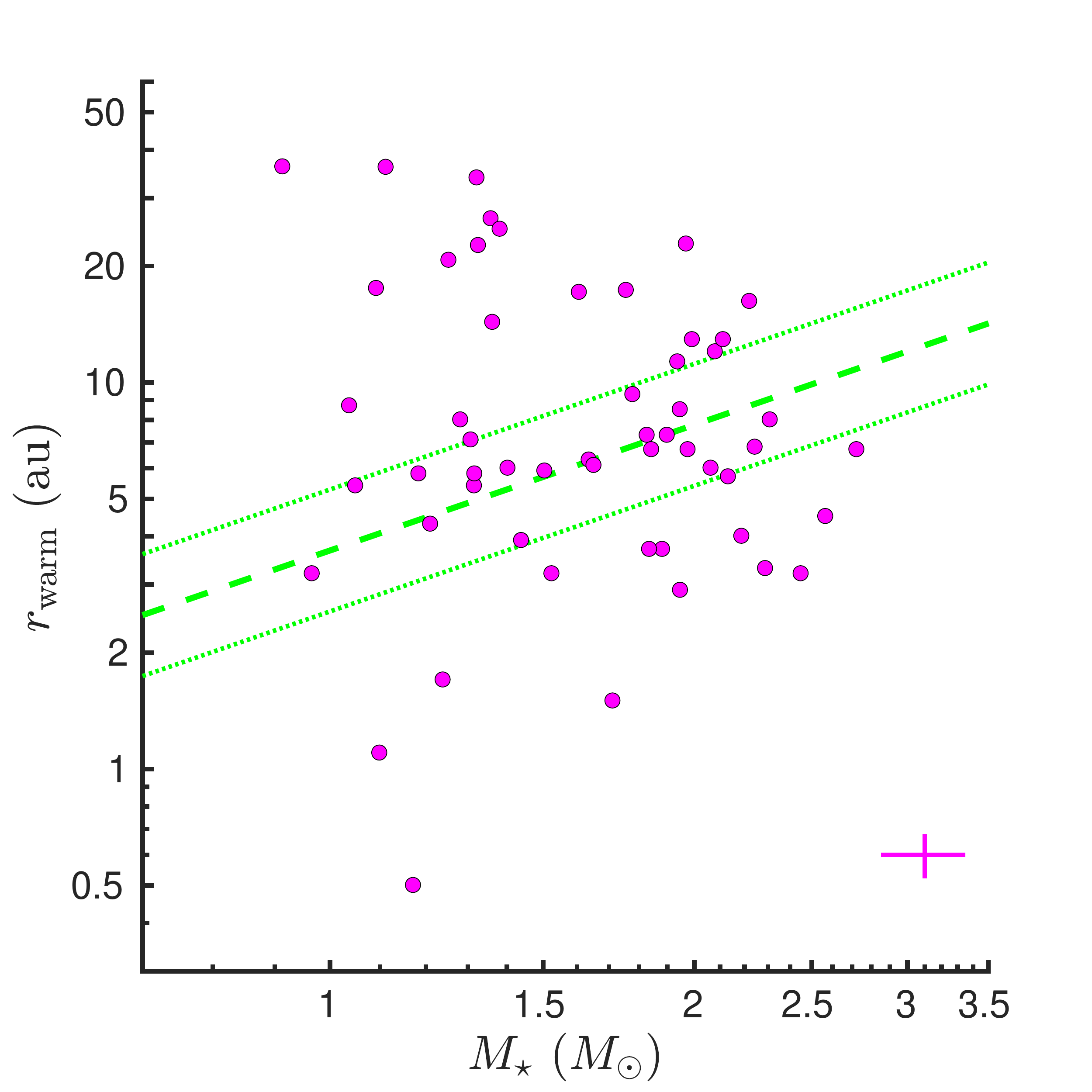}
\caption{Location of the warm dust vs, stellar mass for the 54 targets with two dust components. The dashed green line is the best fit trend line derived for the single-component systems (as in Figure \ref{fig:scatteronecomp}), which we found likely arise from asteroid belts with locations set by the primordial snow line. The thin dotted green lines show the measured 1$\sigma$ scatter around the trend ($\pm$ 0.16 dex). We see that many targets are consistent with this trend (suggesting they also are set by the primordial snow line) but there are also many sources located away from the trend. A representative error bar is in the lower-right, showing the typical 8$\%$ uncertainty on $M_\star$ and $13\%$ uncertainty on $r_\text{warm}$.}
\label{fig:scattertwocomp}
\end{figure}

We subtracted the values of $r_\text{warm}$ predicted by the trend line for the single-component systems from the measured $r_\text{warm}$ values in both samples. Fitting a normal distribution to residuals of the single-component systems (using \texttt{fitdist}) gave a mean = 0.0 and $\sigma$=0.16 dex in $\log(r_\text{warm}/\text{au})$. This provided a measure of the inherent scatter around this trend for systems that likely follow the primordial snow line relation. The 1$\sigma$ region around the trend due to this scatter is depicted in Figures \ref{fig:scatteronecomp} and \ref{fig:scattertwocomp} with thin dotted green lines.

\begin{figure}
\epsscale{1.2}
\plotone{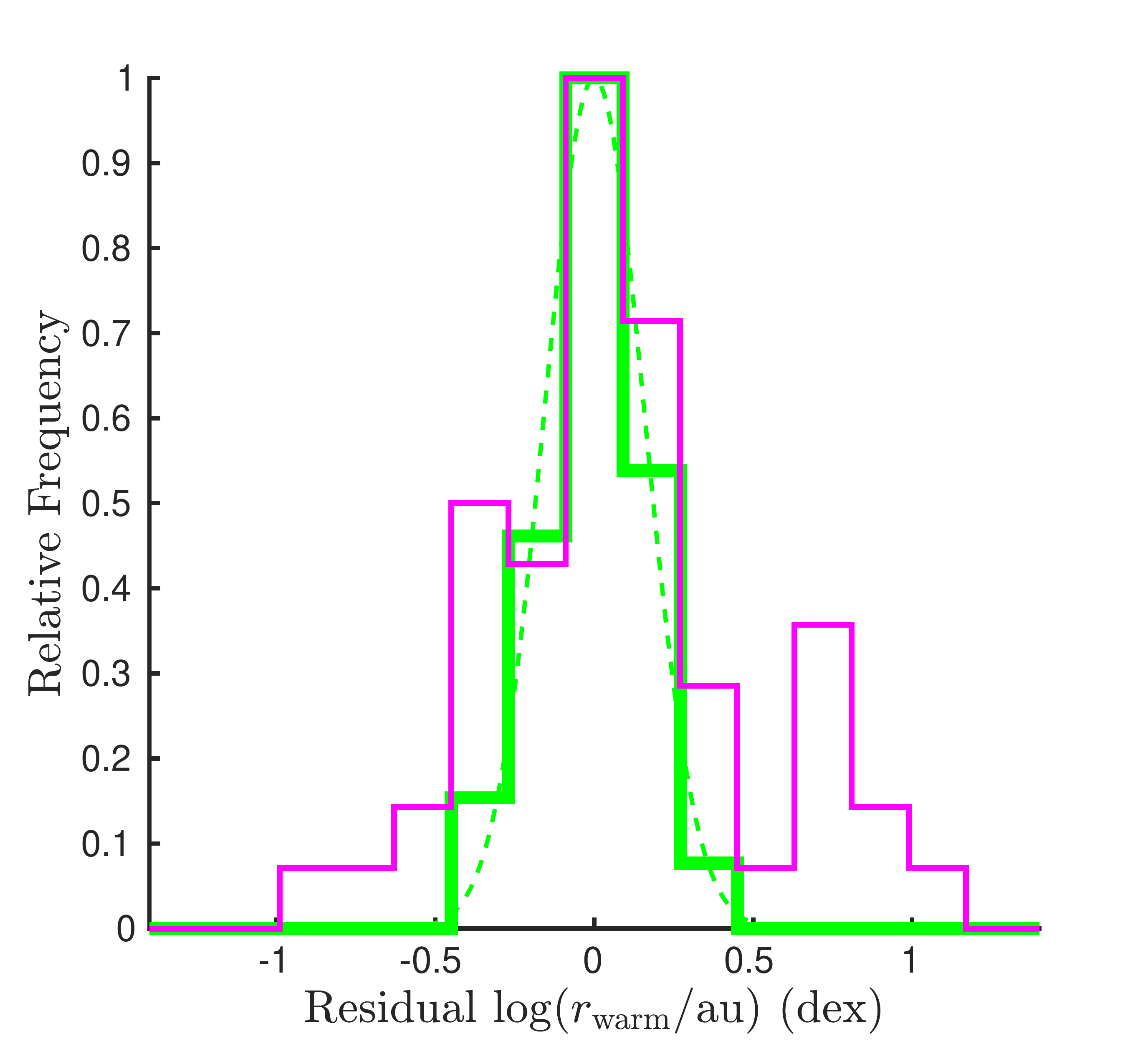}
\caption{Distributions of the warm dust residual locations for the single-component (green) and two-component (magenta) systems around the trend line found for the single-component systems. The single-component systems show a symmetric distribution of residuals with mean = 0.0 and $\sigma$ = 0.16 dex. A normal distribution with this mean and $\sigma$ is shown in the dotted green line. All three curves are unity normalized in order to better compare their shapes. The two-component systems show a peak centered on the trend, a tail of systems below the trend, and a separate population of systems above the trend.}
\label{fig:residualhisto}
\end{figure}

Figure \ref{fig:residualhisto} shows the distribution of the residual warm dust locations from the trend (green for the single-component systems, magenta for the two-component systems). The distribution for the two-component systems peaks around the trend (no residual). We note that 32/54 = 59\% of these systems fall within the 2$\sigma$ tolerance of the trend for the systems with only warm components. The histogram shows for the two-component systems a decreasing tail of warm dust locations below the trend (negative residual), whereas the warm dust locations above the trend show signs of a separate population of systems. It may be that some warm components in two-component systems have locations set by the primordial snow line while others end up off of this trend for various reasons. Alternatively, these warm components may have locations scattered more or less randomly, with some inevitably falling along the trend set by the single-component systems. In Section \ref{sec:warmdiscussion} we discuss various possibilities for the nature of the warm components in two-component systems.

\section{Discussion}

\subsection{Effect of Model Assumptions}
\label{sec:testassumptions}

\begin{figure}
\epsscale{1.2}
\plotone{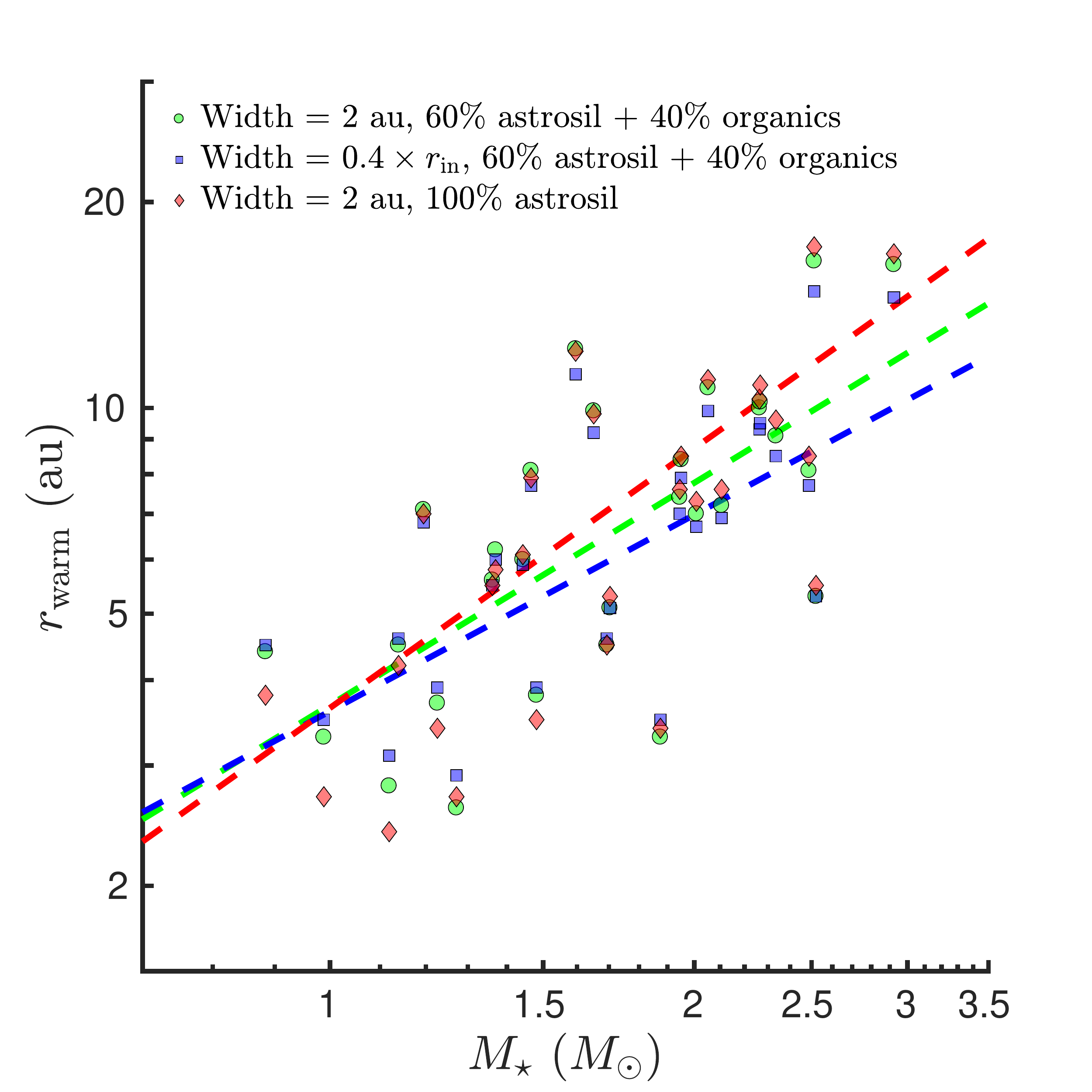}
\caption{Warm dust location vs, stellar mass for the 29 single-component targets, with the warm dust location determined by three different models. The green circles assume the dust is located in a 2 au wide ring and is composed of a mixture of astronomical silicates and organic refractory material. These results were presented in Section \ref{sec:analysis} and are the same points shown in Figure \ref{fig:scatteronecomp}. The blue squares assume the same dust composition, but use a ring width that scales as 0.4 times the ring's location. The red diamonds again assume 2 au wide belts, but change the composition of the dust to pure astronomical silicates. The results are nearly identical for the three different sets of model assumptions. Therefore our conclusion that these warm belts are likely set by the primordial snow line and not by the current snow line is robust against our particular choice of model.}
\label{fig:testassumptions}
\end{figure}

An essential feature of our analysis is that we compare the trend of warm dust location with stellar mass to the predicted trend for the two possible snow lines. By comparing trends rather than absolute values, we expect that systematic errors due to our choice of model disk geometry or grain composition (optical constants) will not influence our results. To test what role our model assumptions may have had on our results, we repeated our analysis twice with different assumptions. First, we re-fit the SEDs with dust belts with fractional widths of 0.4 $r_\text{in}$ (in contrast to the fixed belt width of 2 au we used in the main analysis). The results agreed almost exactly with those of our main analysis: the single-component systems showed a clear trend in belt location with stellar mass while the two-component systems showed considerable scatter. A bootstrap analysis of the single-component power-law index yielded $b = 0.95 \pm 0.19$. Second, instead of using a mixture of astronomical silicates and refractory organic material for the dust composition, we used 100\% astronomical silicates (a common first-order assumption in debris disk modeling). Again, we found very similar results to our main analysis with $b = 1.26 \pm 0.21$. The $r_\text{warm}$ vs, $M_\star$ trends for all three sets of model assumptions are plotted together in Figure \ref{fig:testassumptions}. Thus neither of these modifications to our model assumptions affected our main conclusion that the locations of single-component systems are consistent with the primordial snow line and inconsistent with the current snow line.

\subsection{Origin of Warm Dust in Two-component Systems}
\label{sec:warmdiscussion}

We have shown that warm dust in the single-component systems resides at locations consistent with being set by the primordial snow line and not by the current snow line, and thus is likely to originate from exo-asteroid belts. The warm dust locations in two-component systems are much more scattered, although many reside at similar locations to those in single-component systems. Here we turn our attention to the nature of the two-component systems, especially those with warm components that clearly do not trace the primordial snow line location.

\subsubsection{Planets?}

Planets are known to sculpt debris disks, and the gap between warm and cold components is often ascribed to the presence of planets. Thus the scatter in warm dust locations may simply reflect a diversity in the locations of planet formation or in their dynamical histories. The warm dust may arise from the in-situ collisional processing of a parent body belt, but its location may not be set by the primordial snow line. However, this scenario makes no clear prediction as to why more scatter would arise in systems with two components than in systems with only a warm component, unless systems without cold components also have fewer planets. 

\subsubsection{Inward Transport?}

As discussed earlier, the inward transport of material from the cold outer belt is expected to result in warm dust at a preferential location---the current snow line. However, these warm components span a large range of locations for a given stellar mass, so it is unlikely that inward transport could explain all of these systems.

\begin{figure}
\epsscale{1.2}
\plotone{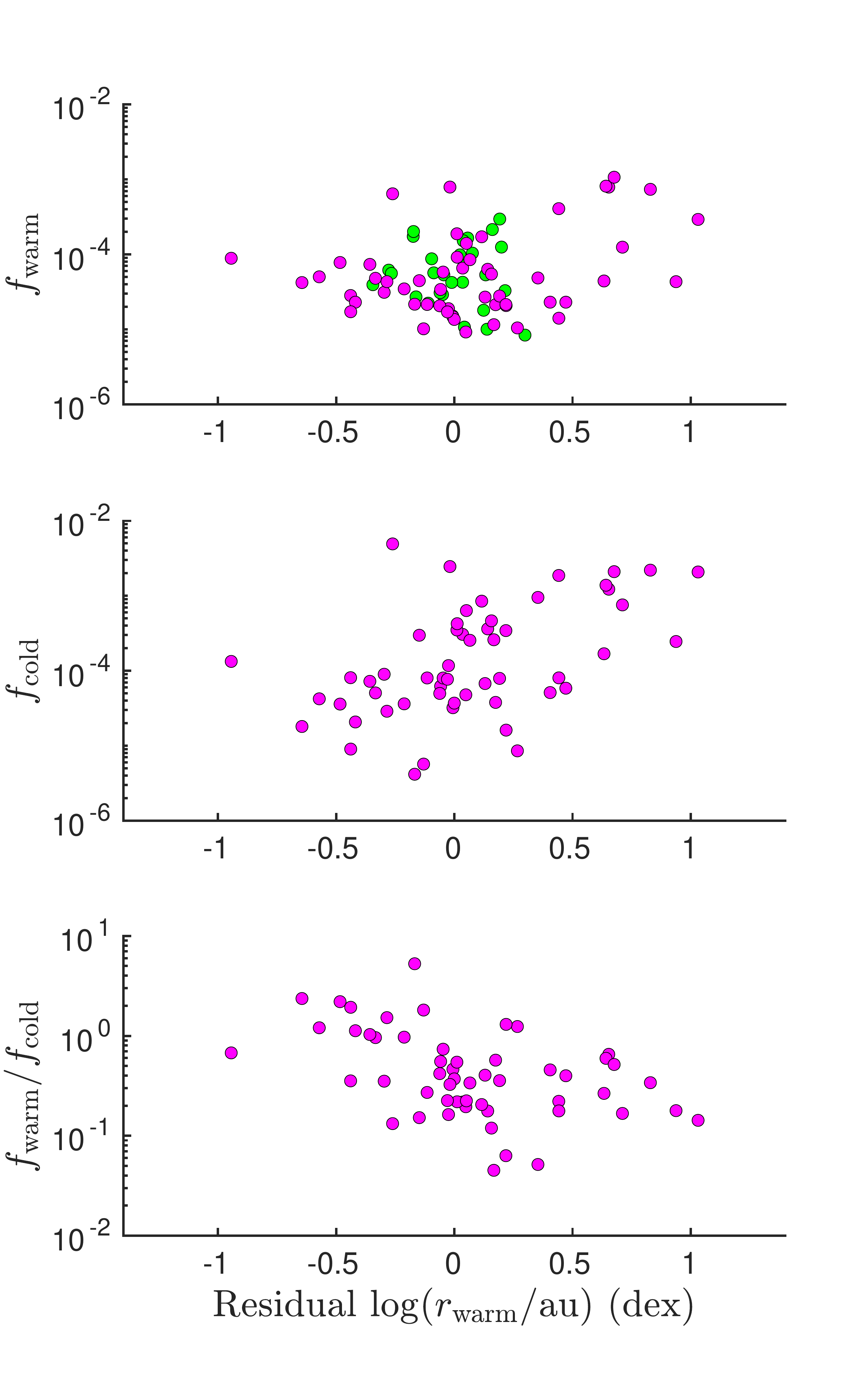}
\caption{Fractional luminosity (brightness) of the warm components (top panel) and the cold components (middle panel), and the ratio of the two (bottom panel) vs, the residual warm dust locations relative to the best fit trend. The single-component systems are in green and the two-component systems are in magenta. The systems above the trend (on the right side of these plots) tend to have brighter than average warm and cold components, but the ratio of their brightnesses are in line with the sample as a whole.}  
\label{fig:fraclum}
\end{figure}

The specific scenario of inward transport by drag forces predicts that the warm component of dragged in material should be much fainter than the cold reservoir from which it originates \citep{kennedy2015}. We looked for this in our sample by plotting the fractional luminosities of the warm\footnote{Note from the top panel of Figure \ref{fig:fraclum} that the two-component systems (magenta points) near the trend have a similar brightness distribution as the single-component systems (green points), furthering the notion that these warm components arise from a common mechanism.} and cold components---and their ratio---against the residual warm dust location from the trend (Figure \ref{fig:fraclum}). We found that the $f_\text{warm}$/$f_\text{cold}$ ratio (bottom panel) is roughly the same across our entire sample (or perhaps somewhat larger in systems below the trend). Thus this provides no additional support for the drag scenario. In fact, \citet{geiler2017} found that the $f_\text{warm}$/$f_\text{cold}$ ratios for two-component debris disks are consistent with the steady-state collisional evolution of inner and outer parent body belts originating from protoplanetary disks with reasonable radial density profiles.

Later-type stars may be able to drag in a substantial amount of dust to generate a bright warm component if their luminosity is low enough such that no grains are blown out of the system by radiation pressure (there is no $a_\text{BOS}$). Drag is also enhanced in later-type stars by stronger stellar winds. With no significant radiation force on the grains, there would also be no pile-up of grains when the icy constituents sublimate, so the warm components would not trace any snow line location. The systems in our sample, however, have sufficiently high luminosities that they should be able to blow grains out, so we deem this situation unlikely.

\subsubsection{Contamination by Exozodiacal Dust?}

The warm components that fall below the single-component trend line may result from exozodiacal dust components, which are considered a separate component from the traditional warm components \citep{su2014}. We purposely discarded targets with exozodiacal dust from our sample (Section \ref{sec:targets}), so perhaps these systems should be ignored for the same reason. Two of the four systems with warm dust located more than 3$\sigma$ below the trend (HIP 1481 and HIP 77432) show clear silicate features in their best fit model spectra, and silicate features are a signature of exozodiacal dust \citep{ballering2014}. \citet{ballering2014} found that all systems with exozodiacal dust also had outer components, consistent with finding such systems only in our two-component sample.

\subsubsection{Contamination by Outer Dust?}

For the systems with warm component locations above the trend, it is possible that we are not seeing dust interior to the cold component; rather, there is dust with a range of temperatures located within the outer belt. While \citet{kennedy2014} argued that this is not possible for most two-component systems, HIP 95270 (HD 181327), which we found to have $r_\text{warm}$ above the trend, was an exception. In fact, \citet{lebreton2012} fit the entire SED of this system with dust from a single outer belt.

It is also possible that we are seeing emission from the spatially unresolved blowout halo of small grains \textit{beyond} the cold parent body belt. These grains must be small---potentially below the blowout size---so are warm despite their large stellocentric distance. From the middle panel of Figure \ref{fig:fraclum} we see that the systems above the trend tend to have slightly higher than average cold component fractional luminosities, consistent with systems capable of generating large halos. Detailed studies of disks with such halos seen in resolved images have found that the halo component's contribution to the SED often peaks at wavelengths shorter than the cold component but longer than a typical warm component (as is seen for the systems above the trend in our sample), although this varies among specific systems. For example, the halo of $\gamma$ Ophiuchi peaks at nearly the same wavelength as the cold belt \citep[Figure 3 of][]{su2008}, the halo of HR 8799 peaks at a shorter wavelength \citep[Figure 9 of][]{su2009}, and the halo of $\beta$ Pictoris peaks at an even shorter wavelength \citep[Figure 14 of][]{ballering2016}. In many systems with halos, the halo signal can simply blend with the cold belt in the SED and be fit as a single component. This may be especially true if there is also an inner warm component in the system. In fact, HR 8799 (HIP 114189) and $\gamma$ Ophiuchi (HIP 87108)---which are known to have halos---are in our sample, but their halo components are not detected in our fitting separately from their cold components. 

HIP 11847  (HD 15745), a system with a large $r_\text{warm}$ value, has a fan-shaped outer component detected in scattered light images out to 450 au \citep{kalas2007,schneider2014}, which may be a halo component, but no detailed models have shown that this halo could give rise to the warm part of the SED. Another such system, HIP 36948 (HD 61005 a.k.a. ``The Moth") is seen in scattered light to have an outer belt with wings of dust swept back due to interactions with the ISM \citep{hines2007,buenzli2010,schneider2014}. Studies of this system's SED that modeled the outer component with a modified blackbody \citep{kennedy2010} or with a narrow belt \citep{ricarte2013} found the warm component to be fairly cold ($\sim$100--125K)---consistent with the large value of $r_\text{warm}$ found here (36.1 au). In contrast, \citet{olofsson2016} modeled this SED with a wide outer belt plus a fainter and hotter ($\sim$220 K) warm component. This suggests the outer belt comprises grains with a range of temperatures, which our model accounts for with a warm component placed far from the star. This system demonstrates how cold components can add uncertainty to the derived locations of warm components.

\subsubsection{Conclusion}

We can identify no definitive source for the scatter in the warm dust locations of these two-component systems. Planets in a diversity of arrangements offer one explanation for this scatter. It may also be that the scatter arises from multiple mechanisms operating together, potentially including exo-asteroid belts (as is favored for the single-component systems), the presence of exozodiacal dust, contamination from warm dust in a halo or outer belt, and the inward transport of material by comets (although we deem inward transport by drag forces to be unlikely).        

\section{Summary}

\begin{enumerate}
\item Warm components of debris disks have been observed in the spatially unresolved SEDs of many stars, but the nature and origin of the dust is not known. There are two plausible hypotheses for its origin: the in-situ production of dust via collisions in an asteroid belt-like population of parent body planetesimals, or the inward transport of material from an outer reservoir.
\item The first hypothesis predicts the dust to be located at the primordial snow line, while the second hypothesis predicts the dust to be located at the current snow line. The location of the primordial snow line follows a shallower power-law relation with stellar mass than does the current snow line, providing a means to distinguish between the two.
\item We located the warm dust in 83 debris disk systems observed with \textit{Spitzer}/IRS (29 with a single warm component, 54 that also possess a cold component) by fitting model dust belt emission spectra to their SEDs.
\item We found that the $r_\text{warm}$--${M_\star}$ trend for the single-component systems is consistent with  the primordial snow line and not consistent with the current snow line. We thus favor the in-situ dust production scenario for these systems. Many of the two-component systems are also consistent with this relation. Hence we conclude that the collisional processing of exo-asteroid belts is a common mechanism to produce warm debris disk components.
\item We are not able to definitively explain the scatter of warm component locations in the two-component systems. Warm planetesimal belts with locations set by planets in a diversity of architectures offer a single mechanism to explain the scatter. Or the scatter could result from a mixture of systems with warm dust locations set by the primordial snow line and those set by other mechanisms. The inward transport of material by comets remains possible in these systems, and could contribute to the scatter. Warm belts nearer the star than the snow line may be exozodiacal dust, while those farther from the star may be warm dust co-located with the cold dust or even beyond the cold dust in a halo component. 
\end{enumerate}

\acknowledgments
We thank Glenn Schneider and the anonymous referee for many helpful comments on this paper. This work is based on observations made with the {\it Spitzer Space Telescope}, which is operated by the Jet Propulsion Laboratory, California Institute of Technology, under a contract with NASA.

\facility{\textit{Spitzer} (IRS, MIPS).}
\software{MATLAB, miex.}

\bibliographystyle{aasjournal}

\begin{thebibliography}{}
\expandafter\ifx\csname natexlab\endcsname\relax\def\natexlab#1{#1}\fi

\bibitem[{{Absil} {et~al.}(2013){Absil}, {Defr{\`e}re}, {Coud{\'e} du Foresto},
  {Di Folco}, {M{\'e}rand}, {Augereau}, {Ertel}, {Hanot}, {Kervella},
  {Mollier}, {Scott}, {Che}, {Monnier}, {Thureau}, {Tuthill}, {ten Brummelaar},
  {McAlister}, {Sturmann}, {Sturmann}, \& {Turner}}]{absil2013}
{Absil}, O., {Defr{\`e}re}, D., {Coud{\'e} du Foresto}, V., {et~al.} 2013,
  \aap, 555, A104

\bibitem[{{Acke} {et~al.}(2012){Acke}, {Min}, {Dominik}, {Vandenbussche},
  {Sibthorpe}, {Waelkens}, {Olofsson}, {Degroote}, {Smolders}, {Pantin},
  {Barlow}, {Blommaert}, {Brandeker}, {De Meester}, {Dent}, {Exter}, {Di
  Francesco}, {Fridlund}, {Gear}, {Glauser}, {Greaves}, {Harvey}, {Henning},
  {Hogerheijde}, {Holland}, {Huygen}, {Ivison}, {Jean}, {Liseau}, {Naylor},
  {Pilbratt}, {Polehampton}, {Regibo}, {Royer}, {Sicilia-Aguilar}, \&
  {Swinyard}}]{acke2012}
{Acke}, B., {Min}, M., {Dominik}, C., {et~al.} 2012, \aap, 540, A125

\bibitem[{{Augereau} {et~al.}(2001){Augereau}, {Nelson}, {Lagrange},
  {Papaloizou}, \& {Mouillet}}]{augereau2001}
{Augereau}, J.~C., {Nelson}, R.~P., {Lagrange}, A.~M., {Papaloizou}, J.~C.~B.,
  \& {Mouillet}, D. 2001, \aap, 370, 447

\bibitem[{{Ballering} {et~al.}(2014){Ballering}, {Rieke}, \&
  {G{\'a}sp{\'a}r}}]{ballering2014}
{Ballering}, N.~P., {Rieke}, G.~H., \& {G{\'a}sp{\'a}r}, A. 2014, \apj, 793, 57

\bibitem[{{Ballering} {et~al.}(2013){Ballering}, {Rieke}, {Su}, \&
  {Montiel}}]{ballering2013}
{Ballering}, N.~P., {Rieke}, G.~H., {Su}, K.~Y.~L., \& {Montiel}, E. 2013,
  \apj, 775, 55

\bibitem[{{Ballering} {et~al.}(2016){Ballering}, {Su}, {Rieke}, \&
  {G{\'a}sp{\'a}r}}]{ballering2016}
{Ballering}, N.~P., {Su}, K.~Y.~L., {Rieke}, G.~H., \& {G{\'a}sp{\'a}r}, A.
  2016, \apj, 823, 108

\bibitem[{{Boley} {et~al.}(2012){Boley}, {Payne}, {Corder}, {Dent}, {Ford}, \&
  {Shabram}}]{boley2012}
{Boley}, A.~C., {Payne}, M.~J., {Corder}, S., {et~al.} 2012, \apjl, 750, L21

\bibitem[{{Bonsor} {et~al.}(2012){Bonsor}, {Augereau}, \&
  {Th{\'e}bault}}]{bonsor2012}
{Bonsor}, A., {Augereau}, J.-C., \& {Th{\'e}bault}, P. 2012, \aap, 548, A104

\bibitem[{{Bonsor} {et~al.}(2014){Bonsor}, {Raymond}, {Augereau}, \&
  {Ormel}}]{bonsor2014}
{Bonsor}, A., {Raymond}, S.~N., {Augereau}, J.-C., \& {Ormel}, C.~W. 2014,
  \mnras, 441, 2380

\bibitem[{{Booth} {et~al.}(2013){Booth}, {Kennedy}, {Sibthorpe}, {Matthews},
  {Wyatt}, {Duch{\^e}ne}, {Kavelaars}, {Rodriguez}, {Greaves}, {Koning},
  {Vican}, {Rieke}, {Su}, {Moro-Mart{\'{\i}}n}, \& {Kalas}}]{booth2013}
{Booth}, M., {Kennedy}, G., {Sibthorpe}, B., {et~al.} 2013, \mnras, 428, 1263

\bibitem[{{Booth} {et~al.}(2016){Booth}, {Jord{\'a}n}, {Casassus}, {Hales},
  {Dent}, {Faramaz}, {Matr{\`a}}, {Barkats}, {Brahm}, \& {Cuadra}}]{booth2016}
{Booth}, M., {Jord{\'a}n}, A., {Casassus}, S., {et~al.} 2016, \mnras, 460, L10

\bibitem[{{Brauer} {et~al.}(2008){Brauer}, {Henning}, \&
  {Dullemond}}]{brauer2008}
{Brauer}, F., {Henning}, T., \& {Dullemond}, C.~P. 2008, \aap, 487, L1

\bibitem[{{Buenzli} {et~al.}(2010){Buenzli}, {Thalmann}, {Vigan}, {Boccaletti},
  {Chauvin}, {Augereau}, {Meyer}, {M{\'e}nard}, {Desidera}, {Messina},
  {Henning}, {Carson}, {Montagnier}, {Beuzit}, {Bonavita}, {Eggenberger},
  {Lagrange}, {Mesa}, {Mouillet}, \& {Quanz}}]{buenzli2010}
{Buenzli}, E., {Thalmann}, C., {Vigan}, A., {et~al.} 2010, \aap, 524, L1

\bibitem[{{Calvet} {et~al.}(2004){Calvet}, {Muzerolle}, {Brice{\~n}o},
  {Hern{\'a}ndez}, {Hartmann}, {Saucedo}, \& {Gordon}}]{calvet2004}
{Calvet}, N., {Muzerolle}, J., {Brice{\~n}o}, C., {et~al.} 2004, \aj, 128, 1294

\bibitem[{{Carpenter} {et~al.}(2005){Carpenter}, {Wolf}, {Schreyer},
  {Launhardt}, \& {Henning}}]{carpenter2005}
{Carpenter}, J.~M., {Wolf}, S., {Schreyer}, K., {Launhardt}, R., \& {Henning},
  T. 2005, \aj, 129, 1049

\bibitem[{{Castelli} \& {Kurucz}(2004)}]{castelli2004}
{Castelli}, F., \& {Kurucz}, R.~L. 2004, ArXiv Astrophysics e-prints,
  arXiv:astro-ph/0405087

\bibitem[{{Chen} {et~al.}(2014){Chen}, {Mittal}, {Kuchner}, {Forrest}, {Lisse},
  {Manoj}, {Sargent}, \& {Watson}}]{chen2014}
{Chen}, C.~H., {Mittal}, T., {Kuchner}, M., {et~al.} 2014, \apjs, 211, 25

\bibitem[{{Corder} {et~al.}(2009){Corder}, {Carpenter}, {Sargent}, {Zauderer},
  {Wright}, {White}, {Woody}, {Teuben}, {Scott}, {Pound}, {Plambeck}, {Lamb},
  {Koda}, {Hodges}, {Hawkins}, \& {Bock}}]{corder2009}
{Corder}, S., {Carpenter}, J.~M., {Sargent}, A.~I., {et~al.} 2009, \apjl, 690,
  L65

\bibitem[{{Cuzzi} \& {Zahnle}(2004)}]{cuzzi2004}
{Cuzzi}, J.~N., \& {Zahnle}, K.~J. 2004, \apj, 614, 490

\bibitem[{{Donaldson} {et~al.}(2013){Donaldson}, {Lebreton}, {Roberge},
  {Augereau}, \& {Krivov}}]{donaldson2013}
{Donaldson}, J.~K., {Lebreton}, J., {Roberge}, A., {Augereau}, J.-C., \&
  {Krivov}, A.~V. 2013, \apj, 772, 17

\bibitem[{{Donaldson} {et~al.}(2012){Donaldson}, {Roberge}, {Chen}, {Augereau},
  {Dent}, {Eiroa}, {Krivov}, {Mathews}, {Meeus}, {M{\'e}nard},
  {Riviere-Marichalar}, \& {Sandell}}]{donaldson2012}
{Donaldson}, J.~K., {Roberge}, A., {Chen}, C.~H., {et~al.} 2012, \apj, 753, 147

\bibitem[{{Draper} {et~al.}(2016{\natexlab{a}}){Draper}, {Matthews}, {Kennedy},
  {Wyatt}, {Venn}, \& {Sibthorpe}}]{draper2016a}
{Draper}, Z.~H., {Matthews}, B.~C., {Kennedy}, G.~M., {et~al.}
  2016{\natexlab{a}}, \mnras, 456, 459

\bibitem[{{Draper} {et~al.}(2016{\natexlab{b}}){Draper}, {Duch{\^e}ne},
  {Millar-Blanchaer}, {Matthews}, {Wang}, {Kalas}, {Graham}, {Padgett},
  {Ammons}, {Bulger}, {Chen}, {Chilcote}, {Doyon}, {Fitzgerald}, {Follette},
  {Gerard}, {Greenbaum}, {Hibon}, {Hinkley}, {Macintosh}, {Ingraham},
  {Lafreni{\`e}re}, {Marchis}, {Marois}, {Nielsen}, {Oppenheimer}, {Patel},
  {Patience}, {Perrin}, {Pueyo}, {Rajan}, {Rameau}, {Sivaramakrishnan}, {Vega},
  {Ward-Duong}, \& {Wolff}}]{draper2016b}
{Draper}, Z.~H., {Duch{\^e}ne}, G., {Millar-Blanchaer}, M.~A., {et~al.}
  2016{\natexlab{b}}, \apj, 826, 147

\bibitem[{{Eiroa} {et~al.}(2013){Eiroa}, {Marshall}, {Mora}, {Montesinos},
  {Absil}, {Augereau}, {Bayo}, {Bryden}, {Danchi}, {del Burgo}, {Ertel},
  {Fridlund}, {Heras}, {Krivov}, {Launhardt}, {Liseau}, {L{\"o}hne},
  {Maldonado}, {Pilbratt}, {Roberge}, {Rodmann}, {Sanz-Forcada}, {Solano},
  {Stapelfeldt}, {Th{\'e}bault}, {Wolf}, {Ardila}, {Ar{\'e}valo}, {Beichmann},
  {Faramaz}, {Gonz{\'a}lez-Garc{\'{\i}}a}, {Guti{\'e}rrez}, {Lebreton},
  {Mart{\'{\i}}nez-Arn{\'a}iz}, {Meeus}, {Montes}, {Olofsson}, {Su}, {White},
  {Barrado}, {Fukagawa}, {Gr{\"u}n}, {Kamp}, {Lorente}, {Morbidelli},
  {M{\"u}ller}, {Mutschke}, {Nakagawa}, {Ribas}, \& {Walker}}]{eiroa2013}
{Eiroa}, C., {Marshall}, J.~P., {Mora}, A., {et~al.} 2013, \aap, 555, A11

\bibitem[{{Eker} {et~al.}(2015){Eker}, {Soydugan}, {Soydugan}, {Bilir}, {Yaz
  G{\"o}k{\c c}e}, {Steer}, {T{\"u}ys{\"u}z}, {{\c S}eny{\"u}z}, \&
  {Demircan}}]{eker2015}
{Eker}, Z., {Soydugan}, F., {Soydugan}, E., {et~al.} 2015, \aj, 149, 131

\bibitem[{{Ertel} {et~al.}(2012){Ertel}, {Wolf}, {Marshall}, {Eiroa},
  {Augereau}, {Krivov}, {L{\"o}hne}, {Absil}, {Ardila}, {Ar{\'e}valo}, {Bayo},
  {Bryden}, {del Burgo}, {Greaves}, {Kennedy}, {Lebreton}, {Liseau},
  {Maldonado}, {Montesinos}, {Mora}, {Pilbratt}, {Sanz-Forcada}, {Stapelfeldt},
  \& {White}}]{ertel2012b}
{Ertel}, S., {Wolf}, S., {Marshall}, J.~P., {et~al.} 2012, \aap, 541, A148

\bibitem[{{Ertel} {et~al.}(2014){Ertel}, {Absil}, {Defr{\`e}re}, {Le Bouquin},
  {Augereau}, {Marion}, {Blind}, {Bonsor}, {Bryden}, {Lebreton}, \&
  {Milli}}]{ertel2014}
{Ertel}, S., {Absil}, O., {Defr{\`e}re}, D., {et~al.} 2014, \aap, 570, A128

\bibitem[{{G{\'a}sp{\'a}r} {et~al.}(2012){G{\'a}sp{\'a}r}, {Psaltis}, {Rieke},
  \& {{\"O}zel}}]{gaspar2012}
{G{\'a}sp{\'a}r}, A., {Psaltis}, D., {Rieke}, G.~H., \& {{\"O}zel}, F. 2012,
  \apj, 754, 74

\bibitem[{{Geiler} \& {Krivov}(2017)}]{geiler2017}
{Geiler}, F., \& {Krivov}, A.~V. 2017, \mnras, 468, 959

\bibitem[{{Greaves} {et~al.}(2012){Greaves}, {Hales}, {Mason}, \&
  {Matthews}}]{greaves2012}
{Greaves}, J.~S., {Hales}, A.~S., {Mason}, B.~S., \& {Matthews}, B.~C. 2012,
  \mnras, 423, L70

\bibitem[{{Hillenbrand} {et~al.}(2008){Hillenbrand}, {Carpenter}, {Kim},
  {Meyer}, {Backman}, {Moro-Mart{\'{\i}}n}, {Hollenbach}, {Hines}, {Pascucci},
  \& {Bouwman}}]{hillenbrand2008}
{Hillenbrand}, L.~A., {Carpenter}, J.~M., {Kim}, J.~S., {et~al.} 2008, \apj,
  677, 630

\bibitem[{{Hines} {et~al.}(2007){Hines}, {Schneider}, {Hollenbach}, {Mamajek},
  {Hillenbrand}, {Metchev}, {Meyer}, {Carpenter}, {Moro-Mart{\'{\i}}n},
  {Silverstone}, {Kim}, {Henning}, {Bouwman}, \& {Wolf}}]{hines2007}
{Hines}, D.~C., {Schneider}, G., {Hollenbach}, D., {et~al.} 2007, \apjl, 671,
  L165

\bibitem[{{Houck} {et~al.}(2004){Houck}, {Roellig}, {van Cleve}, {Forrest},
  {Herter}, {Lawrence}, {Matthews}, {Reitsema}, {Soifer}, {Watson}, {Weedman},
  {Huisjen}, {Troeltzsch}, {Barry}, {Bernard-Salas}, {Blacken}, {Brandl},
  {Charmandaris}, {Devost}, {Gull}, {Hall}, {Henderson}, {Higdon}, {Pirger},
  {Schoenwald}, {Sloan}, {Uchida}, {Appleton}, {Armus}, {Burgdorf},
  {Fajardo-Acosta}, {Grillmair}, {Ingalls}, {Morris}, \& {Teplitz}}]{houck2004}
{Houck}, J.~R., {Roellig}, T.~L., {van Cleve}, J., {et~al.} 2004, \apjs, 154,
  18

\bibitem[{{Hughes} {et~al.}(2011){Hughes}, {Wilner}, {Andrews}, {Williams},
  {Su}, {Murray-Clay}, \& {Qi}}]{hughes2011}
{Hughes}, A.~M., {Wilner}, D.~J., {Andrews}, S.~M., {et~al.} 2011, \apj, 740,
  38

\bibitem[{{Kalas} {et~al.}(2007){Kalas}, {Duchene}, {Fitzgerald}, \&
  {Graham}}]{kalas2007}
{Kalas}, P., {Duchene}, G., {Fitzgerald}, M.~P., \& {Graham}, J.~R. 2007,
  \apjl, 671, L161

\bibitem[{{Kalas} {et~al.}(2005){Kalas}, {Graham}, \& {Clampin}}]{kalas2005}
{Kalas}, P., {Graham}, J.~R., \& {Clampin}, M. 2005, \nat, 435, 1067

\bibitem[{{Kennedy} \& {Kenyon}(2008)}]{kennedy2008}
{Kennedy}, G.~M., \& {Kenyon}, S.~J. 2008, \apj, 673, 502

\bibitem[{{Kennedy} \& {Piette}(2015)}]{kennedy2015}
{Kennedy}, G.~M., \& {Piette}, A. 2015, \mnras, 449, 2304

\bibitem[{{Kennedy} \& {Wyatt}(2010)}]{kennedy2010}
{Kennedy}, G.~M., \& {Wyatt}, M.~C. 2010, \mnras, 405, 1253

\bibitem[{{Kennedy} \& {Wyatt}(2013)}]{kennedy2013}
---. 2013, \mnras, 433, 2334

\bibitem[{{Kennedy} \& {Wyatt}(2014)}]{kennedy2014}
---. 2014, \mnras, 444, 3164

\bibitem[{{Kobayashi} {et~al.}(2008){Kobayashi}, {Watanabe}, {Kimura}, \&
  {Yamamoto}}]{kobayashi2008}
{Kobayashi}, H., {Watanabe}, S.-i., {Kimura}, H., \& {Yamamoto}, T. 2008,
  Icarus, 195, 871

\bibitem[{{Kretke} \& {Lin}(2007)}]{kretke2007}
{Kretke}, K.~A., \& {Lin}, D.~N.~C. 2007, \apjl, 664, L55

\bibitem[{{Lebouteiller} {et~al.}(2011){Lebouteiller}, {Barry}, {Spoon},
  {Bernard-Salas}, {Sloan}, {Houck}, \& {Weedman}}]{lebouteiller2011}
{Lebouteiller}, V., {Barry}, D.~J., {Spoon}, H.~W.~W., {et~al.} 2011, \apjs,
  196, 8

\bibitem[{{Lebreton} {et~al.}(2012){Lebreton}, {Augereau}, {Thi}, {Roberge},
  {Donaldson}, {Schneider}, {Maddison}, {M{\'e}nard}, {Riviere-Marichalar},
  {Mathews}, {Kamp}, {Pinte}, {Dent}, {Barrado}, {Duch{\^e}ne}, {Gonzalez},
  {Grady}, {Meeus}, {Pantin}, {Williams}, \& {Woitke}}]{lebreton2012}
{Lebreton}, J., {Augereau}, J.-C., {Thi}, W.-F., {et~al.} 2012, \aap, 539, A17

\bibitem[{{Levison} \& {Duncan}(1997)}]{levison1997}
{Levison}, H.~F., \& {Duncan}, M.~J. 1997, \icarus, 127, 13

\bibitem[{{Lieman-Sifry} {et~al.}(2016){Lieman-Sifry}, {Hughes}, {Carpenter},
  {Gorti}, {Hales}, \& {Flaherty}}]{lieman-sifry2016}
{Lieman-Sifry}, J., {Hughes}, A.~M., {Carpenter}, J.~M., {et~al.} 2016, \apj,
  828, 25

\bibitem[{{Liseau} {et~al.}(2008){Liseau}, {Risacher}, {Brandeker}, {Eiroa},
  {Fridlund}, {Nilsson}, {Olofsson}, {Pilbratt}, \&
  {Th{\'e}bault}}]{liseau2008}
{Liseau}, R., {Risacher}, C., {Brandeker}, A., {et~al.} 2008, \aap, 480, L47

\bibitem[{{MacGregor} {et~al.}(2015){MacGregor}, {Wilner}, {Andrews}, \&
  {Hughes}}]{macgregor2015}
{MacGregor}, M.~A., {Wilner}, D.~J., {Andrews}, S.~M., \& {Hughes}, A.~M. 2015,
  \apj, 801, 59

\bibitem[{{MacGregor} {et~al.}(2016){MacGregor}, {Wilner}, {Chandler}, {Ricci},
  {Maddison}, {Cranmer}, {Andrews}, {Hughes}, \& {Steele}}]{macgregor2016}
{MacGregor}, M.~A., {Wilner}, D.~J., {Chandler}, C., {et~al.} 2016, \apj, 823,
  79

\bibitem[{{MacGregor} {et~al.}(2017){MacGregor}, {Matra}, {Kalas}, {Wilner},
  {Pan}, {Kennedy}, {Wyatt}, {Duchene}, {Hughes}, {Rieke}, {Clampin},
  {Fitzgerald}, {Graham}, {Holland}, {Panic}, {Shannon}, \&
  {Su}}]{macgregor2017}
{MacGregor}, M.~A., {Matra}, L., {Kalas}, P., {et~al.} 2017, ArXiv e-prints,
  arXiv:1705.05867

\bibitem[{{Maness} {et~al.}(2008){Maness}, {Fitzgerald}, {Paladini}, {Kalas},
  {Duchene}, \& {Graham}}]{maness2008}
{Maness}, H.~L., {Fitzgerald}, M.~P., {Paladini}, R., {et~al.} 2008, \apjl,
  686, L25

\bibitem[{{Marino} {et~al.}(2016){Marino}, {Matr{\`a}}, {Stark}, {Wyatt},
  {Casassus}, {Kennedy}, {Rodriguez}, {Zuckerman}, {Perez}, {Dent}, {Kuchner},
  {Hughes}, {Schneider}, {Steele}, {Roberge}, {Donaldson}, \&
  {Nesvold}}]{marino2016}
{Marino}, S., {Matr{\`a}}, L., {Stark}, C., {et~al.} 2016, \mnras, 460, 2933

\bibitem[{{Martin} \& {Livio}(2013{\natexlab{a}})}]{martin2013}
{Martin}, R.~G., \& {Livio}, M. 2013{\natexlab{a}}, \mnras, 434, 633

\bibitem[{{Martin} \& {Livio}(2013{\natexlab{b}})}]{martin2013b}
---. 2013{\natexlab{b}}, \mnras, 428, L11

\bibitem[{{Matthews} {et~al.}(2014{\natexlab{a}}){Matthews}, {Kennedy},
  {Sibthorpe}, {Booth}, {Wyatt}, {Broekhoven-Fiene}, {Macintosh}, \&
  {Marois}}]{matthews2014b}
{Matthews}, B., {Kennedy}, G., {Sibthorpe}, B., {et~al.} 2014{\natexlab{a}},
  \apj, 780, 97

\bibitem[{{Matthews} {et~al.}(2014{\natexlab{b}}){Matthews}, {Krivov}, {Wyatt},
  {Bryden}, \& {Eiroa}}]{matthews2014}
{Matthews}, B.~C., {Krivov}, A.~V., {Wyatt}, M.~C., {Bryden}, G., \& {Eiroa},
  C. 2014{\natexlab{b}}, ArXiv e-prints, arXiv:1401.0743

\bibitem[{{McDonald} {et~al.}(2012){McDonald}, {Zijlstra}, \&
  {Boyer}}]{mcdonald2012}
{McDonald}, I., {Zijlstra}, A.~A., \& {Boyer}, M.~L. 2012, \mnras, 427, 343

\bibitem[{{Meeus} {et~al.}(2012){Meeus}, {Montesinos}, {Mendigut{\'{\i}}a},
  {Kamp}, {Thi}, {Eiroa}, {Grady}, {Mathews}, {Sandell}, {Martin-Za{\"i}di},
  {Brittain}, {Dent}, {Howard}, {M{\'e}nard}, {Pinte}, {Roberge},
  {Vandenbussche}, \& {Williams}}]{meeus2012}
{Meeus}, G., {Montesinos}, B., {Mendigut{\'{\i}}a}, I., {et~al.} 2012, \aap,
  544, A78

\bibitem[{{Min} {et~al.}(2011){Min}, {Dullemond}, {Kama}, \&
  {Dominik}}]{min2011}
{Min}, M., {Dullemond}, C.~P., {Kama}, M., \& {Dominik}, C. 2011, Icarus, 212,
  416

\bibitem[{{Mo{\'o}r} {et~al.}(2006){Mo{\'o}r}, {{\'A}brah{\'a}m}, {Derekas},
  {Kiss}, {Kiss}, {Apai}, {Grady}, \& {Henning}}]{moor2006}
{Mo{\'o}r}, A., {{\'A}brah{\'a}m}, P., {Derekas}, A., {et~al.} 2006, \apj, 644,
  525

\bibitem[{{Mo{\'o}r} {et~al.}(2009){Mo{\'o}r}, {Apai}, {Pascucci},
  {{\'A}brah{\'a}m}, {Grady}, {Henning}, {Juh{\'a}sz}, {Kiss}, \&
  {K{\'o}sp{\'a}l}}]{moor2009}
{Mo{\'o}r}, A., {Apai}, D., {Pascucci}, I., {et~al.} 2009, \apjl, 700, L25

\bibitem[{{Mo{\'o}r} {et~al.}(2011){Mo{\'o}r}, {Pascucci}, {K{\'o}sp{\'a}l},
  {{\'A}brah{\'a}m}, {Csengeri}, {Kiss}, {Apai}, {Grady}, {Henning}, {Kiss},
  {Bayliss}, {Juh{\'a}sz}, {Kov{\'a}cs}, \& {Szalai}}]{moor2011}
{Mo{\'o}r}, A., {Pascucci}, I., {K{\'o}sp{\'a}l}, {\'A}., {et~al.} 2011, \apjs,
  193, 4

\bibitem[{{Mo{\'o}r} {et~al.}(2015{\natexlab{a}}){Mo{\'o}r}, {Henning},
  {Juh{\'a}sz}, {{\'A}brah{\'a}m}, {Balog}, {K{\'o}sp{\'a}l}, {Pascucci},
  {Szab{\'o}}, {Vavrek}, {Cur{\'e}}, {Csengeri}, {Grady}, {G{\"u}sten}, \&
  {Kiss}}]{moor2015b}
{Mo{\'o}r}, A., {Henning}, T., {Juh{\'a}sz}, A., {et~al.} 2015{\natexlab{a}},
  \apj, 814, 42

\bibitem[{{Mo{\'o}r} {et~al.}(2015{\natexlab{b}}){Mo{\'o}r}, {K{\'o}sp{\'a}l},
  {{\'A}brah{\'a}m}, {Apai}, {Balog}, {Grady}, {Henning}, {Juh{\'a}sz}, {Kiss},
  {Krivov}, {Pawellek}, \& {Szab{\'o}}}]{moor2015}
{Mo{\'o}r}, A., {K{\'o}sp{\'a}l}, {\'A}., {{\'A}brah{\'a}m}, P., {et~al.}
  2015{\natexlab{b}}, \mnras, 447, 577

\bibitem[{{Morales} {et~al.}(2013){Morales}, {Bryden}, {Werner}, \&
  {Stapelfeldt}}]{morales2013}
{Morales}, F.~Y., {Bryden}, G., {Werner}, M.~W., \& {Stapelfeldt}, K.~R. 2013,
  \apj, 776, 111

\bibitem[{{Morales} {et~al.}(2011){Morales}, {Rieke}, {Werner}, {Bryden},
  {Stapelfeldt}, \& {Su}}]{morales2011}
{Morales}, F.~Y., {Rieke}, G.~H., {Werner}, M.~W., {et~al.} 2011, \apjl, 730,
  L29

\bibitem[{{Muzerolle} {et~al.}(2005){Muzerolle}, {Luhman}, {Brice{\~n}o},
  {Hartmann}, \& {Calvet}}]{muzerolle2005}
{Muzerolle}, J., {Luhman}, K.~L., {Brice{\~n}o}, C., {Hartmann}, L., \&
  {Calvet}, N. 2005, \apj, 625, 906

\bibitem[{{Natta} {et~al.}(2006){Natta}, {Testi}, \& {Randich}}]{natta2006}
{Natta}, A., {Testi}, L., \& {Randich}, S. 2006, \aap, 452, 245

\bibitem[{{Nesvorn{\'y}} {et~al.}(2010){Nesvorn{\'y}}, {Jenniskens}, {Levison},
  {Bottke}, {Vokrouhlick{\'y}}, \& {Gounelle}}]{nesvorny2010}
{Nesvorn{\'y}}, D., {Jenniskens}, P., {Levison}, H.~F., {et~al.} 2010, \apj,
  713, 816

\bibitem[{{Nilsson} {et~al.}(2009){Nilsson}, {Liseau}, {Brandeker}, {Olofsson},
  {Risacher}, {Fridlund}, \& {Pilbratt}}]{nilsson2009}
{Nilsson}, R., {Liseau}, R., {Brandeker}, A., {et~al.} 2009, \aap, 508, 1057

\bibitem[{{Nilsson} {et~al.}(2010){Nilsson}, {Liseau}, {Brandeker}, {Olofsson},
  {Pilbratt}, {Risacher}, {Rodmann}, {Augereau}, {Bergman}, {Eiroa},
  {Fridlund}, {Th{\'e}bault}, \& {White}}]{nilsson2010}
---. 2010, \aap, 518, A40

\bibitem[{{Olofsson} {et~al.}(2016){Olofsson}, {Samland}, {Avenhaus},
  {Caceres}, {Henning}, {Mo{\'o}r}, {Milli}, {Canovas}, {Quanz}, {Schreiber},
  {Augereau}, {Bayo}, {Bazzon}, {Beuzit}, {Boccaletti}, {Buenzli}, {Casassus},
  {Chauvin}, {Dominik}, {Desidera}, {Feldt}, {Gratton}, {Janson}, {Lagrange},
  {Langlois}, {Lannier}, {Maire}, {Mesa}, {Pinte}, {Rouan}, {Salter},
  {Thalmann}, \& {Vigan}}]{olofsson2016}
{Olofsson}, J., {Samland}, M., {Avenhaus}, H., {et~al.} 2016, \aap, 591, A108

\bibitem[{{Pani{\'c}} {et~al.}(2013){Pani{\'c}}, {Holland}, {Wyatt}, {Kennedy},
  {Matthews}, {Lestrade}, {Sibthorpe}, {Greaves}, {Marshall}, {Phillips}, \&
  {Tottle}}]{panic2013}
{Pani{\'c}}, O., {Holland}, W.~S., {Wyatt}, M.~C., {et~al.} 2013, \mnras, 435,
  1037

\bibitem[{{Pascual} {et~al.}(2016){Pascual}, {Montesinos}, {Meeus}, {Marshall},
  {Mendigut{\'{\i}}a}, \& {Sandell}}]{pascual2016}
{Pascual}, N., {Montesinos}, B., {Meeus}, G., {et~al.} 2016, \aap, 586, A6

\bibitem[{{Pawellek} {et~al.}(2014){Pawellek}, {Krivov}, {Marshall},
  {Montesinos}, {{\'A}brah{\'a}m}, {Mo{\'o}r}, {Bryden}, \&
  {Eiroa}}]{pawellek2014}
{Pawellek}, N., {Krivov}, A.~V., {Marshall}, J.~P., {et~al.} 2014, \apj, 792,
  65

\bibitem[{{Pecaut} \& {Mamajek}(2013)}]{pecaut2013}
{Pecaut}, M.~J., \& {Mamajek}, E.~E. 2013, \apjs, 208, 9

\bibitem[{{Petit} {et~al.}(2001){Petit}, {Morbidelli}, \&
  {Chambers}}]{petit2001}
{Petit}, J.-M., {Morbidelli}, A., \& {Chambers}, J. 2001, Icarus, 153, 338

\bibitem[{{Ricarte} {et~al.}(2013){Ricarte}, {Moldvai}, {Hughes},
  {Duch{\^e}ne}, {Williams}, {Andrews}, \& {Wilner}}]{ricarte2013}
{Ricarte}, A., {Moldvai}, N., {Hughes}, A.~M., {et~al.} 2013, \apj, 774, 80

\bibitem[{{Ricci} {et~al.}(2015{\natexlab{a}}){Ricci}, {Carpenter}, {Fu},
  {Hughes}, {Corder}, \& {Isella}}]{ricci2015}
{Ricci}, L., {Carpenter}, J.~M., {Fu}, B., {et~al.} 2015{\natexlab{a}}, \apj,
  798, 124

\bibitem[{{Ricci} {et~al.}(2015{\natexlab{b}}){Ricci}, {Maddison}, {Wilner},
  {MacGregor}, {Ubach}, {Carpenter}, \& {Testi}}]{ricci2015b}
{Ricci}, L., {Maddison}, S.~T., {Wilner}, D., {et~al.} 2015{\natexlab{b}},
  \apj, 813, 138

\bibitem[{{Rickman} {et~al.}(2017){Rickman}, {Gabryszewski}, {Wajer},
  {Wi{\'s}niowski}, {W{\'o}jcikowski}, {Szutowicz}, {Valsecchi}, \&
  {Morbidelli}}]{rickman2017}
{Rickman}, H., {Gabryszewski}, R., {Wajer}, P., {et~al.} 2017, \aap, 598, A110

\bibitem[{{Rieke} {et~al.}(2016){Rieke}, {G{\'a}sp{\'a}r}, \&
  {Ballering}}]{rieke2016}
{Rieke}, G.~H., {G{\'a}sp{\'a}r}, A., \& {Ballering}, N.~P. 2016, \apj, 816, 50

\bibitem[{{Rieke} {et~al.}(2004){Rieke}, {Young}, {Engelbracht}, {Kelly},
  {Low}, {Haller}, {Beeman}, {Gordon}, {Stansberry}, {Misselt}, {Cadien},
  {Morrison}, {Rivlis}, {Latter}, {Noriega-Crespo}, {Padgett}, {Stapelfeldt},
  {Hines}, {Egami}, {Muzerolle}, {Alonso-Herrero}, {Blaylock}, {Dole}, {Hinz},
  {Le Floc'h}, {Papovich}, {P{\'e}rez-Gonz{\'a}lez}, {Smith}, {Su}, {Bennett},
  {Frayer}, {Henderson}, {Lu}, {Masci}, {Pesenson}, {Rebull}, {Rho}, {Keene},
  {Stolovy}, {Wachter}, {Wheaton}, {Werner}, \& {Richards}}]{rieke2004}
{Rieke}, G.~H., {Young}, E.~T., {Engelbracht}, C.~W., {et~al.} 2004, \apjs,
  154, 25

\bibitem[{{Rieke} {et~al.}(2005){Rieke}, {Su}, {Stansberry}, {Trilling},
  {Bryden}, {Muzerolle}, {White}, {Gorlova}, {Young}, {Beichman},
  {Stapelfeldt}, \& {Hines}}]{rieke2005}
{Rieke}, G.~H., {Su}, K.~Y.~L., {Stansberry}, J.~A., {et~al.} 2005, \apj, 620,
  1010

\bibitem[{{Riviere-Marichalar} {et~al.}(2014){Riviere-Marichalar}, {Barrado},
  {Montesinos}, {Duch{\^e}ne}, {Bouy}, {Pinte}, {Menard}, {Donaldson}, {Eiroa},
  {Krivov}, {Kamp}, {Mendigut{\'{\i}}a}, {Dent}, \&
  {Lillo-Box}}]{riviere-marichalar2014}
{Riviere-Marichalar}, P., {Barrado}, D., {Montesinos}, B., {et~al.} 2014, \aap,
  565, A68

\bibitem[{{Roccatagliata} {et~al.}(2009){Roccatagliata}, {Henning}, {Wolf},
  {Rodmann}, {Corder}, {Carpenter}, {Meyer}, \& {Dowell}}]{roccatagliata2009}
{Roccatagliata}, V., {Henning}, T., {Wolf}, S., {et~al.} 2009, \aap, 497, 409

\bibitem[{{Schneider} {et~al.}(2006){Schneider}, {Silverstone}, {Hines},
  {Augereau}, {Pinte}, {M{\'e}nard}, {Krist}, {Clampin}, {Grady}, {Golimowski},
  {Ardila}, {Henning}, {Wolf}, \& {Rodmann}}]{schneider2006}
{Schneider}, G., {Silverstone}, M.~D., {Hines}, D.~C., {et~al.} 2006, \apj,
  650, 414

\bibitem[{{Schneider} {et~al.}(2014){Schneider}, {Grady}, {Hines}, {Stark},
  {Debes}, {Carson}, {Kuchner}, {Perrin}, {Weinberger}, {Wisniewski},
  {Silverstone}, {Jang-Condell}, {Henning}, {Woodgate}, {Serabyn},
  {Moro-Martin}, {Tamura}, {Hinz}, \& {Rodigas}}]{schneider2014}
{Schneider}, G., {Grady}, C.~A., {Hines}, D.~C., {et~al.} 2014, \aj, 148, 59

\bibitem[{{Sch{\"u}tz} {et~al.}(2005){Sch{\"u}tz}, {Meeus}, \&
  {Sterzik}}]{schutz2005}
{Sch{\"u}tz}, O., {Meeus}, G., \& {Sterzik}, M.~F. 2005, \aap, 431, 175

\bibitem[{{Sierchio} {et~al.}(2014){Sierchio}, {Rieke}, {Su}, \&
  {G{\'a}sp{\'a}r}}]{sierchio2014}
{Sierchio}, J.~M., {Rieke}, G.~H., {Su}, K.~Y.~L., \& {G{\'a}sp{\'a}r}, A.
  2014, \apj, 785, 33

\bibitem[{{Stapelfeldt} {et~al.}(2004){Stapelfeldt}, {Holmes}, {Chen}, {Rieke},
  {Su}, {Hines}, {Werner}, {Beichman}, {Jura}, {Padgett}, {Stansberry},
  {Bendo}, {Cadien}, {Marengo}, {Thompson}, {Velusamy}, {Backus}, {Blaylock},
  {Egami}, {Engelbracht}, {Frayer}, {Gordon}, {Keene}, {Latter}, {Megeath},
  {Misselt}, {Morrison}, {Muzerolle}, {Noriega-Crespo}, {Van Cleve}, \&
  {Young}}]{stapelfeldt2004}
{Stapelfeldt}, K.~R., {Holmes}, E.~K., {Chen}, C., {et~al.} 2004, \apjs, 154,
  458

\bibitem[{{Steele} {et~al.}(2016){Steele}, {Hughes}, {Carpenter}, {Ricarte},
  {Andrews}, {Wilner}, \& {Chiang}}]{steele2016}
{Steele}, A., {Hughes}, A.~M., {Carpenter}, J., {et~al.} 2016, \apj, 816, 27

\bibitem[{{Stevenson} \& {Lunine}(1988)}]{stevenson1988}
{Stevenson}, D.~J., \& {Lunine}, J.~I. 1988, \icarus, 75, 146

\bibitem[{{Su} \& {Rieke}(2014)}]{su2014}
{Su}, K.~Y.~L., \& {Rieke}, G.~H. 2014, in IAU Symposium, Vol. 299, IAU
  Symposium, ed. M.~{Booth}, B.~C. {Matthews}, \& J.~R. {Graham}, 318--321

\bibitem[{{Su} {et~al.}(2016){Su}, {Rieke}, {Defr{\'e}re}, {Wang}, {Lai},
  {Wilner}, {van Lieshout}, \& {Lee}}]{su2016}
{Su}, K.~Y.~L., {Rieke}, G.~H., {Defr{\'e}re}, D., {et~al.} 2016, \apj, 818, 45

\bibitem[{{Su} {et~al.}(2008){Su}, {Rieke}, {Stapelfeldt}, {Smith}, {Bryden},
  {Chen}, \& {Trilling}}]{su2008}
{Su}, K.~Y.~L., {Rieke}, G.~H., {Stapelfeldt}, K.~R., {et~al.} 2008, \apjl,
  679, L125

\bibitem[{{Su} {et~al.}(2005){Su}, {Rieke}, {Misselt}, {Stansberry},
  {Moro-Martin}, {Stapelfeldt}, {Werner}, {Trilling}, {Bendo}, {Gordon},
  {Hines}, {Wyatt}, {Holland}, {Marengo}, {Megeath}, \& {Fazio}}]{su2005}
{Su}, K.~Y.~L., {Rieke}, G.~H., {Misselt}, K.~A., {et~al.} 2005, \apj, 628, 487

\bibitem[{{Su} {et~al.}(2006){Su}, {Rieke}, {Stansberry}, {Bryden},
  {Stapelfeldt}, {Trilling}, {Muzerolle}, {Beichman}, {Moro-Martin}, {Hines},
  \& {Werner}}]{su2006}
{Su}, K.~Y.~L., {Rieke}, G.~H., {Stansberry}, J.~A., {et~al.} 2006, \apj, 653,
  675

\bibitem[{{Su} {et~al.}(2009){Su}, {Rieke}, {Stapelfeldt}, {Malhotra},
  {Bryden}, {Smith}, {Misselt}, {Moro-Martin}, \& {Williams}}]{su2009}
{Su}, K.~Y.~L., {Rieke}, G.~H., {Stapelfeldt}, K.~R., {et~al.} 2009, \apj, 705,
  314

\bibitem[{{Su} {et~al.}(2013){Su}, {Rieke}, {Malhotra}, {Stapelfeldt},
  {Hughes}, {Bonsor}, {Wilner}, {Balog}, {Watson}, {Werner}, \&
  {Misselt}}]{su2013}
{Su}, K.~Y.~L., {Rieke}, G.~H., {Malhotra}, R., {et~al.} 2013, \apj, 763, 118

\bibitem[{{Tanner} {et~al.}(2009){Tanner}, {Beichman}, {Bryden}, {Lisse}, \&
  {Lawler}}]{tanner2009}
{Tanner}, A., {Beichman}, C., {Bryden}, G., {Lisse}, C., \& {Lawler}, S. 2009,
  \apj, 704, 109

\bibitem[{{Thureau} {et~al.}(2014){Thureau}, {Greaves}, {Matthews}, {Kennedy},
  {Phillips}, {Booth}, {Duch{\^e}ne}, {Horner}, {Rodriguez}, {Sibthorpe}, \&
  {Wyatt}}]{thureau2014}
{Thureau}, N.~D., {Greaves}, J.~S., {Matthews}, B.~C., {et~al.} 2014, \mnras,
  445, 2558

\bibitem[{{van Lieshout} {et~al.}(2014){van Lieshout}, {Dominik}, {Kama}, \&
  {Min}}]{vanlieshout2014}
{van Lieshout}, R., {Dominik}, C., {Kama}, M., \& {Min}, M. 2014, \aap, 571,
  A51

\bibitem[{{Vican} {et~al.}(2016){Vican}, {Schneider}, {Bryden}, {Melis},
  {Zuckerman}, {Rhee}, \& {Song}}]{vican2016}
{Vican}, L., {Schneider}, A., {Bryden}, G., {et~al.} 2016, \apj, 833, 263

\bibitem[{{Williams} \& {Andrews}(2006)}]{williams2006}
{Williams}, J.~P., \& {Andrews}, S.~M. 2006, \apj, 653, 1480

\bibitem[{{Williams} {et~al.}(2004){Williams}, {Najita}, {Liu}, {Bottinelli},
  {Carpenter}, {Hillenbrand}, {Meyer}, \& {Soderblom}}]{williams2004}
{Williams}, J.~P., {Najita}, J., {Liu}, M.~C., {et~al.} 2004, \apj, 604, 414

\bibitem[{{Wolf} \& {Voshchinnikov}(2004)}]{wolf2004}
{Wolf}, S., \& {Voshchinnikov}, N.~V. 2004, Computer Physics Communications,
  162, 113

\bibitem[{{Wyatt}(2005)}]{wyatt2005}
{Wyatt}, M.~C. 2005, \aap, 433, 1007

\bibitem[{{Wyatt}(2008)}]{wyatt2008}
---. 2008, \araa, 46, 339

\bibitem[{{Wyatt} {et~al.}(2007){Wyatt}, {Smith}, {Greaves}, {Beichman},
  {Bryden}, \& {Lisse}}]{wyatt2007b}
{Wyatt}, M.~C., {Smith}, R., {Greaves}, J.~S., {et~al.} 2007, \apj, 658, 569

\end{thebibliography}

\onecolumngrid
\clearpage


\begin{figure*}
\epsscale{1.2}
\plotone{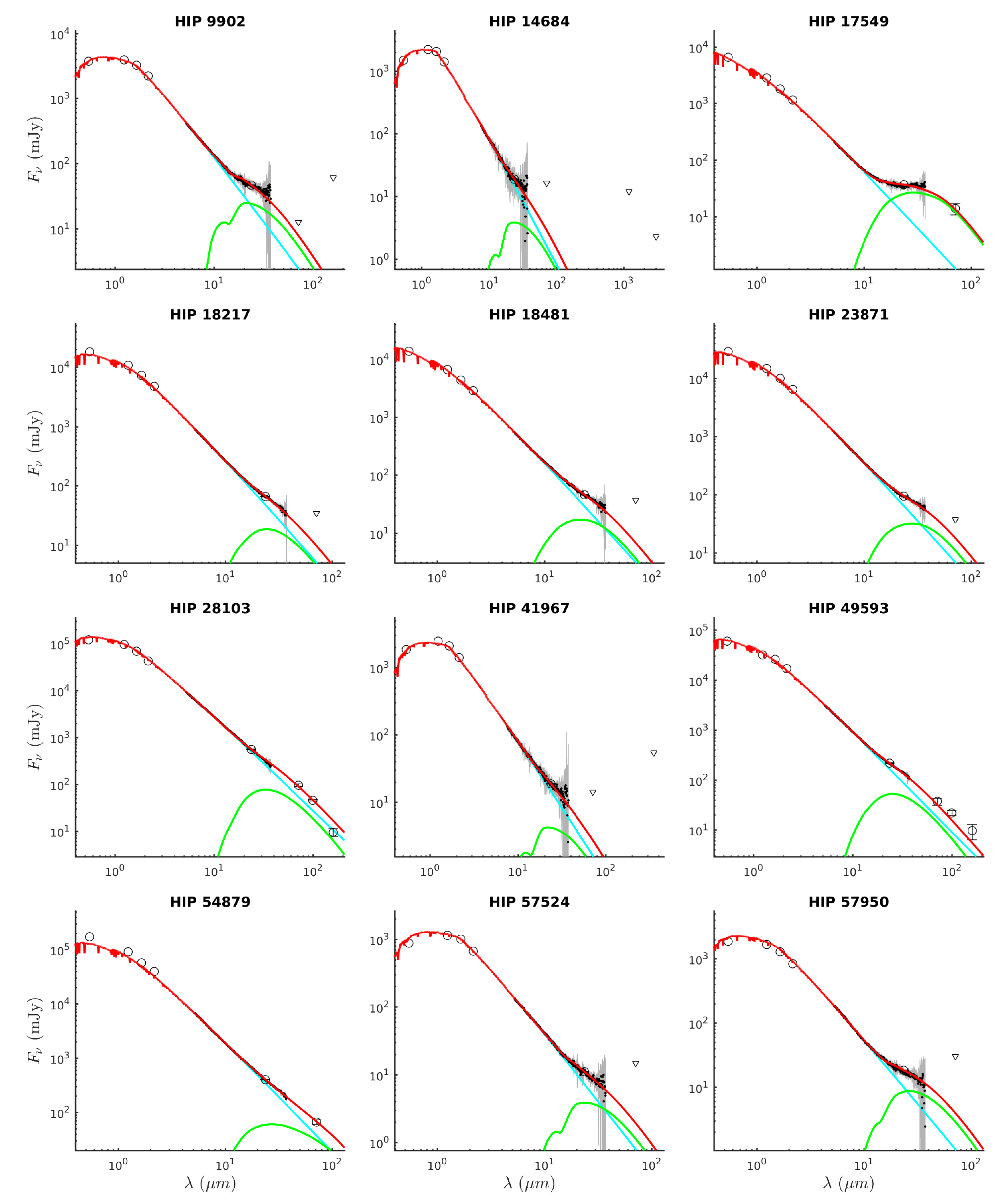}
\caption{Best fits to the 29 single-component systems. Small black points are the IRS data (with gray error bars), large black circles are photometry data, and the black triangles are photometric upper limits. The stellar photosphere is cyan, the warm dust belt model is green, and the total model is red.}
\label{fig:onecompfits}
\end{figure*}

\begin{figure*}
\epsscale{1.2}
\figurenum{7}
\plotone{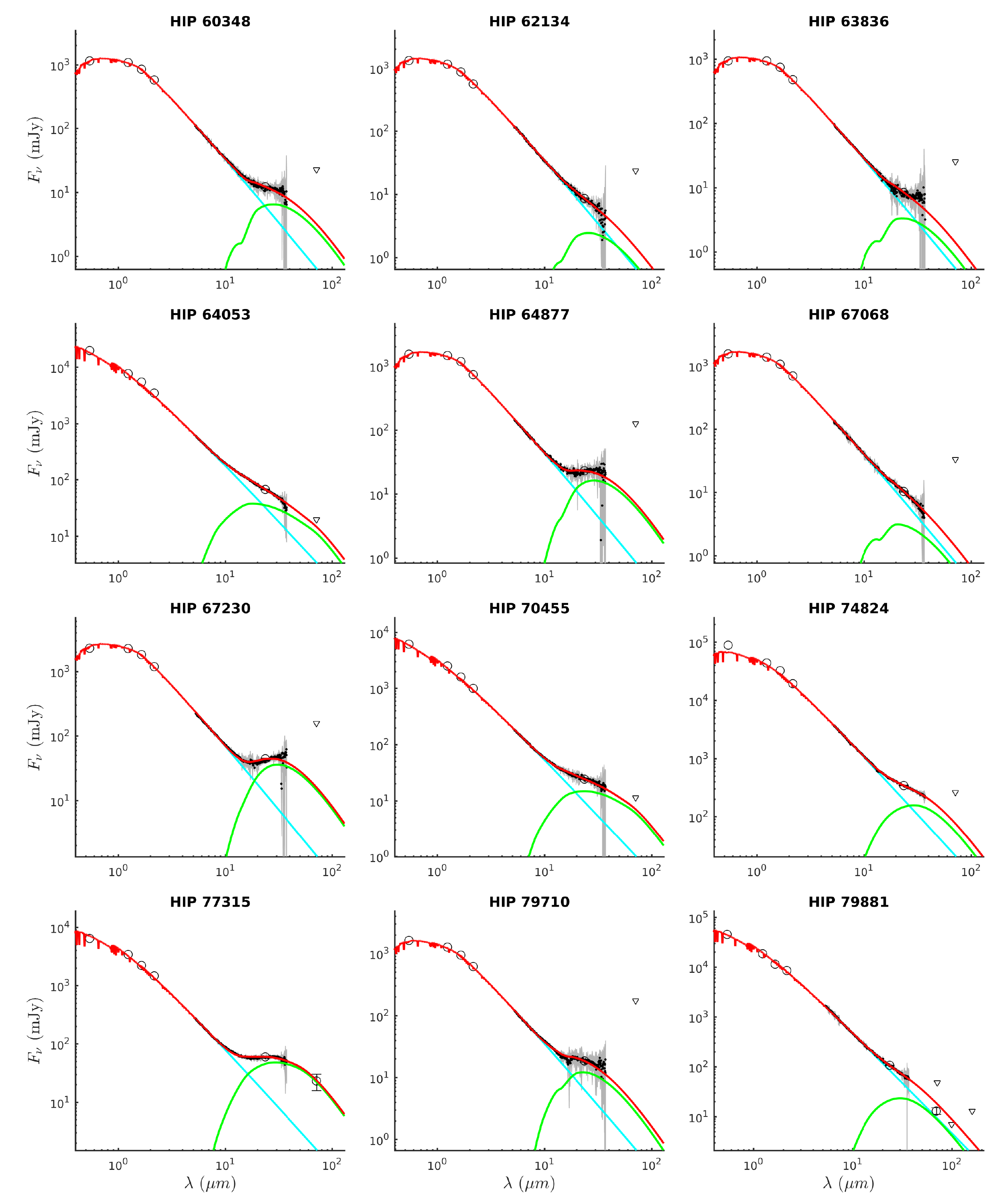}
\caption{(continued).}
\end{figure*}

\begin{figure*}
\epsscale{1.2}
\figurenum{7}
\plotone{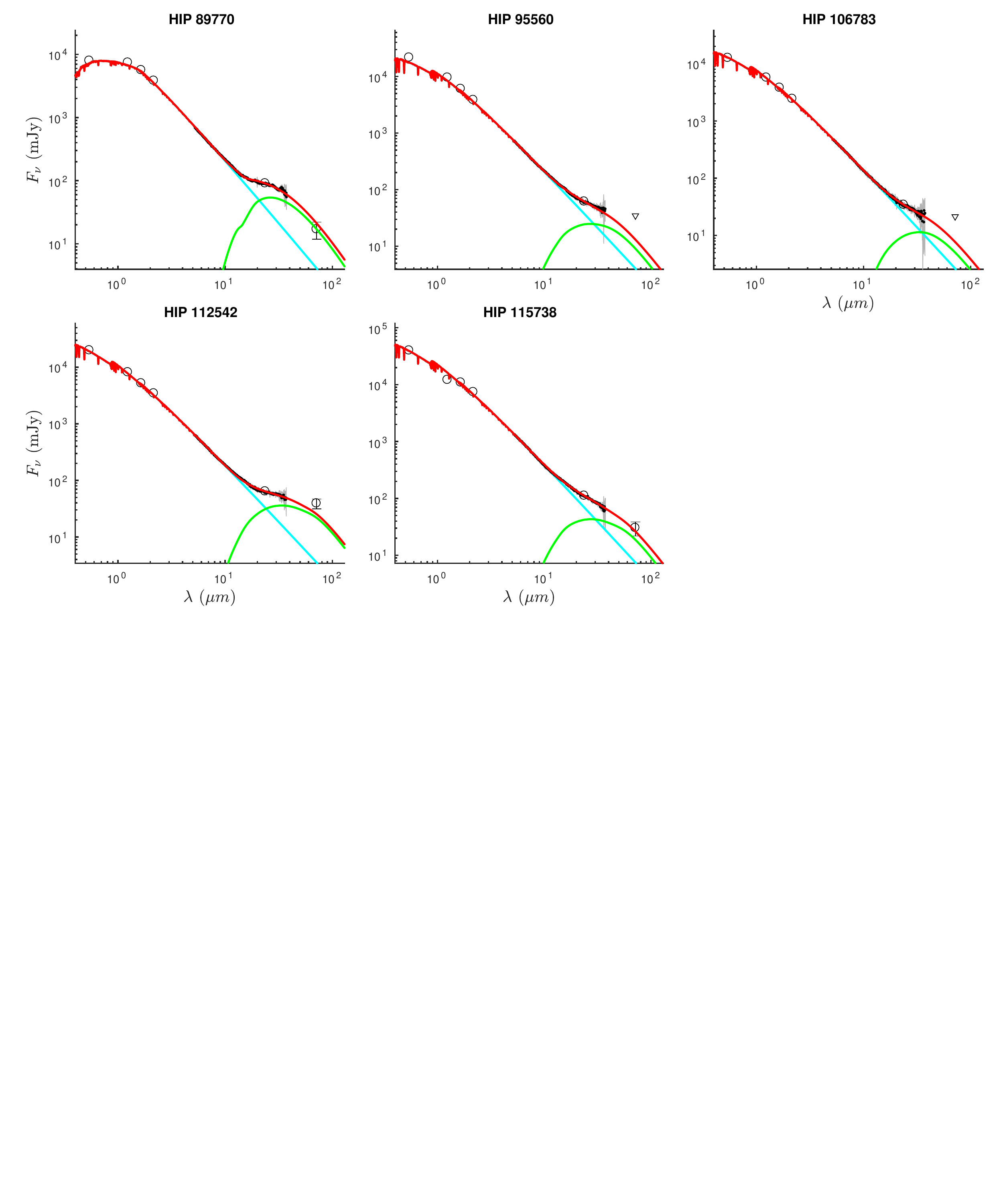}
\caption{(continued).}
\end{figure*}

\begin{figure*}
\epsscale{1.2}
\plotone{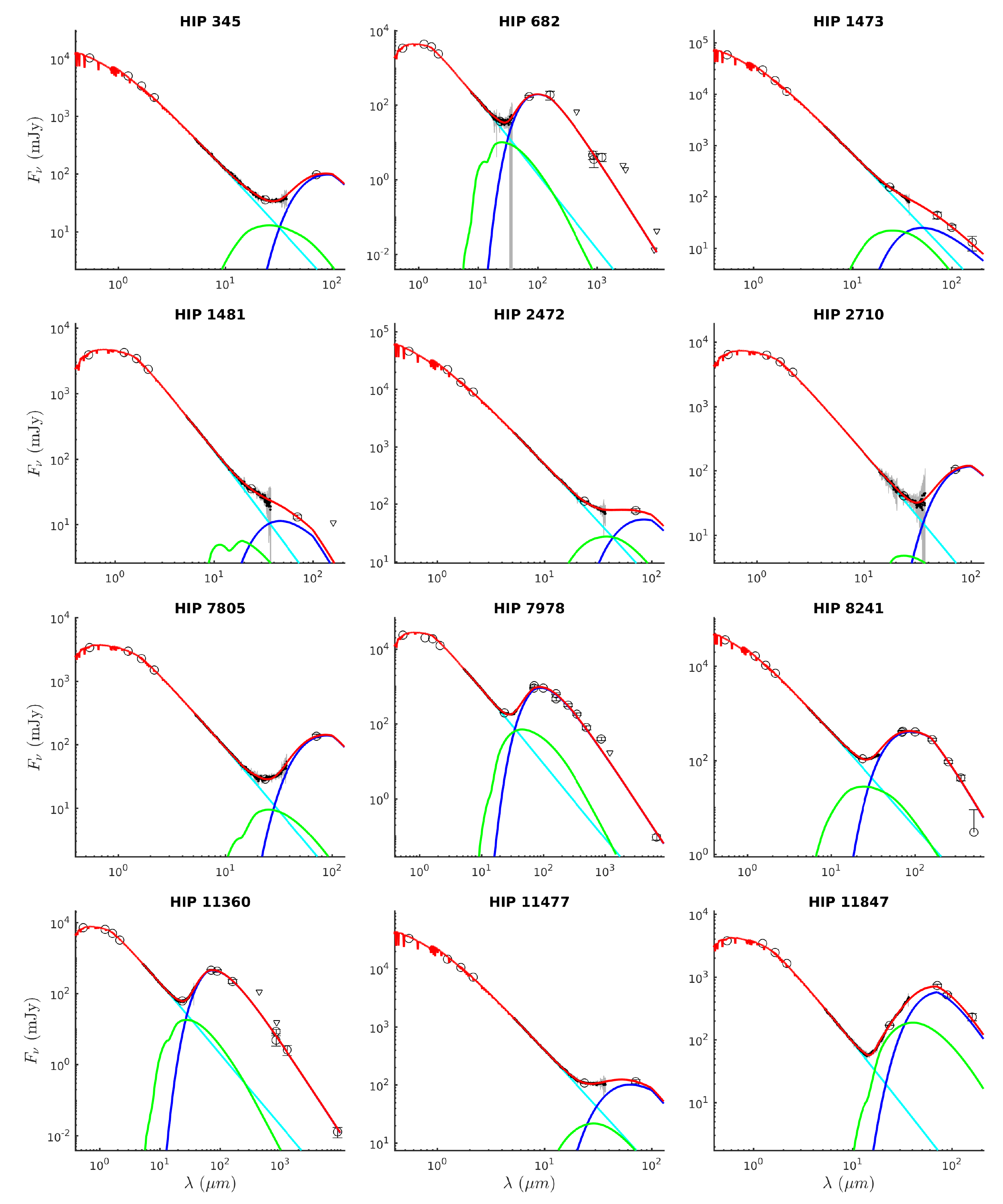}
\caption{Best fits to the 54 two-component systems. Small black points are the IRS data (with gray error bars), large black circles are photometry data, and the black triangles are photometric upper limits. The stellar photosphere is cyan, the warm dust belt model is green, the cold component is blue, and the total model is red.}
\label{fig:twocompfits}
\end{figure*}

\begin{figure*}
\epsscale{1.2}
\figurenum{8}
\plotone{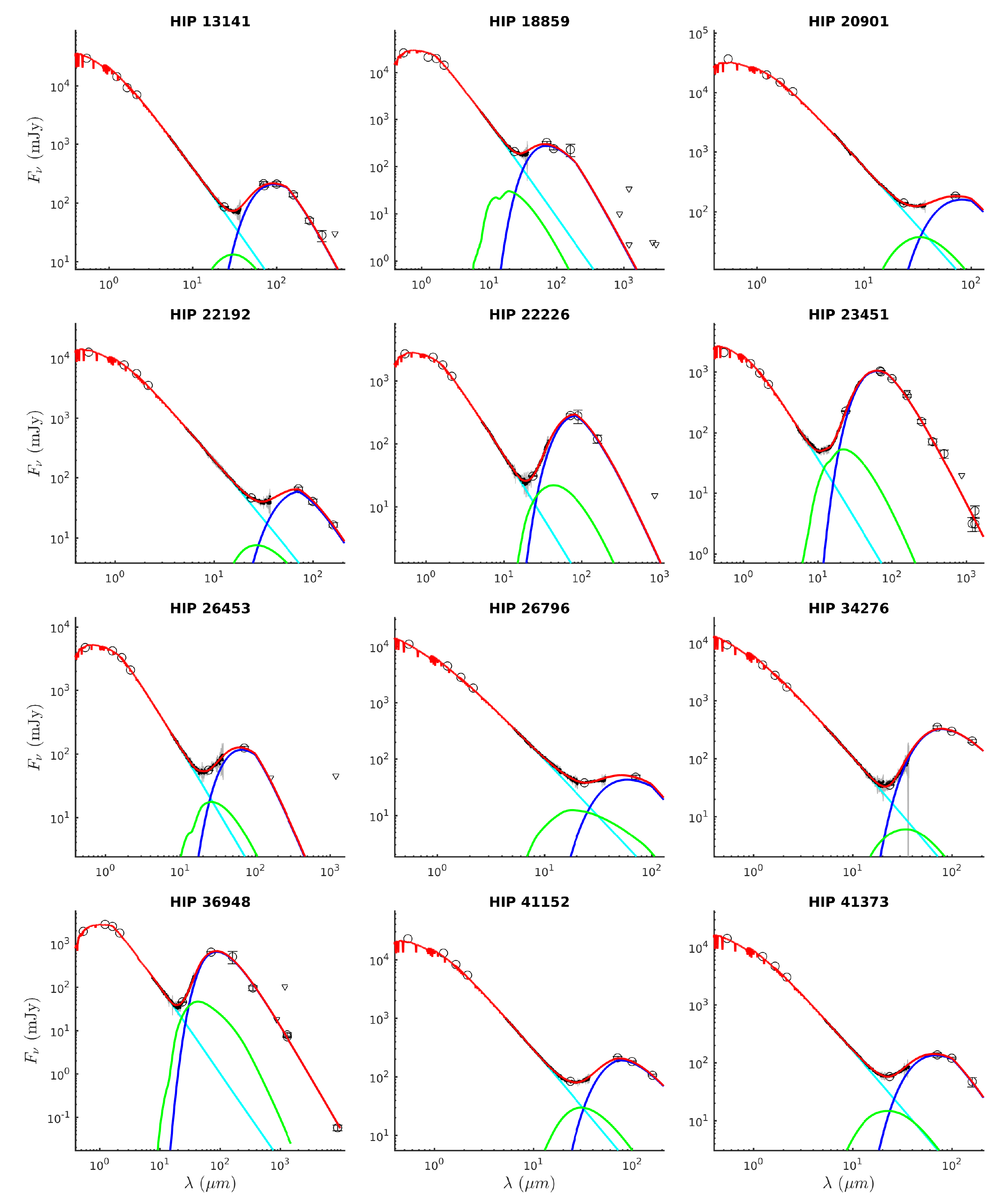}
\caption{(continued).}
\end{figure*}

\begin{figure*}
\epsscale{1.2}
\figurenum{8}
\plotone{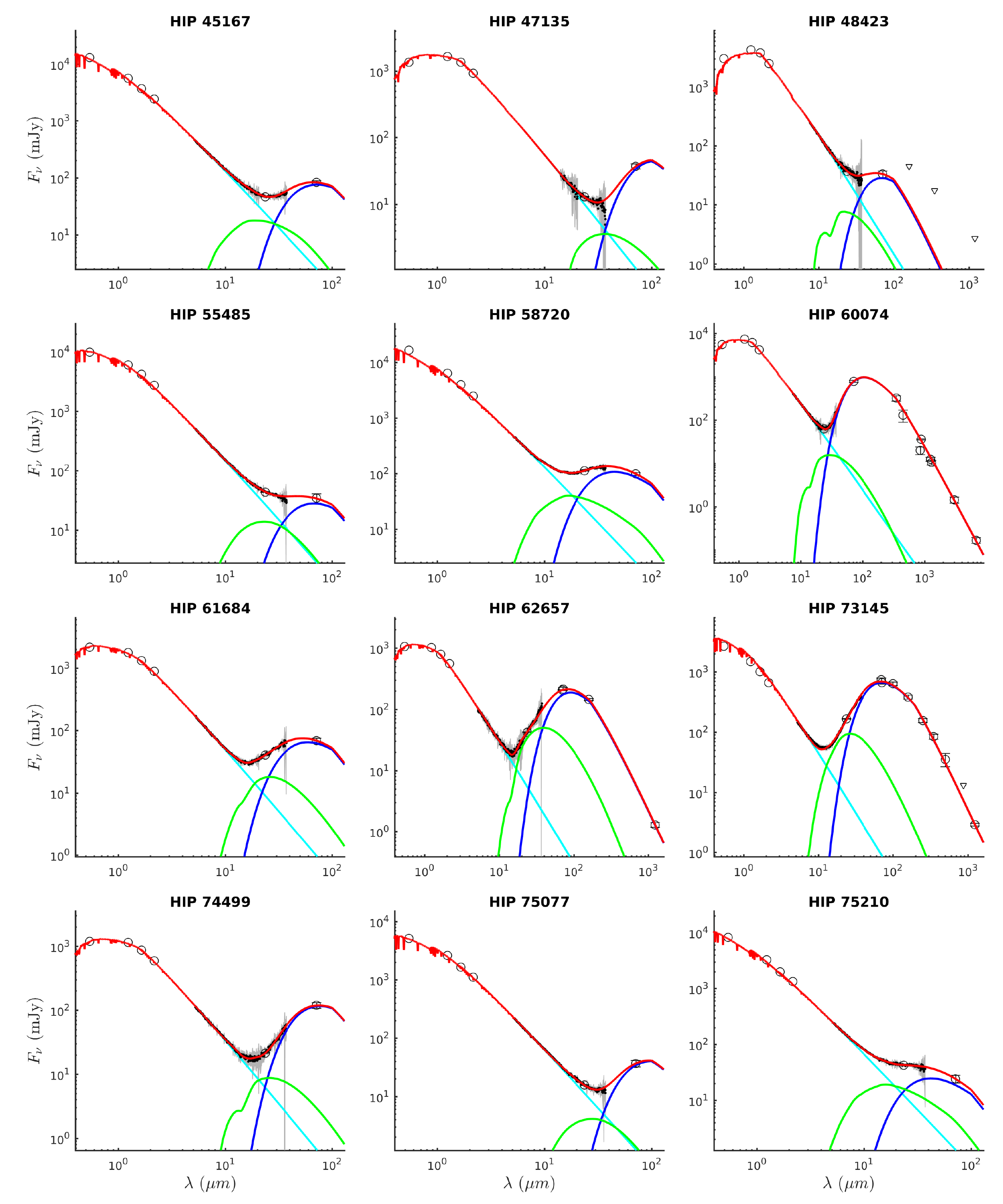}
\caption{(continued).}
\end{figure*}

\begin{figure*}
\epsscale{1.2}
\figurenum{8}
\plotone{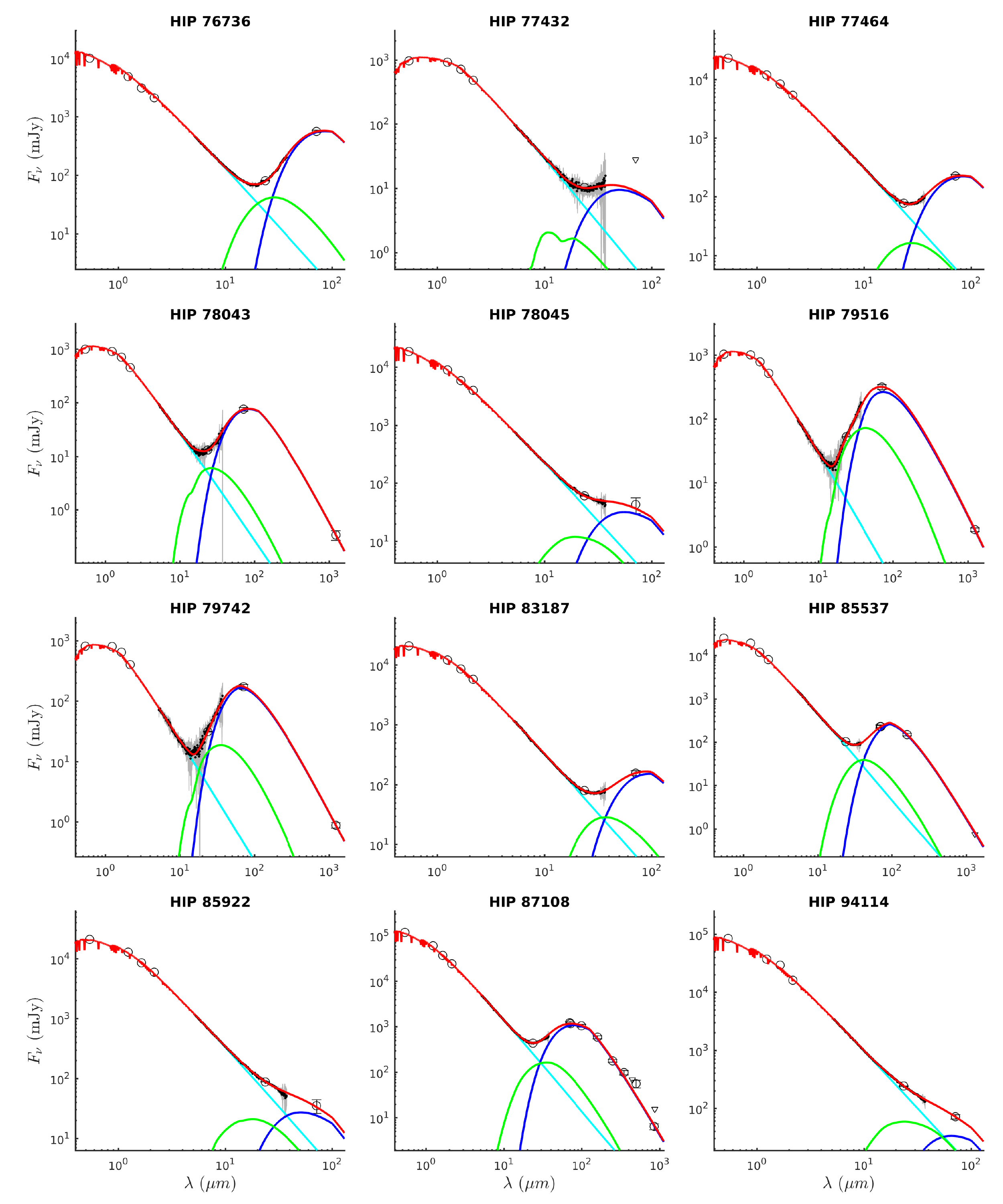}
\caption{(continued).}
\end{figure*}

\begin{figure*}
\epsscale{1.2}
\figurenum{8}
\plotone{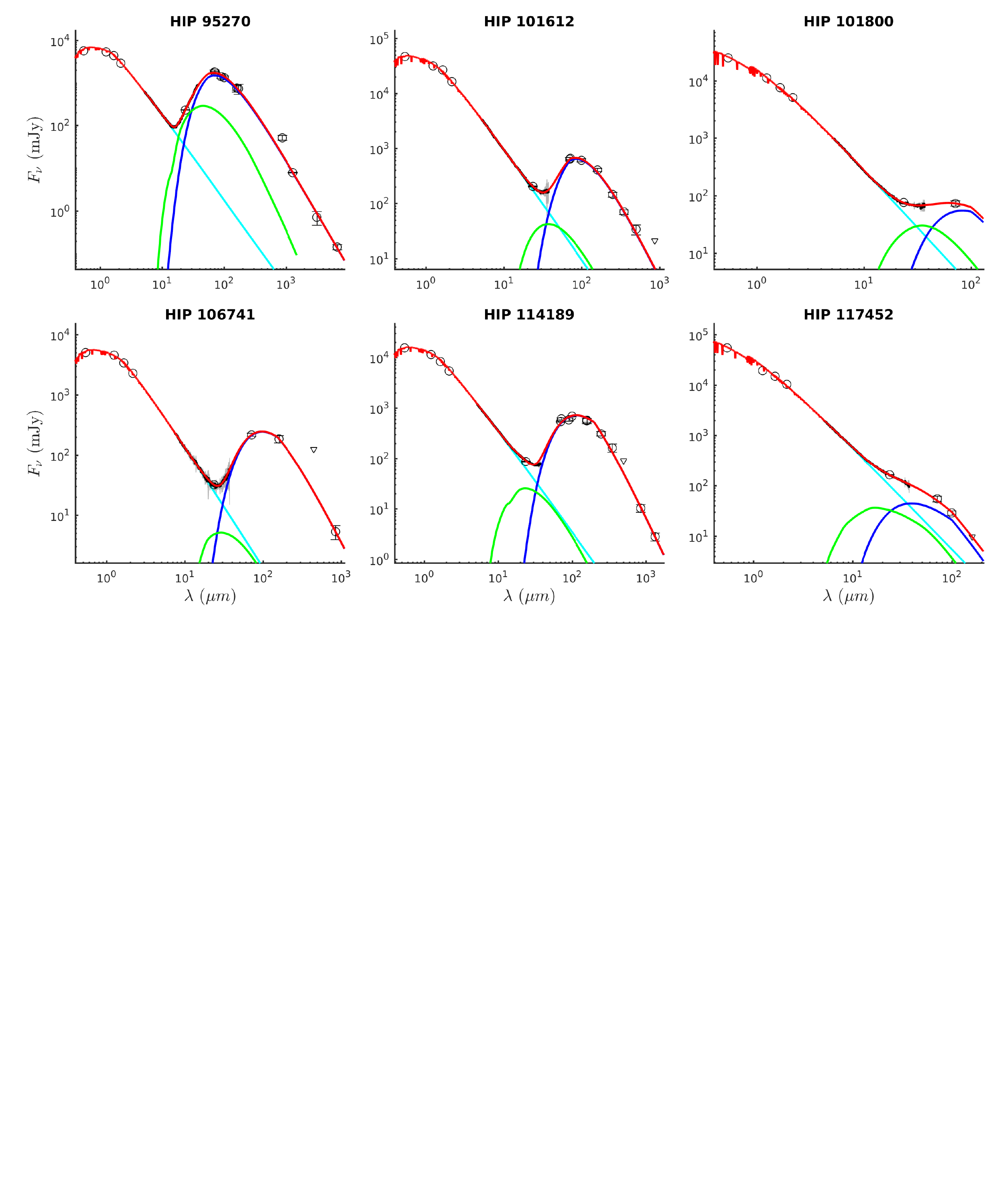}
\caption{(continued).}
\end{figure*}

\clearpage

\begin{deluxetable}{ccccccccccccccc}
\tabletypesize{\scriptsize}
\tablewidth{0pt}
\tablecolumns{15}
\tablecaption{Target Properties \label{table:warmtargetlist}}
\tablehead{\colhead{HIP} & \colhead{Spectral} & \colhead{$V$} & \colhead{$J$} & \colhead{$H$} & \colhead{$K$} & \colhead{$T_\star$} & \colhead{$L_\star$} & \colhead{$M_\star$} & \colhead{$a_\text{BOS}$} & \colhead{$d$} & \colhead{IRS} & \colhead{$x_\text{LL1}$} & \colhead{$x_\text{SL1}$} & \colhead{$x_\text{SL2}$} \\ \colhead{Identifier} & \colhead{Type} & \colhead{(mag)} & \colhead{(mag)} & \colhead{(mag)} & \colhead{(mag)} & \colhead{(K)} & \colhead{($L_\sun$)} & \colhead{($M_\sun$)} & \colhead{($\micron$)} & \colhead{(pc)} & \colhead{AOR} & \colhead{} & \colhead{} & \colhead{}}
\startdata
\cutinhead{Single-component Systems}
HIP 9902 & F8V & 7.5 & 6.53 & 6.3 & 6.2 & 6294 & 1.62 & 1.12 & 1 & 44.17 & 13621504 & 1.00 & 1.07 & 1.05 \\
HIP 14684 & G0V & 8.49 & 7.16 & 6.79 & 6.7 & 5528 & 0.52 & 0.88 & 0.5 & 37.41 & 5340672 & 1.00 & 1.05 & \nodata \\
HIP 17549 & A0 & 6.88 & 6.88 & 6.93 & 6.92 & 9820 & 34.15 & 2.26 & 7.5 & 140.84 & 14148096 & 0.94 & 1.19 & 1.08 \\
HIP 18217 & A5m & 5.79 & 5.45 & 5.41 & 5.37 & 7667 & 9.73 & 1.69 & 3.1 & 50.51 & 14160640 & 1.00 & 1.02 & 1.02 \\
HIP 18481 & A2Vn & 6.08 & 5.97 & 5.97 & 5.92 & 8801 & 15.12 & 1.88 & 4.2 & 70.22 & 14139648 & 0.99 & 1.00 & 0.99 \\
HIP 23871 & A5V & 5.28 & 5.11 & 5.08 & 5.05 & 8308 & 20.32 & 2.01 & 5.2 & 58.11 & 14139904 & 1.00 & 1.10 & 1.06 \\
HIP 28103\tablenotemark{a} & F2V & 3.72 & 3.06 & 2.98 & 2.99 & 7207 & 7.88 & 1.60 & 2.7 & 14.88 & 15998720 & 1.00 & 1.04 & 1.04 \\
HIP 41967 & G5V & 8.26 & 7.05 & 6.77 & 6.7 & 5875 & 0.89 & 0.99 & 0.7 & 45.07 & 6596864 & 0.94 & 1.00 & \nodata \\
HIP 49593 & A7V & 4.49 & 4.27 & 4.05 & 4 & 8000 & 9.98 & 1.70 & 3.2 & 28.24 & 14141184 & 1.00 & 1.06 & 1.00 \\
HIP 54879 & A2V & 3.32 & 3.12 & 3.19 & 3.08 & 8080 & 92.43 & 2.92 & 15.5 & 50.61 & 16177152 & 1.00 & 1.00 & 1.00 \\
HIP 57524 & G4Vp & 9.07 & 7.89 & 7.57 & 7.51 & 5887 & 1.75 & 1.14 & 1 & 91.66 & 21799936 & 1.00 & 1.06 & 1.04 \\
HIP 57950 & F2IV/V & 8.25 & 7.47 & 7.31 & 7.28 & 6730 & 3.89 & 1.37 & 1.7 & 98.14 & 21800448 & 0.95 & 1.12 & 1.10 \\
HIP 60348 & F5V & 8.78 & 7.95 & 7.76 & 7.67 & 6515 & 2.15 & 1.19 & 1.2 & 93.72 & 21801728 & 0.97 & 1.12 & 1.11 \\
HIP 62134 & F2V & 8.63 & 7.88 & 7.73 & 7.71 & 6658 & 3.79 & 1.36 & 1.7 & 115.61 & 21802240 & 1.00 & 1.12 & 1.11 \\
HIP 63836 & F7 & 9 & 8.09 & 7.89 & 7.87 & 6501 & 2.41 & 1.23 & 1.3 & 107.41 & 21803008 & 1.00 & 1.13 & 1.11 \\
HIP 64053 & B8V & 5.7 & 5.82 & 5.75 & 5.73 & 9800 & 51.38 & 2.52 & 10.1 & 100.1 & 22803712 & 1.00 & 1.03 & 1.02 \\
HIP 64877 & F5V & 8.47 & 7.62 & 7.41 & 7.41 & 6448 & 5.21 & 1.47 & 2.1 & 125 & 22806784 & 1.00 & 1.07 & 1.07 \\
HIP 67068 & F3V & 8.46 & 7.69 & 7.52 & 7.47 & 6787 & 2.82 & 1.27 & 1.4 & 91.58 & 26361856 & 1.00 & 0.99 & 0.96 \\
HIP 67230 & F5V & 8.03 & 7.14 & 6.93 & 6.89 & 6476 & 8.73 & 1.65 & 2.9 & 131.75 & 22802688 & 1.00 & 1.08 & 1.08 \\
HIP 70455 & B8V & 6.96 & 7.03 & 7.07 & 7.08 & 10460 & 48.73 & 2.49 & 9.7 & 165.02 & 14148608 & 1.00 & 1.02 & 1.00 \\
HIP 74824 & A3Va & 4.07 & 3.93 & 3.81 & 3.88 & 7694 & 17.73 & 1.95 & 4.8 & 30.55 & 14141952 & 1.00 & 1.00 & 1.00 \\
HIP 77315 & A0V & 6.92 & 6.69 & 6.72 & 6.67 & 8880 & 34.38 & 2.27 & 7.6 & 147.28 & 22804992 & 1.00 & 1.05 & 1.05 \\
HIP 79710 & F0V & 8.4 & 7.77 & 7.65 & 7.61 & 7000 & 5.45 & 1.48 & 2.1 & 127.39 & 26314752 & 1.00 & 0.98 & 0.98 \\
HIP 79881 & A0V & 4.79 & 4.86 & 4.94 & 4.74 & 9210 & 17.93 & 1.95 & 4.7 & 41.29 & 21809408 & 1.00 & 1.09 & 1.15 \\
HIP 89770 & F5 & 6.68 & 5.85 & 5.7 & 5.62 & 6599 & 4.87 & 1.44 & 2 & 53.22 & 15016960 & 1.00 & 1.00 & 0.99 \\
HIP 95560 & A0V & 5.59 & 5.58 & 5.63 & 5.61 & 8673 & 25.02 & 2.11 & 6.1 & 72.89 & 14143488 & 0.98 & 1.04 & 1.04 \\
HIP 106783 & A2V & 6.18 & 6.13 & 6.13 & 6.11 & 9158 & 22.37 & 2.05 & 5.6 & 87.64 & 21812992 & 0.99 & 1.10 & 1.11 \\
HIP 112542 & B9V & 5.68 & 5.75 & 5.79 & 5.73 & 10016 & 50.69 & 2.51 & 10 & 97.37 & 14144000 & 1.00 & 1.05 & 1.05 \\
HIP 115738\tablenotemark{a} & A0p... & 4.93 & 5.32 & 4.98 & 4.9 & 9836 & 37.72 & 2.34 & 8.1 & 47.06 & 14144256 & 0.99 & 1.11 & 1.05 \\
\cutinhead{Two-component Systems}
HIP 345 & A0V & 6.39 & 6.28 & 6.25 & 6.26 & 8936 & 37.34 & 2.31 & 8.1 & 124.84 & 12720128 & 1.00 & 1.01 & 1.01 \\
HIP 682 & G2V & 7.59 & 6.42 & 6.15 & 6.12 & 5962 & 1.23 & 1.05 & 0.8 & 39.08 & 5268736 & 1.00 & 1.07 & \nodata \\
HIP 1473 & A2V & 4.52 & 4.34 & 4.42 & 4.46 & 9005 & 22.94 & 2.06 & 5.7 & 41.32 & 14160384 & 1.00 & 1.03 & 1.03 \\
HIP 1481 & F9V & 7.46 & 6.46 & 6.25 & 6.15 & 6273 & 1.50 & 1.10 & 0.9 & 41.55 & 21788160 & 0.97 & 1.07 & 1.06 \\
HIP 2472\tablenotemark{a} & A0V & 4.77 & 4.67 & 4.77 & 4.7 & 9459 & 29.59 & 2.22 & 6.7 & 52.97 & 21788672 & 1.00 & 1.09 & 1.11 \\
HIP 2710 & F2 & 6.91 & 6.04 & 5.85 & 5.75 & 6400 & 2.28 & 1.21 & 1.2 & 40.92 & 25673472 & 1.02 & \nodata & \nodata \\
HIP 7805 & F2IV/V & 7.61 & 6.84 & 6.69 & 6.63 & 6796 & 3.27 & 1.32 & 1.5 & 67.25 & 21789952 & 1.00 & 1.13 & 1.11 \\
HIP 7978 & F9V & 5.52 & 4.79 & 4.4 & 4.34 & 6000 & 1.58 & 1.11 & 1 & 17.43 & 16029952 & 0.99 & 1.05 & 1.06 \\
HIP 8241 & A1V & 5.04 & 4.99 & 5.03 & 4.96 & 9478 & 32.98 & 2.25 & 7.4 & 62.03 & 14139392 & 1.00 & 1.06 & 1.04 \\
HIP 11360 & F2 & 6.8 & 6.03 & 5.86 & 5.82 & 6500 & 3.18 & 1.31 & 1.5 & 45.23 & 10885632 & 1.00 & 1.18 & 1.08 \\
HIP 11477 & A2V & 5.13 & 5.12 & 5.03 & 4.94 & 8900 & 16.00 & 1.90 & 4.4 & 46.6 & 21790976 & 1.00 & 1.11 & 1.13 \\
HIP 11847 & F2V & 7.49 & 6.7 & 6.61 & 6.55 & 6922 & 3.38 & 1.33 & 1.5 & 63.49 & 10886144 & 0.99 & 1.07 & 1.08 \\
HIP 13141\tablenotemark{a} & A2V & 5.26 & 5.14 & 5.16 & 4.97 & 8390 & 17.48 & 1.95 & 4.7 & 50.45 & 21791744 & 1.00 & 1.11 & 1.11 \\
HIP 18859\tablenotemark{a} & F5V & 5.38 & 4.71 & 4.34 & 4.18 & 6315 & 2.59 & 1.24 & 1.3 & 18.83 & 5308416 & 1.00 & 1.01 & \nodata \\
HIP 20901 & A7V & 5.01 & 4.79 & 4.66 & 4.53 & 7592 & 17.44 & 1.94 & 4.7 & 48.85 & 14146816 & 1.00 & 1.03 & 1.03 \\
HIP 22192 & A3IV & 6.18 & 5.8 & 5.73 & 5.71 & 7901 & 8.41 & 1.64 & 2.8 & 56.18 & 14147584 & 0.99 & 1.12 & 1.06 \\
HIP 22226 & F3V & 7.86 & 7.1 & 6.95 & 6.89 & 6828 & 3.75 & 1.36 & 1.7 & 80.26 & 10887424 & 1.02 & 1.16 & 1.12 \\
HIP 23451 & A0 & 8.14 & 7.69 & 7.62 & 7.59 & 7957 & 6.19 & 1.53 & 2.3 & 112.36 & 15903744 & 1.00 & 1.05 & 1.12 \\
HIP 26453 & F3V & 7.25 & 6.47 & 6.29 & 6.28 & 6758 & 3.28 & 1.32 & 1.5 & 56.79 & 5306880 & 1.00 & 1.08 & \nodata \\
HIP 26796 & A0V & 6.34 & 6.41 & 6.45 & 6.43 & 10130 & 69.94 & 2.72 & 12.6 & 153.61 & 12713728 & 0.99 & 1.10 & 1.07 \\
HIP 34276 & A0V & 6.52 & 6.49 & 6.48 & 6.48 & 9400 & 23.76 & 2.08 & 5.8 & 102.35 & 21795584 & 1.00 & 1.07 & 1.07 \\
HIP 36948 & G8Vk+? & 8.22 & 6.91 & 6.58 & 6.46 & 5598 & 0.61 & 0.91 & 0.5 & 35.35 & 5267200 & 1.00 & 1.02 & \nodata \\
HIP 41152 & A3V & 5.54 & 5.25 & 5.29 & 5.25 & 8077 & 12.05 & 1.78 & 3.6 & 50.43 & 14140416 & 1.00 & 1.10 & 1.06 \\
HIP 41373 & A0V & 6.05 & 5.94 & 5.91 & 5.89 & 8839 & 15.39 & 1.88 & 4.3 & 69.44 & 14140672 & 1.00 & 1.12 & 1.08 \\
HIP 45167 & A0V & 6.14 & 6.14 & 6.16 & 6.12 & 9152 & 29.56 & 2.19 & 6.8 & 99.3 & 12710400 & 1.00 & 1.04 & 1.03 \\
HIP 47135 & G2V & 8.59 & 7.47 & 7.24 & 7.16 & 6050 & 1.46 & 1.09 & 0.9 & 67.98 & 25677056 & 1.00 & \nodata & \nodata \\
HIP 48423 & G5 & 7.73 & 6.47 & 6.14 & 6.09 & 5199 & 0.80 & 0.97 & 0.6 & 32.8 & 5399808 & 1.00 & 1.04 & \nodata \\
HIP 55485 & A7Vn & 6.44 & 6.08 & 6.02 & 5.99 & 8000 & 13.83 & 1.84 & 4 & 80.84 & 14141440 & 1.00 & 1.08 & 1.03 \\
HIP 58720 & B9V & 5.88 & 6.01 & 6.08 & 6.08 & 10000 & 55.31 & 2.57 & 10.6 & 105.71 & 22799360 & 0.99 & 1.08 & 1.07 \\
HIP 60074\tablenotemark{a} & G2V & 7.07 & 5.87 & 5.61 & 5.54 & 5809 & 1.11 & 1.04 & 0.8 & 27.46 & 5312256 & 1.00 & 0.97 & \nodata \\
HIP 61684 & A9V & 8.09 & 7.41 & 7.27 & 7.2 & 7000 & 5.84 & 1.51 & 2.2 & 111.86 & 22801152 & 1.00 & 1.06 & 1.04 \\
HIP 62657 & F5/F6V & 8.87 & 8 & 7.83 & 7.72 & 6417 & 2.65 & 1.25 & 1.3 & 108.58 & 13621248 & 1.01 & 1.06 & 1.05 \\
HIP 73145 & A2IV & 7.86 & 7.6 & 7.56 & 7.52 & 8281 & 8.75 & 1.65 & 2.9 & 122.7 & 22803968 & 0.98 & 1.14 & 1.15 \\
HIP 74499 & F4V & 8.74 & 7.88 & 7.73 & 7.65 & 6545 & 2.07 & 1.18 & 1.1 & 89.93 & 21806336 & 0.98 & 1.08 & 1.06 \\
HIP 75077 & A1V & 7.16 & 6.98 & 7.04 & 6.96 & 8599 & 18.99 & 1.98 & 5 & 131.58 & 14151168 & 1.00 & 0.97 & 0.97 \\
HIP 75210 & B8/B9V & 6.64 & 6.75 & 6.83 & 6.76 & 10642 & 45.95 & 2.45 & 9.3 & 136.24 & 14147840 & 1.00 & 1.05 & 1.04 \\
HIP 76736 & A1V & 6.42 & 6.3 & 6.34 & 6.27 & 8769 & 13.56 & 1.83 & 3.9 & 78.49 & 14142208 & 1.00 & 1.10 & 1.12 \\
HIP 77432 & F5V & 8.97 & 8.11 & 7.94 & 7.87 & 6594 & 1.98 & 1.17 & 1.1 & 96.34 & 21808128 & 1.00 & 1.12 & 1.10 \\
HIP 77464 & A5IV & 5.53 & 5.33 & 5.27 & 5.26 & 8248 & 14.07 & 1.84 & 4 & 54.02 & 14142720 & 1.00 & 1.03 & 1.03 \\
HIP 78043 & F3V & 8.95 & 8.15 & 7.97 & 7.94 & 6639 & 4.31 & 1.40 & 1.8 & 144.3 & 26313728 & 1.00 & 0.99 & 0.99 \\
HIP 78045 & A3V & 5.75 & 5.65 & 5.66 & 5.57 & 8777 & 17.84 & 1.95 & 4.7 & 66.01 & 14142464 & 1.00 & 1.05 & 1.05 \\
HIP 79516 & F5V & 8.9 & 8.02 & 7.85 & 7.79 & 6495 & 4.04 & 1.38 & 1.8 & 133.69 & 15554560 & 1.00 & 1.05 & 1.04 \\
HIP 79742 & F5 & 9.16 & 8.28 & 8.06 & 8.06 & 6516 & 3.80 & 1.36 & 1.7 & 146.2 & 15555328 & 1.00 & 1.00 & 0.98 \\
HIP 83187 & A5IV-V & 5.65 & 5.32 & 5.26 & 5.19 & 7800 & 11.42 & 1.76 & 3.5 & 51.81 & 14160896 & 1.00 & 1.03 & 1.06 \\
HIP 85537 & A8V & 5.42 & 4.81 & 4.88 & 4.8 & 7201 & 18.72 & 1.97 & 5 & 59.63 & 27224064 & 1.00 & 0.98 & 1.00 \\
HIP 85922 & A5V & 5.62 & 5.25 & 5.25 & 5.14 & 7800 & 10.23 & 1.71 & 3.2 & 48.1 & 14142976 & 1.00 & 1.06 & 1.03 \\
HIP 87108 & A0V & 3.75 & 3.59 & 3.66 & 3.62 & 8517 & 25.39 & 2.11 & 6.1 & 31.52 & 4931328 & 1.00 & 1.01 & 1.02 \\
HIP 94114 & A2Va & 4.1 & 4.09 & 3.92 & 4.05 & 8400 & 26.47 & 2.13 & 6.3 & 38.43 & 14145536 & 1.00 & 1.05 & 1.01 \\
HIP 95270 & F5/F6V & 7.04 & 6.2 & 5.98 & 5.91 & 6502 & 3.34 & 1.32 & 1.5 & 51.81 & 3564032 & 1.00 & 1.05 & 1.07 \\
HIP 101612\tablenotemark{a} & F0V & 4.76 & 4.28 & 4.02 & 4.04 & 7233 & 8.10 & 1.61 & 2.8 & 27.79 & 21812224 & 1.00 & 1.10 & 1.11 \\
HIP 101800 & A2V & 5.43 & 5.41 & 5.37 & 5.3 & 9131 & 19.67 & 1.99 & 5.1 & 57.94 & 14161408 & 0.98 & 1.08 & 1.08 \\
HIP 106741 & F4IV & 7.17 & 6.38 & 6.25 & 6.18 & 6786 & 2.92 & 1.28 & 1.4 & 51.81 & 15022592 & 1.00 & 1.10 & \nodata \\
HIP 114189 & A5V & 5.95 & 5.38 & 5.28 & 5.24 & 7033 & 4.82 & 1.44 & 2 & 39.4 & 28889856 & 1.00 & 1.00 & 1.00 \\
HIP 117452\tablenotemark{a} & A0V & 4.58 & 4.8 & 4.64 & 4.53 & 9673 & 33.98 & 2.29 & 7.4 & 42.14 & 14161664 & 1.00 & 1.02 & 1.02 \\
\enddata
\tablenotetext{a}{These targets not listed in \citet{mcdonald2012}; we inferred their stellar properties from their $V-K$ color.}
\end{deluxetable}

\newpage
\begin{deluxetable}{ccccc}
\tabletypesize{\scriptsize}
\tablewidth{0pt}
\tablecolumns{5}
\tablecaption{Target Photometry \label{table:photometry}}
\tablehead{\colhead{HIP} & \colhead{$\lambda$} & \colhead{$F_\nu$} & \colhead{Instrument} & \colhead{References} \\ \colhead{Identifier} & \colhead{($\micron$)} & \colhead{(mJy)} & \colhead{} & \colhead{}}
\startdata
\cutinhead{Single-component Systems}
HIP 9902 & 24 & 46.67 $\pm$ 0.47 & \textit{Spitzer}/MIPS & \citet{ballering2013} \\
HIP 9902 & 70 & $<$12.56 & \textit{Spitzer}/MIPS & \citet{ballering2013} \\
HIP 9902 & 160 & $<$60.60 & \textit{Spitzer}/MIPS & \citet{moor2009} \\
HIP 14684 & 24 & 18.68 $\pm$ 0.22 & \textit{Spitzer}/MIPS & \citet{ballering2013} \\
HIP 14684 & 70 & $<$16.38 & \textit{Spitzer}/MIPS & \citet{ballering2013} \\
HIP 14684 & 1200 & $<$12.00 & SEST & \citet{carpenter2005} \\
HIP 14684 & 3000 & $<$2.30 & OVRO & \citet{carpenter2005} \\
HIP 17549 & 24 & 36.57 $\pm$ 0.40 & \textit{Spitzer}/MIPS & \citet{ballering2013} \\
HIP 17549 & 70 & 13.95 $\pm$ 3.20 & \textit{Spitzer}/MIPS & \citet{ballering2013} \\
HIP 18217 & 24 & 66.06 $\pm$ 0.68 & \textit{Spitzer}/MIPS & \citet{ballering2013} \\
HIP 18217 & 70 & $<$34.44 & \textit{Spitzer}/MIPS & \citet{ballering2013} \\
HIP 18481 & 24 & 45.72 $\pm$ 0.49 & \textit{Spitzer}/MIPS & \citet{ballering2013} \\
HIP 18481 & 70 & $<$37.02 & \textit{Spitzer}/MIPS & \citet{ballering2013} \\
HIP 23871 & 24 & 94.01 $\pm$ 0.98 & \textit{Spitzer}/MIPS & \citet{ballering2013} \\
HIP 23871 & 70 & $<$36.90 & \textit{Spitzer}/MIPS & \citet{ballering2013} \\
HIP 28103 & 24 & 558.50 $\pm$ 5.67 & \textit{Spitzer}/MIPS & \citet{ballering2013} \\
HIP 28103 & 70 & 95.96 $\pm$ 6.16 & \textit{Spitzer}/MIPS & \citet{ballering2013} \\
HIP 28103 & 100 & 45.46 $\pm$ 1.42 & \textit{Herschel}/PACS & \citet{eiroa2013} \\
HIP 28103 & 160 & 9.37 $\pm$ 1.84 & \textit{Herschel}/PACS & \citet{eiroa2013} \\
HIP 41967 & 24 & 18.70 $\pm$ 0.22 & \textit{Spitzer}/MIPS & \citet{ballering2013} \\
HIP 41967 & 70 & $<$14.16 & \textit{Spitzer}/MIPS & \citet{ballering2013} \\
HIP 41967 & 350 & $<$54.00 & CSO & \citet{roccatagliata2009} \\
HIP 49593 & 24 & 218.20 $\pm$ 2.14 & \textit{Spitzer}/MIPS & \citet{ballering2013} \\
HIP 49593 & 70 & 37.46 $\pm$ 5.99 & \textit{Spitzer}/MIPS & \citet{ballering2013} \\
HIP 49593 & 100 & 22.07 $\pm$ 2.36 & \textit{Herschel}/PACS & \citet{thureau2014} \\
HIP 49593 & 160 & 9.65 $\pm$ 3.40 & \textit{Herschel}/PACS & \citet{thureau2014} \\
HIP 54879 & 24 & 401.10 $\pm$ 4.00 & \textit{Spitzer}/MIPS & \citet{ballering2013} \\
HIP 54879 & 70 & 66.17 $\pm$ 6.28 & \textit{Spitzer}/MIPS & \citet{ballering2013} \\
HIP 57524 & 24 & 11.11 $\pm$ 0.17 & \textit{Spitzer}/MIPS & \citet{ballering2013} \\
HIP 57524 & 70 & $<$14.80 & \textit{Spitzer}/MIPS & \citet{ballering2013} \\
HIP 57950 & 24 & 18.41 $\pm$ 0.22 & \textit{Spitzer}/MIPS & \citet{ballering2013} \\
HIP 57950 & 70 & $<$30.51 & \textit{Spitzer}/MIPS & \citet{ballering2013} \\
HIP 60348 & 24 & 12.18 $\pm$ 0.18 & \textit{Spitzer}/MIPS & \citet{ballering2013} \\
HIP 60348 & 70 & $<$22.70 & \textit{Spitzer}/MIPS & \citet{ballering2013} \\
HIP 62134 & 24 & 8.59 $\pm$ 0.16 & \textit{Spitzer}/MIPS & \citet{ballering2013} \\
HIP 62134 & 70 & $<$23.62 & \textit{Spitzer}/MIPS & \citet{ballering2013} \\
HIP 63836 & 24 & 8.40 $\pm$ 0.16 & \textit{Spitzer}/MIPS & \citet{ballering2013} \\
HIP 63836 & 70 & $<$25.19 & \textit{Spitzer}/MIPS & \citet{ballering2013} \\
HIP 64053 & 24 & 67.32 $\pm$ 0.69 & \textit{Spitzer}/MIPS & \citet{ballering2013} \\
HIP 64053 & 70 & $<$19.81 & \textit{Spitzer}/MIPS & \citet{ballering2013} \\
HIP 64877 & 24 & 22.88 $\pm$ 0.99 & \textit{Spitzer}/MIPS & \citet{ballering2013} \\
HIP 64877 & 70 & $<$126.60 & \textit{Spitzer}/MIPS & \citet{ballering2013} \\
HIP 67068 & 24 & 10.25 $\pm$ 0.17 & \textit{Spitzer}/MIPS & \citet{ballering2013} \\
HIP 67068 & 70 & $<$33.48 & \textit{Spitzer}/MIPS & \citet{ballering2013} \\
HIP 67230 & 24 & 44.09 $\pm$ 1.24 & \textit{Spitzer}/MIPS & \citet{ballering2013} \\
HIP 67230 & 70 & $<$157.98 & \textit{Spitzer}/MIPS & \citet{ballering2013} \\
HIP 70455 & 24 & 24.25 $\pm$ 0.33 & \textit{Spitzer}/MIPS & \citet{ballering2013} \\
HIP 70455 & 70 & $<$11.27 & \textit{Spitzer}/MIPS & \citet{ballering2013} \\
HIP 74824 & 24 & 341.80 $\pm$ 3.46 & \textit{Spitzer}/MIPS & \citet{ballering2013} \\
HIP 74824 & 70 & $<$263.58 & \textit{Spitzer}/MIPS & \citet{ballering2013} \\
HIP 77315 & 24 & 59.61 $\pm$ 0.60 & \textit{Spitzer}/MIPS & \citet{ballering2013} \\
HIP 77315 & 70 & 22.92 $\pm$ 7.22 & \textit{Spitzer}/MIPS & \citet{ballering2013} \\
HIP 79710 & 24 & 18.61 $\pm$ 1.07 & \textit{Spitzer}/MIPS & \citet{ballering2013} \\
HIP 79710 & 70 & $<$173.46 & \textit{Spitzer}/MIPS & \citet{ballering2013} \\
HIP 79881 & 24 & 106.60 $\pm$ 1.09 & \textit{Spitzer}/MIPS & \citet{ballering2013} \\
HIP 79881 & 70 & $<$47.88 & \textit{Spitzer}/MIPS & \citet{ballering2013} \\
HIP 79881 & 70 & 13.00 $\pm$ 2.00 & \textit{Herschel}/PACS & \citet{riviere-marichalar2014} \\
HIP 79881 & 100 & $<$7.00 & \textit{Herschel}/PACS & \citet{riviere-marichalar2014} \\
HIP 79881 & 160 & $<$13.00 & \textit{Herschel}/PACS & \citet{riviere-marichalar2014} \\
HIP 89770 & 24 & 90.66 $\pm$ 0.90 & \textit{Spitzer}/MIPS & \citet{ballering2013} \\
HIP 89770 & 70 & 16.99 $\pm$ 5.08 & \textit{Spitzer}/MIPS & \citet{ballering2013} \\
HIP 95560 & 24 & 61.84 $\pm$ 0.65 & \textit{Spitzer}/MIPS & \citet{ballering2013} \\
HIP 95560 & 70 & $<$34.77 & \textit{Spitzer}/MIPS & \citet{ballering2013} \\
HIP 106783 & 24 & 34.38 $\pm$ 0.37 & \textit{Spitzer}/MIPS & \citet{ballering2013} \\
HIP 106783 & 70 & $<$21.42 & \textit{Spitzer}/MIPS & \citet{ballering2013} \\
HIP 112542 & 24 & 63.92 $\pm$ 0.66 & \textit{Spitzer}/MIPS & \citet{ballering2013} \\
HIP 112542 & 70 & 39.17 $\pm$ 7.65 & \textit{Spitzer}/MIPS & \citet{ballering2013} \\
HIP 115738 & 24 & 111.70 $\pm$ 1.12 & \textit{Spitzer}/MIPS & \citet{ballering2013} \\
HIP 115738 & 70 & 30.27 $\pm$ 8.38 & \textit{Spitzer}/MIPS & \citet{ballering2013} \\
\cutinhead{Two-component Systems}
HIP 345 & 24 & 35.70 $\pm$ 0.38 & \textit{Spitzer}/MIPS & \citet{ballering2013} \\
HIP 345 & 70 & 97.24 $\pm$ 5.35 & \textit{Spitzer}/MIPS & \citet{ballering2013} \\
HIP 682 & 24 & 36.49 $\pm$ 0.39 & \textit{Spitzer}/MIPS & \citet{ballering2013} \\
HIP 682 & 70 & 170.60 $\pm$ 10.62 & \textit{Spitzer}/MIPS & \citet{ballering2013} \\
HIP 682 & 160 & 187.50 $\pm$ 50.40 & \textit{Spitzer}/MIPS & \citet{hillenbrand2008} \\
HIP 682 & 450 & $<$66.00 & JCMT/SCUBA-2 & \citet{panic2013} \\
HIP 682 & 850 & 4.60 $\pm$ 1.20 & JCMT/SCUBA-2 & \citet{panic2013} \\
HIP 682 & 880 & 3.50 $\pm$ 1.40 & SMA & \citet{steele2016} \\
HIP 682 & 1200 & 4.00 $\pm$ 1.00 & IRAM & \citet{roccatagliata2009} \\
HIP 682 & 2700 & $<$2.40 & OVRO & \citet{carpenter2005} \\
HIP 682 & 3000 & $<$1.83 & OVRO & \citet{carpenter2005} \\
HIP 682 & 9000 & $<$0.01 & VLA & \citet{macgregor2016} \\
HIP 682 & 10000 & $<$0.04 & GBT & \citet{greaves2012} \\
HIP 1473 & 24 & 154.40 $\pm$ 1.56 & \textit{Spitzer}/MIPS & \citet{ballering2013} \\
HIP 1473 & 70 & 43.82 $\pm$ 6.50 & \textit{Spitzer}/MIPS & \citet{ballering2013} \\
HIP 1473 & 100 & 25.48 $\pm$ 2.65 & \textit{Herschel}/PACS & \citet{thureau2014} \\
HIP 1473 & 160 & 12.94 $\pm$ 4.10 & \textit{Herschel}/PACS & \citet{thureau2014} \\
HIP 1481 & 24 & 34.76 $\pm$ 0.36 & \textit{Spitzer}/MIPS & \citet{ballering2013} \\
HIP 1481 & 70 & $<$ & \textit{Spitzer}/MIPS & \citet{ballering2013} \\
HIP 1481 & 70 & 13.00 $\pm$ 0.90 & \textit{Herschel}/PACS & \citet{donaldson2012} \\
HIP 1481 & 160 & $<$10.60 & \textit{Herschel}/PACS & \citet{donaldson2012} \\
HIP 2472 & 24 & 112.50 $\pm$ 1.12 & \textit{Spitzer}/MIPS & \citet{ballering2013} \\
HIP 2472 & 70 & 76.98 $\pm$ 6.54 & \textit{Spitzer}/MIPS & \citet{ballering2013} \\
HIP 2710 & 24 & 40.62 $\pm$ 0.44 & \textit{Spitzer}/MIPS & \citet{ballering2013} \\
HIP 2710 & 70 & 104.70 $\pm$ 6.50 & \textit{Spitzer}/MIPS & \citet{ballering2013} \\
HIP 7805 & 24 & 28.74 $\pm$ 0.31 & \textit{Spitzer}/MIPS & \citet{ballering2013} \\
HIP 7805 & 70 & 136.10 $\pm$ 9.30 & \textit{Spitzer}/MIPS & \citet{ballering2013} \\
HIP 7978 & 24 & 196.20 $\pm$ 1.96 & \textit{Spitzer}/MIPS & \citet{ballering2013} \\
HIP 7978 & 70 & 1035.00 $\pm$ 52.07 & \textit{Spitzer}/MIPS & \citet{ballering2013} \\
HIP 7978 & 70 & 896.20 $\pm$ 26.90 & \textit{Herschel}/PACS & \citet{eiroa2013} \\
HIP 7978 & 100 & 897.10 $\pm$ 26.90 & \textit{Herschel}/PACS & \citet{eiroa2013} \\
HIP 7978 & 160 & 635.90 $\pm$ 31.80 & \textit{Herschel}/PACS & \citet{eiroa2013} \\
HIP 7978 & 160 & 462.00 $\pm$ 50.00 & \textit{Spitzer}/MIPS & \citet{tanner2009} \\
HIP 7978 & 250 & 312.30 $\pm$ 25.60 & \textit{Herschel}/SPIRE & \citet{eiroa2013} \\
HIP 7978 & 350 & 179.90 $\pm$ 14.60 & \textit{Herschel}/SPIRE & \citet{eiroa2013} \\
HIP 7978 & 500 & 78.40 $\pm$ 9.80 & \textit{Herschel}/SPIRE & \citet{eiroa2013} \\
HIP 7978 & 870 & 39.40 $\pm$ 4.10 & APEX/LABOCA & \citet{liseau2008} \\
HIP 7978 & 1200 & $<$17.00 & SEST/SIMBA & \citet{schutz2005} \\
HIP 7978 & 6800 & 0.09 $\pm$ 0.02 & ATCA & \citet{ricci2015b} \\
HIP 8241 & 24 & 108.30 $\pm$ 1.09 & \textit{Spitzer}/MIPS & \citet{ballering2013} \\
HIP 8241 & 70 & 413.80 $\pm$ 21.54 & Spitzer/MIPS & \citet{ballering2013} \\
HIP 8241 & 70 & 396.00 $\pm$ 28.00 & \textit{Herschel}/PACS & \citet{moor2015} \\
HIP 8241 & 100 & 403.00 $\pm$ 28.00 & \textit{Herschel}/PACS & \citet{moor2015} \\
HIP 8241 & 160 & 277.00 $\pm$ 20.00 & \textit{Herschel}/PACS & \citet{moor2015} \\
HIP 8241 & 250 & 94.00 $\pm$ 7.00 & \textit{Herschel}/SPIRE & \citet{moor2015} \\
HIP 8241 & 350 & 43.00 $\pm$ 6.00 & \textit{Herschel}/SPIRE & \citet{moor2015} \\
HIP 8241 & 500 & 3.00 $\pm$ 6.00 & \textit{Herschel}/SPIRE & \citet{moor2015} \\
HIP 11360 & 24 & 60.87 $\pm$ 0.64 & \textit{Spitzer}/MIPS & \citet{ballering2013} \\
HIP 11360 & 70 & 454.30 $\pm$ 24.37 & \textit{Spitzer}/MIPS & \citet{ballering2013} \\
HIP 11360 & 90 & 427.00 $\pm$ 30.00 & \textit{ISO} & \citet{moor2006} \\
HIP 11360 & 160 & 217.30 $\pm$ 27.80 & \textit{Spitzer}/MIPS & \citet{moor2011} \\
HIP 11360 & 450 & $<$111.00 & JCMT/SCUBA-2 & \citet{panic2013} \\
HIP 11360 & 850 & 8.50 $\pm$ 1.20 & JCMT/SCUBA-2 & \citet{panic2013} \\
HIP 11360 & 850 & 4.90 $\pm$ 1.60 & JCMT/SCUBA & \citet{williams2006} \\
HIP 11360 & 870 & $<$15.30 & APEX/LABOCA & \citet{nilsson2009} \\
HIP 11360 & 1300 & 2.60 $\pm$ 0.80 & SMA & \citet{macgregor2015} \\
HIP 11360 & 9000 & 0.01 $\pm$ 0.00 & VLA & \citet{macgregor2016} \\
HIP 11477 & 24 & 108.60 $\pm$ 1.06 & \textit{Spitzer}/MIPS & \citet{ballering2013} \\
HIP 11477 & 70 & 114.70 $\pm$ 8.81 & \textit{Spitzer}/MIPS & \citet{ballering2013} \\
HIP 11847 & 24 & 170.10 $\pm$ 1.71 & \textit{Spitzer}/MIPS & \citet{ballering2013} \\
HIP 11847 & 70 & 722.90 $\pm$ 36.80 & \textit{Spitzer}/MIPS & \citet{ballering2013} \\
HIP 11847 & 90 & 515.00 $\pm$ 36.00 & \textit{ISO} & \citet{moor2006} \\
HIP 11847 & 160 & 230.80 $\pm$ 29.90 & \textit{Spitzer}/MIPS & \citet{moor2011} \\
HIP 13141 & 24 & 86.22 $\pm$ 0.88 & \textit{Spitzer}/MIPS & \citet{ballering2013} \\
HIP 13141 & 70 & 197.10 $\pm$ 11.61 & \textit{Spitzer}/MIPS & \citet{ballering2013} \\
HIP 13141 & 70 & 213.00 $\pm$ 17.00 & \textit{Herschel}/PACS & \citet{moor2015} \\
HIP 13141 & 100 & 210.00 $\pm$ 18.00 & \textit{Herschel}/PACS & \citet{moor2015} \\
HIP 13141 & 160 & 138.00 $\pm$ 11.00 & \textit{Herschel}/PACS & \citet{moor2015} \\
HIP 13141 & 250 & 50.00 $\pm$ 5.00 & \textit{Herschel}/SPIRE & \citet{moor2015} \\
HIP 13141 & 350 & 28.00 $\pm$ 6.00 & \textit{Herschel}/SPIRE & \citet{moor2015} \\
HIP 13141 & 500 & $<$30.00 & \textit{Herschel}/SPIRE & \citet{moor2015} \\
HIP 18859 & 24 & 207.90 $\pm$ 2.02 & \textit{Spitzer}/MIPS & \citet{ballering2013} \\
HIP 18859 & 70 & 321.70 $\pm$ 16.92 & \textit{Spitzer}/MIPS & \citet{ballering2013} \\
HIP 18859 & 90 & 242.00 $\pm$ 18.00 & \textit{ISO} & \citet{moor2006} \\
HIP 18859 & 160 & 229.40 $\pm$ 67.50 & \textit{Spitzer}/MIPS & \citet{hillenbrand2008} \\
HIP 18859 & 870 & $<$9.90 & APEX/LABOCA & \citet{nilsson2010} \\
HIP 18859 & 1200 & $<$34.00 & SEST & \citet{carpenter2005} \\
HIP 18859 & 1200 & $<$2.20 & IRAM & \citet{roccatagliata2009} \\
HIP 18859 & 2700 & $<$2.48 & OVRO & \citet{carpenter2005} \\
HIP 18859 & 3000 & $<$2.23 & OVRO & \citet{carpenter2005} \\
HIP 20901 & 24 & 140.60 $\pm$ 1.40 & \textit{Spitzer}/MIPS & \citet{ballering2013} \\
HIP 20901 & 70 & 182.50 $\pm$ 11.21 & \textit{Spitzer}/MIPS & \citet{ballering2013} \\
HIP 22192 & 24 & 46.17 $\pm$ 0.47 & \textit{Spitzer}/MIPS & \citet{ballering2013} \\
HIP 22192 & 70 & 65.54 $\pm$ 3.92 & \textit{Spitzer}/MIPS & \citet{ballering2013} \\
HIP 22192 & 100 & 40.20 $\pm$ 3.80 & \textit{Herschel}/PACS & \citet{draper2016a} \\
HIP 22192 & 160 & 16.40 $\pm$ 1.50 & \textit{Herschel}/PACS & \citet{draper2016a} \\
HIP 22226 & 24 & 30.61 $\pm$ 0.30 & \textit{Spitzer}/MIPS & \citet{ballering2013} \\
HIP 22226 & 70 & 283.30 $\pm$ 15.33 & \textit{Spitzer}/MIPS & \citet{ballering2013} \\
HIP 22226 & 90 & 277.00 $\pm$ 67.00 & \textit{ISO} & \citet{moor2006} \\
HIP 22226 & 160 & 120.30 $\pm$ 17.60 & \textit{Spitzer}/MIPS & \citet{moor2011} \\
HIP 22226 & 870 & $<$15.00 & APEX/LABOCA & \citet{nilsson2010} \\
HIP 23451 & 24 & 226.20 $\pm$ 2.26 & \textit{Spitzer}/MIPS & \citet{ballering2013} \\
HIP 23451 & 70 & 1003.00 $\pm$ 50.55 & \textit{Spitzer}/MIPS & \citet{ballering2013} \\
HIP 23451 & 70 & 1038.00 $\pm$ 29.00 & \textit{Herschel}/PACS & \citet{donaldson2013} \\
HIP 23451 & 100 & 770.00 $\pm$ 22.00 & \textit{Herschel}/PACS & \citet{donaldson2013} \\
HIP 23451 & 160 & 403.00 $\pm$ 20.00 & \textit{Herschel}/PACS & \citet{donaldson2013} \\
HIP 23451 & 160 & $<$460.00 & \textit{Spitzer}/MIPS & \citet{maness2008} \\
HIP 23451 & 250 & 153.00 $\pm$ 12.00 & \textit{Herschel}/SPIRE & \citet{donaldson2013} \\
HIP 23451 & 350 & 71.00 $\pm$ 8.00 & \textit{Herschel}/SPIRE & \citet{donaldson2013} \\
HIP 23451 & 500 & 45.00 $\pm$ 7.00 & \textit{Herschel}/SPIRE & \citet{donaldson2013} \\
HIP 23451 & 870 & $<$19.50 & APEX/LABOCA & \citet{nilsson2010} \\
HIP 23451 & 1200 & 3.14 $\pm$ 0.82 & IRAM/MAMBO2 & \citet{meeus2012} \\
HIP 23451 & 1300 & 3.10 $\pm$ 0.74 & SMA & \citet{meeus2012} \\
HIP 23451 & 1300 & 5.10 $\pm$ 1.10 & CARMA & \citet{maness2008} \\
HIP 26453 & 24 & 55.16 $\pm$ 0.54 & \textit{Spitzer}/MIPS & \citet{ballering2013} \\
HIP 26453 & 70 & 123.30 $\pm$ 7.80 & \textit{Spitzer}/MIPS & \citet{ballering2013} \\
HIP 26453 & 160 & $<$42.00 & \textit{Spitzer}/MIPS & \citet{hillenbrand2008} \\
HIP 26453 & 1200 & $<$45.00 & SEST & \citet{carpenter2005} \\
HIP 26796 & 24 & 37.66 $\pm$ 0.39 & \textit{Spitzer}/MIPS & \citet{ballering2013} \\
HIP 26796 & 70 & 47.92 $\pm$ 4.19 & \textit{Spitzer}/MIPS & \citet{ballering2013} \\
HIP 34276 & 24 & 34.76 $\pm$ 0.37 & \textit{Spitzer}/MIPS & \citet{ballering2013} \\
HIP 34276 & 70 & 348.70 $\pm$ 18.23 & \textit{Spitzer}/MIPS & \citet{ballering2013} \\
HIP 34276 & 100 & 297.00 $\pm$ 8.90 & \textit{Herschel}/PACS & \citet{vican2016} \\
HIP 34276 & 160 & 200.00 $\pm$ 11.00 & \textit{Herschel}/PACS & \citet{vican2016} \\
HIP 36948 & 24 & 45.24 $\pm$ 0.47 & \textit{Spitzer}/MIPS & \citet{ballering2013} \\
HIP 36948 & 70 & 636.20 $\pm$ 32.40 & \textit{Spitzer}/MIPS & \citet{ballering2013} \\
HIP 36948 & 160 & 502.60 $\pm$ 160.10 & \textit{Spitzer}/MIPS & \citet{hillenbrand2008} \\
HIP 36948 & 350 & 95.00 $\pm$ 12.00 & CSO & \citet{roccatagliata2009} \\
HIP 36948 & 870 & $<$18.00 & APEX/LABOCA & \citet{nilsson2010} \\
HIP 36948 & 1200 & $<$102.00 & SEST & \citet{carpenter2005} \\
HIP 36948 & 1300 & 7.98 $\pm$ 0.80 & SMA & \citet{steele2016} \\
HIP 36948 & 1300 & 7.20 $\pm$ 0.30 & SMA & \citet{ricarte2013} \\
HIP 36948 & 9000 & 0.06 $\pm$ 0.01 & VLA & \citet{macgregor2016} \\
HIP 41152 & 24 & 83.30 $\pm$ 0.82 & \textit{Spitzer}/MIPS & \citet{ballering2013} \\
HIP 41152 & 70 & 209.70 $\pm$ 12.17 & \textit{Spitzer}/MIPS & \citet{ballering2013} \\
HIP 41152 & 100 & 181.30 $\pm$ 4.80 & \textit{Herschel}/PACS & \citet{morales2013} \\
HIP 41152 & 160 & 106.70 $\pm$ 3.90 & \textit{Herschel}/PACS & \citet{morales2013} \\
HIP 41373 & 24 & 57.73 $\pm$ 0.59 & \textit{Spitzer}/MIPS & \citet{ballering2013} \\
HIP 41373 & 70 & 136.70 $\pm$ 11.69 & \textit{Spitzer}/MIPS & \citet{ballering2013} \\
HIP 41373 & 100 & 120.50 $\pm$ 4.10 & \textit{Herschel}/PACS & \citet{morales2013} \\
HIP 41373 & 160 & 46.90 $\pm$ 8.70 & \textit{Herschel}/PACS & \citet{morales2013} \\
HIP 45167 & 24 & 46.51 $\pm$ 0.49 & \textit{Spitzer}/MIPS & \citet{ballering2013} \\
HIP 45167 & 70 & 82.85 $\pm$ 5.37 & \textit{Spitzer}/MIPS & \citet{ballering2013} \\
HIP 47135 & 24 & 12.93 $\pm$ 0.13 & \textit{Spitzer}/MIPS & \citet{ballering2013} \\
HIP 47135 & 70 & 36.90 $\pm$ 3.27 & \textit{Spitzer}/MIPS & \citet{ballering2013} \\
HIP 48423 & 24 & 36.88 $\pm$ 0.38 & \textit{Spitzer}/MIPS & \citet{ballering2013} \\
HIP 48423 & 70 & 33.09 $\pm$ 4.36 & \textit{Spitzer}/MIPS & \citet{ballering2013} \\
HIP 48423 & 160 & $<$44.80 & \textit{Spitzer}/MIPS & \citet{hillenbrand2008} \\
HIP 48423 & 350 & $<$17.40 & CSO & \citet{roccatagliata2009} \\
HIP 48423 & 1200 & $<$2.70 & IRAM & \citet{roccatagliata2009} \\
HIP 55485 & 24 & 43.52 $\pm$ 0.45 & \textit{Spitzer}/MIPS & \citet{ballering2013} \\
HIP 55485 & 70 & 34.45 $\pm$ 6.17 & \textit{Spitzer}/MIPS & \citet{ballering2013} \\
HIP 58720 & 24 & 111.70 $\pm$ 1.10 & \textit{Spitzer}/MIPS & \citet{ballering2013} \\
HIP 58720 & 70 & 98.53 $\pm$ 6.32 & \textit{Spitzer}/MIPS & \citet{ballering2013} \\
HIP 60074 & 24 & 62.86 $\pm$ 0.64 & \textit{Spitzer}/MIPS & \citet{ballering2013} \\
HIP 60074 & 70 & 782.20 $\pm$ 39.62 & \textit{Spitzer}/MIPS & \citet{ballering2013} \\
HIP 60074 & 350 & 319.00 $\pm$ 45.00 & CSO & \citet{corder2009} \\
HIP 60074 & 450 & 130.00 $\pm$ 40.00 & JCMT/SCUBA & \citet{williams2004} \\
HIP 60074 & 850 & 20.00 $\pm$ 4.00 & JCMT/SCUBA & \citet{williams2004} \\
HIP 60074 & 880 & 36.00 $\pm$ 1.00 & SMA & \citet{hughes2011} \\
HIP 60074 & 1250 & 12.50 $\pm$ 1.30 & ALMA & \citet{ricci2015} \\
HIP 60074 & 1300 & 10.40 $\pm$ 1.40 & CARMA & \citet{corder2009} \\
HIP 60074 & 3000 & 1.42 $\pm$ 0.23 & OVRO & \citet{carpenter2005} \\
HIP 60074 & 6800 & 0.17 $\pm$ 0.03 & ATCA & \citet{ricci2015b} \\
HIP 61684 & 24 & 40.59 $\pm$ 0.42 & \textit{Spitzer}/MIPS & \citet{ballering2013} \\
HIP 61684 & 70 & 69.02 $\pm$ 6.09 & \textit{Spitzer}/MIPS & \citet{ballering2013} \\
HIP 62657 & 24 & 42.34 $\pm$ 0.43 & \textit{Spitzer}/MIPS & \citet{ballering2013} \\
HIP 62657 & 70 & 214.80 $\pm$ 13.39 & \textit{Spitzer}/MIPS & \citet{ballering2013} \\
HIP 62657 & 70 & 205.00 $\pm$ 4.00 & \textit{Herschel}/PACS & \citet{draper2016b} \\
HIP 62657 & 160 & 145.00 $\pm$ 6.00 & \textit{Herschel}/PACS & \citet{draper2016b} \\
HIP 62657 & 1240 & 1.29 $\pm$ 0.11 & ALMA & \citet{lieman-sifry2016} \\
HIP 73145 & 24 & 166.60 $\pm$ 1.67 & \textit{Spitzer}/MIPS & \citet{ballering2013} \\
HIP 73145 & 70 & 659.20 $\pm$ 33.40 & \textit{Spitzer}/MIPS & \citet{ballering2013} \\
HIP 73145 & 70 & 738.70 $\pm$ 52.50 & \textit{Herschel}/PACS & \citet{moor2015b} \\
HIP 73145 & 100 & 637.00 $\pm$ 45.50 & \textit{Herschel}/PACS & \citet{moor2015b} \\
HIP 73145 & 160 & 382.30 $\pm$ 27.90 & \textit{Herschel}/PACS & \citet{moor2015b} \\
HIP 73145 & 250 & 156.40 $\pm$ 11.50 & \textit{Herschel}/SPIRE & \citet{moor2015b} \\
HIP 73145 & 350 & 84.30 $\pm$ 8.30 & \textit{Herschel}/SPIRE & \citet{moor2015b} \\
HIP 73145 & 500 & 35.40 $\pm$ 8.90 & \textit{Herschel}/SPIRE & \citet{moor2015b} \\
HIP 73145 & 870 & $<$13.20 & APEX/LABOCA & \citet{nilsson2010} \\
HIP 73145 & 1240 & 2.90 $\pm$ 0.15 & ALMA & \citet{lieman-sifry2016} \\
HIP 74499 & 24 & 21.42 $\pm$ 0.27 & \textit{Spitzer}/MIPS & \citet{ballering2013} \\
HIP 74499 & 70 & 118.80 $\pm$ 11.85 & \textit{Spitzer}/MIPS & \citet{ballering2013} \\
HIP 75077 & 24 & 15.64 $\pm$ 0.24 & \textit{Spitzer}/MIPS & \citet{ballering2013} \\
HIP 75077 & 70 & 36.80 $\pm$ 4.64 & \textit{Spitzer}/MIPS & \citet{ballering2013} \\
HIP 75210 & 24 & 42.33 $\pm$ 0.46 & \textit{Spitzer}/MIPS & \citet{ballering2013} \\
HIP 75210 & 70 & 24.20 $\pm$ 3.74 & \textit{Spitzer}/MIPS & \citet{ballering2013} \\
HIP 76736 & 24 & 80.67 $\pm$ 0.82 & \textit{Spitzer}/MIPS & \citet{ballering2013} \\
HIP 76736 & 70 & 560.60 $\pm$ 29.29 & \textit{Spitzer}/MIPS & \citet{ballering2013} \\
HIP 77432 & 24 & 10.28 $\pm$ 0.19 & \textit{Spitzer}/MIPS & \citet{ballering2013} \\
HIP 77432 & 70 & $<$27.93 & \textit{Spitzer}/MIPS & \citet{ballering2013} \\
HIP 77464 & 24 & 78.32 $\pm$ 0.81 & \textit{Spitzer}/MIPS & \citet{ballering2013} \\
HIP 77464 & 70 & 224.90 $\pm$ 14.34 & \textit{Spitzer}/MIPS & \citet{ballering2013} \\
HIP 78043 & 24 & 13.14 $\pm$ 0.16 & \textit{Spitzer}/MIPS & \citet{ballering2013} \\
HIP 78043 & 70 & 75.36 $\pm$ 6.32 & \textit{Spitzer}/MIPS & \citet{ballering2013} \\
HIP 78043 & 1240 & 0.34 $\pm$ 0.07 & ALMA & \citet{lieman-sifry2016} \\
HIP 78045 & 24 & 60.25 $\pm$ 0.63 & \textit{Spitzer}/MIPS & \citet{ballering2013} \\
HIP 78045 & 70 & 42.95 $\pm$ 13.02 & \textit{Spitzer}/MIPS & \citet{ballering2013} \\
HIP 79516 & 24 & 52.88 $\pm$ 0.57 & \textit{Spitzer}/MIPS & \citet{ballering2013} \\
HIP 79516 & 70 & 317.20 $\pm$ 25.71 & \textit{Spitzer}/MIPS & \citet{ballering2013} \\
HIP 79516 & 1240 & 1.85 $\pm$ 0.12 & ALMA & \citet{lieman-sifry2016} \\
HIP 79742 & 24 & 31.15 $\pm$ 0.34 & \textit{Spitzer}/MIPS & \citet{ballering2013} \\
HIP 79742 & 70 & 173.00 $\pm$ 16.28 & \textit{Spitzer}/MIPS & \citet{ballering2013} \\
HIP 79742 & 1240 & 0.88 $\pm$ 0.09 & ALMA & \citet{lieman-sifry2016} \\
HIP 83187 & 24 & 80.34 $\pm$ 0.80 & \textit{Spitzer}/MIPS & \citet{ballering2013} \\
HIP 83187 & 70 & 156.60 $\pm$ 13.89 & \textit{Spitzer}/MIPS & \citet{ballering2013} \\
HIP 85537 & 24 & 103.10 $\pm$ 1.03 & \textit{Spitzer}/MIPS & \citet{ballering2013} \\
HIP 85537 & 70 & 229.10 $\pm$ 11.90 & \textit{Spitzer}/MIPS & \citet{ballering2013} \\
HIP 85537 & 70 & 230.00 $\pm$ 20.00 & \textit{Herschel}/PACS & \citet{pascual2016} \\
HIP 85537 & 160 & 150.00 $\pm$ 10.00 & \textit{Herschel}/PACS & \citet{pascual2016} \\
HIP 85537 & 1300 & $<$0.75 & SMA & \citet{meeus2012} \\
HIP 85922 & 24 & 87.64 $\pm$ 0.88 & \textit{Spitzer}/MIPS & \citet{ballering2013} \\
HIP 85922 & 70 & 35.21 $\pm$ 9.07 & \textit{Spitzer}/MIPS & \citet{ballering2013} \\
HIP 87108 & 24 & 434.10 $\pm$ 4.34 & \textit{Spitzer}/MIPS & \citet{ballering2013} \\
HIP 87108 & 70 & 1166.00 $\pm$ 58.37 & \textit{Spitzer}/MIPS & \citet{ballering2013} \\
HIP 87108 & 70 & 1222.00 $\pm$ 85.00 & \textit{Herschel}/PACS & \citet{moor2015} \\
HIP 87108 & 100 & 1051.00 $\pm$ 73.00 & \textit{Herschel}/PACS & \citet{moor2015} \\
HIP 87108 & 160 & 587.00 $\pm$ 44.00 & \textit{Herschel}/PACS & \citet{moor2015} \\
HIP 87108 & 250 & 177.00 $\pm$ 12.00 & \textit{Herschel}/SPIRE & \citet{moor2015} \\
HIP 87108 & 350 & 98.00 $\pm$ 10.00 & \textit{Herschel}/SPIRE & \citet{moor2015} \\
HIP 87108 & 450 & $<$69.00 & JCMT/SCUBA-2 & \citet{panic2013} \\
HIP 87108 & 500 & 56.00 $\pm$ 11.00 & \textit{Herschel}/SPIRE & \citet{moor2015} \\
HIP 87108 & 850 & 6.40 $\pm$ 1.00 & JCMT/SCUBA-2 & \citet{panic2013} \\
HIP 87108 & 870 & $<$15.60 & APEX/LABOCA & \citet{nilsson2010} \\
HIP 94114 & 24 & 240.00 $\pm$ 2.40 & \textit{Spitzer}/MIPS & \citet{ballering2013} \\
HIP 94114 & 70 & 70.92 $\pm$ 5.92 & \textit{Spitzer}/MIPS & \citet{ballering2013} \\
HIP 95270 & 24 & 230.30 $\pm$ 2.31 & \textit{Spitzer}/MIPS & \citet{ballering2013} \\
HIP 95270 & 70 & 1776.00 $\pm$ 89.31 & \textit{Spitzer}/MIPS & \citet{ballering2013} \\
HIP 95270 & 70 & 1827.00 $\pm$ 183.00 & \textit{Herschel}/PACS & \citet{lebreton2012} \\
HIP 95270 & 90 & 1410.00 $\pm$ 140.00 & \textit{ISO} & \citet{moor2006} \\
HIP 95270 & 100 & 1337.00 $\pm$ 134.00 & \textit{Herschel}/PACS & \citet{lebreton2012} \\
HIP 95270 & 160 & 767.00 $\pm$ 153.00 & \textit{Herschel}/PACS & \citet{lebreton2012} \\
HIP 95270 & 160 & 770.00 $\pm$ 90.00 & \textit{Spitzer}/MIPS & \citet{schneider2006} \\
HIP 95270 & 170 & 736.00 $\pm$ 192.00 & \textit{ISO} & \citet{moor2006} \\
HIP 95270 & 870 & 51.70 $\pm$ 6.20 & APEX/LABOCA & \citet{nilsson2009} \\
HIP 95270 & 1300 & 7.90 $\pm$ 0.20 & ALMA & \citet{marino2016} \\
HIP 95270 & 3190 & 0.72 $\pm$ 0.25 & ATCA & \citet{lebreton2012} \\
HIP 95270 & 6800 & 0.14 $\pm$ 0.02 & ATCA & \citet{ricci2015b} \\
HIP 101612 & 24 & 204.10 $\pm$ 2.05 & \textit{Spitzer}/MIPS & \citet{ballering2013} \\
HIP 101612 & 70 & 654.00 $\pm$ 33.53 & \textit{Spitzer}/MIPS & \citet{ballering2013} \\
HIP 101612 & 70 & 629.00 $\pm$ 44.00 & \textit{Herschel}/PACS & \citet{moor2015} \\
HIP 101612 & 100 & 607.00 $\pm$ 43.00 & \textit{Herschel}/PACS & \citet{moor2015} \\
HIP 101612 & 160 & 405.00 $\pm$ 29.00 & \textit{Herschel}/PACS & \citet{moor2015} \\
HIP 101612 & 250 & 145.00 $\pm$ 14.00 & \textit{Herschel}/SPIRE & \citet{moor2015} \\
HIP 101612 & 350 & 70.00 $\pm$ 7.00 & \textit{Herschel}/SPIRE & \citet{moor2015} \\
HIP 101612 & 500 & 34.00 $\pm$ 7.00 & \textit{Herschel}/SPIRE & \citet{moor2015} \\
HIP 101612 & 870 & $<$21.30 & APEX/LABOCA & \citet{nilsson2010} \\
HIP 101800 & 24 & 76.55 $\pm$ 0.77 & \textit{Spitzer}/MIPS & \citet{ballering2013} \\
HIP 101800 & 70 & 73.32 $\pm$ 7.64 & \textit{Spitzer}/MIPS & \citet{ballering2013} \\
HIP 106741 & 24 & 31.58 $\pm$ 0.34 & \textit{Spitzer}/MIPS & \citet{ballering2013} \\
HIP 106741 & 70 & 217.20 $\pm$ 12.98 & \textit{Spitzer}/MIPS & \citet{ballering2013} \\
HIP 106741 & 160 & 185.60 $\pm$ 26.40 & \textit{Spitzer}/MIPS & \citet{moor2011} \\
HIP 106741 & 450 & $<$125.00 & JCMT/SCUBA-2 & \citet{panic2013} \\
HIP 106741 & 850 & 5.30 $\pm$ 1.40 & JCMT/SCUBA-2 & \citet{panic2013} \\
HIP 114189 & 24 & 86.60 $\pm$ 0.87 & \textit{Spitzer}/MIPS & \citet{ballering2013} \\
HIP 114189 & 70 & 610.00 $\pm$ 30.95 & \textit{Spitzer}/MIPS & \citet{ballering2013} \\
HIP 114189 & 70 & 537.00 $\pm$ 15.00 & \textit{Herschel}/PACS & \citet{matthews2014b} \\
HIP 114189 & 90 & 585.00 $\pm$ 41.00 & \textit{ISO} & \citet{moor2006} \\
HIP 114189 & 100 & 687.00 $\pm$ 20.00 & \textit{Herschel}/PACS & \citet{matthews2014b} \\
HIP 114189 & 160 & 555.00 $\pm$ 66.00 & \textit{Spitzer}/MIPS & \citet{su2009} \\
HIP 114189 & 160 & 570.00 $\pm$ 50.00 & \textit{Herschel}/PACS & \citet{matthews2014b} \\
HIP 114189 & 250 & 309.00 $\pm$ 30.00 & \textit{Herschel}/SPIRE & \citet{matthews2014b} \\
HIP 114189 & 350 & 163.00 $\pm$ 30.00 & \textit{Herschel}/SPIRE & \citet{matthews2014b} \\
HIP 114189 & 500 & $<$90.00 & \textit{Herschel}/SPIRE & \citet{matthews2014b} \\
HIP 114189 & 850 & 10.30 $\pm$ 1.80 & JCMT/SCUBA & \citet{williams2006} \\
HIP 114189 & 1340 & 2.80 $\pm$ 0.50 & ALMA & \citet{booth2016} \\
HIP 117452 & 24 & 165.60 $\pm$ 1.61 & \textit{Spitzer}/MIPS & \citet{ballering2013} \\
HIP 117452 & 70 & 54.80 $\pm$ 7.16 & \textit{Spitzer}/MIPS & \citet{ballering2013} \\
HIP 117452 & 100 & 28.89 $\pm$ 2.85 & \textit{Herschel}/PACS & \citet{thureau2014} \\
HIP 117452 & 160 & $<$9.70 & \textit{Herschel}/PACS & \citet{thureau2014} \\
\enddata
\end{deluxetable}

\clearpage

\begin{deluxetable}{ccccc}
\tabletypesize{\scriptsize}
\tablewidth{0pt}
\tablecolumns{5}
\tablecaption{Single-component Fit Results \label{table:onecompfits}}
\tablehead{\colhead{HIP} & \colhead{$r_\text{warm}$} & \colhead{$M_\text{warm}$} & \colhead{$f_\text{warm}$} & \colhead{$c_\text{IRS}$} \\ \colhead{Identifier} & \colhead{(au)} & \colhead{($\times 10^{-5} M_\earth$)} & \colhead{($\times 10^{-5}$)} & \colhead{}}
\startdata
HIP 9902 & 2.80 & 0.64 & 17.00 & 0.90 \\
HIP 14684 & 4.40 & 0.25 & 5.21 & 0.92 \\
HIP 17549 & 10.00 & 11.02 & 8.33 & 0.92 \\
HIP 18217 & 4.50 & 0.48 & 2.65 & 0.95 \\
HIP 18481 & 3.30 & 0.51 & 3.85 & 0.99 \\
HIP 23871 & 7.00 & 1.55 & 2.84 & 0.91 \\
HIP 28103 & 12.20 & 0.79 & 0.82 & 0.92 \\
HIP 41967 & 3.30 & 0.20 & 5.21 & 0.95 \\
HIP 49593 & 5.10 & 0.50 & 2.18 & 0.94 \\
HIP 54879 & 16.20 & 4.94 & 0.98 & 0.98 \\
HIP 57524 & 4.50 & 0.78 & 9.64 & 0.92 \\
HIP 57950 & 6.20 & 2.10 & 10.16 & 0.98 \\
HIP 60348 & 7.10 & 2.51 & 12.24 & 0.86 \\
HIP 62134 & 5.60 & 0.72 & 4.15 & 0.92 \\
HIP 63836 & 3.70 & 0.60 & 8.53 & 0.87 \\
HIP 64053 & 5.30 & 3.09 & 6.03 & 0.93 \\
HIP 64877 & 8.10 & 8.03 & 20.97 & 0.93 \\
HIP 67068 & 2.60 & 0.23 & 5.46 & 0.99 \\
HIP 67230 & 9.90 & 19.87 & 28.98 & 0.92 \\
HIP 70455 & 8.10 & 5.87 & 5.55 & 0.97 \\
HIP 74824 & 7.40 & 2.35 & 4.13 & 0.93 \\
HIP 77315 & 10.20 & 22.31 & 16.16 & 0.89 \\
HIP 79710 & 3.80 & 2.04 & 19.73 & 1.00 \\
HIP 79881 & 8.40 & 0.75 & 1.05 & 0.90 \\
HIP 89770 & 6.00 & 3.22 & 14.85 & 0.93 \\
HIP 95560 & 7.20 & 1.93 & 3.02 & 0.93 \\
HIP 106783 & 10.70 & 2.19 & 1.76 & 0.92 \\
HIP 112542 & 16.40 & 12.76 & 3.21 & 0.92 \\
HIP 115738 & 9.10 & 1.72 & 1.47 & 0.91 \\
\enddata
\end{deluxetable}

\newpage
\begin{deluxetable}{ccccccccc}
\tabletypesize{\scriptsize}
\tablewidth{0pt}
\tablecolumns{9}
\tablecaption{Two-component Fit Results \label{table:twocompfits}}
\tablehead{\colhead{HIP} & \colhead{$r_\text{warm}$} & \colhead{$M_\text{warm}$} & \colhead{$f_\text{warm}$} & \colhead{$T_\text{cold}$} & \colhead{$f_\text{cold}$} & \colhead{$\lambda_0$} & \colhead{$\tilde{\beta}$} & \colhead{$c_\text{IRS}$} \\ \colhead{Identifier} & \colhead{(au)} & \colhead{($\times 10^{-5} M_\earth$)} & \colhead{($\times 10^{-5}$)} & \colhead{(K)} & \colhead{($\times 10^{-5}$)} & \colhead{} & \colhead{} & \colhead{}}
\startdata
HIP 345 & 8.00 & 3.07 & 3.33 & 56.17 & 6.09 & 100.00 & 1.00 & 0.96 \\
HIP 682 & 5.40 & 0.59 & 6.18 & 51.03 & 35.47 & 160.00 & 0.55 & 0.92 \\
HIP 1473 & 6.00 & 0.44 & 0.99 & 101.53 & 0.56 & 237.43 & 0.11 & 0.91 \\
HIP 1481 & 1.10 & 0.04 & 4.91 & 109.63 & 4.14 & 100.00 & 1.00 & 0.93 \\
HIP 2472 & 16.20 & 3.09 & 1.02 & 60.06 & 0.84 & 100.00 & 1.00 & 0.91 \\
HIP 2710 & 4.30 & 0.16 & 1.84 & 51.24 & 11.48 & 100.00 & 1.00 & 0.93 \\
HIP 7805 & 5.40 & 0.95 & 6.42 & 57.67 & 29.95 & 100.00 & 1.00 & 0.92 \\
HIP 7978 & 36.00 & 16.09 & 4.23 & 46.37 & 24.06 & 70.00 & 0.56 & 0.98 \\
HIP 8241 & 6.80 & 1.37 & 2.10 & 57.79 & 7.85 & 154.25 & 1.20 & 0.94 \\
HIP 11360 & 7.10 & 1.28 & 5.34 & 53.79 & 45.39 & 61.15 & 0.61 & 0.99 \\
HIP 11477 & 7.30 & 0.76 & 1.44 & 80.76 & 3.15 & 100.00 & 1.00 & 0.89 \\
HIP 11847 & 22.60 & 160.33 & 77.27 & 67.11 & 120.06 & 71.42 & 0.80 & 0.91 \\
HIP 13141 & 8.50 & 0.66 & 0.90 & 54.75 & 4.68 & 136.05 & 0.93 & 0.91 \\
HIP 18859 & 1.70 & 0.06 & 2.76 & 71.68 & 7.92 & 196.02 & 0.76 & 0.96 \\
HIP 20901 & 11.30 & 2.53 & 2.08 & 62.27 & 3.71 & 100.00 & 1.00 & 0.91 \\
HIP 22192 & 6.30 & 0.40 & 1.32 & 64.70 & 3.63 & 71.42 & 1.12 & 0.94 \\
HIP 22226 & 26.50 & 37.44 & 12.21 & 58.26 & 74.10 & 83.52 & 0.82 & 0.88 \\
HIP 23451 & 3.20 & 5.29 & 62.92 & 80.41 & 482.70 & 87.34 & 0.47 & 0.96 \\
HIP 26453 & 5.80 & 1.40 & 8.31 & 77.43 & 24.96 & 100.00 & 1.00 & 0.91 \\
HIP 26796 & 6.70 & 3.00 & 3.39 & 85.61 & 3.55 & 100.00 & 1.00 & 0.92 \\
HIP 34276 & 12.00 & 1.78 & 1.13 & 61.12 & 25.38 & 71.42 & 0.16 & 1.01 \\
HIP 36948 & 36.10 & 67.92 & 28.64 & 49.64 & 203.69 & 50.00 & 0.41 & 0.93 \\
HIP 41152 & 9.30 & 1.89 & 2.63 & 57.42 & 6.61 & 71.42 & 0.41 & 0.91 \\
HIP 41373 & 3.70 & 0.50 & 3.05 & 73.92 & 8.80 & 100.77 & 0.97 & 0.89 \\
HIP 45167 & 4.00 & 1.16 & 4.69 & 70.44 & 4.96 & 100.00 & 1.00 & 0.94 \\
HIP 47135 & 17.50 & 3.81 & 4.32 & 45.97 & 16.51 & 100.00 & 1.00 & 0.93 \\
HIP 48423 & 3.20 & 0.18 & 5.68 & 73.49 & 7.84 & 100.00 & 1.00 & 0.96 \\
HIP 55485 & 3.70 & 0.65 & 4.23 & 75.57 & 2.82 & 100.00 & 1.00 & 0.94 \\
HIP 58720 & 4.50 & 2.87 & 7.19 & 112.95 & 7.08 & 100.00 & 1.00 & 0.91 \\
HIP 60074 & 8.70 & 1.05 & 4.74 & 48.28 & 93.20 & 349.97 & 0.71 & 1.02 \\
HIP 61684 & 5.90 & 4.16 & 18.42 & 86.22 & 34.30 & 100.00 & 1.00 & 0.91 \\
HIP 62657 & 20.70 & 125.70 & 79.60 & 56.17 & 135.68 & 166.66 & 0.69 & 0.86 \\
HIP 73145 & 6.10 & 22.66 & 77.36 & 71.83 & 240.75 & 194.79 & 0.73 & 1.00 \\
HIP 74499 & 5.80 & 2.27 & 16.80 & 66.50 & 83.11 & 100.00 & 1.00 & 0.92 \\
HIP 75077 & 6.70 & 1.00 & 2.01 & 50.94 & 4.88 & 100.00 & 1.00 & 1.01 \\
HIP 75210 & 3.20 & 1.60 & 7.62 & 120.77 & 3.52 & 100.00 & 1.00 & 0.91 \\
HIP 76736 & 7.30 & 4.41 & 8.96 & 59.72 & 41.58 & 100.00 & 1.00 & 0.98 \\
HIP 77432 & 0.50 & 0.03 & 8.68 & 101.45 & 13.05 & 100.00 & 1.00 & 0.91 \\
HIP 77464 & 6.70 & 0.71 & 1.67 & 60.46 & 7.54 & 100.00 & 1.00 & 0.97 \\
HIP 78043 & 6.00 & 2.78 & 13.70 & 60.41 & 62.21 & 114.20 & 0.70 & 1.02 \\
HIP 78045 & 2.90 & 0.26 & 2.25 & 91.42 & 2.03 & 100.00 & 1.00 & 0.96 \\
HIP 79516 & 24.90 & 295.37 & 104.69 & 57.92 & 205.73 & 50.53 & 0.48 & 0.91 \\
HIP 79742 & 14.30 & 37.63 & 39.87 & 69.64 & 182.99 & 62.45 & 0.42 & 0.96 \\
HIP 83187 & 17.30 & 5.02 & 2.25 & 53.10 & 5.01 & 100.00 & 1.00 & 0.95 \\
HIP 85537 & 22.80 & 10.67 & 2.25 & 44.59 & 5.73 & 93.26 & 0.98 & 0.95 \\
HIP 85922 & 1.50 & 0.14 & 4.11 & 99.92 & 1.77 & 100.00 & 1.00 & 0.93 \\
HIP 87108 & 12.90 & 5.04 & 2.72 & 64.53 & 7.75 & 127.18 & 1.01 & 0.98 \\
HIP 94114 & 5.70 & 0.91 & 2.12 & 77.84 & 0.41 & 100.00 & 1.00 & 0.96 \\
HIP 95270 & 33.80 & 324.22 & 72.14 & 60.43 & 215.48 & 55.16 & 0.50 & 1.00 \\
HIP 101612 & 17.10 & 2.57 & 1.38 & 43.90 & 7.87 & 71.42 & 0.96 & 0.89 \\
HIP 101800 & 12.90 & 3.36 & 2.03 & 61.59 & 1.58 & 100.00 & 1.00 & 0.92 \\
HIP 106741 & 8.00 & 0.59 & 2.09 & 52.27 & 33.65 & 152.58 & 0.62 & 0.96 \\
HIP 114189 & 3.90 & 0.46 & 4.36 & 44.51 & 29.16 & 200.82 & 1.18 & 1.09 \\
HIP 117452 & 3.30 & 0.32 & 1.68 & 130.00 & 0.88 & 100.00 & 1.00 & 0.94 \\
\enddata
\end{deluxetable}

\end{document}